\documentclass[smallextended]{svjour3}       

\usepackage{moreverb}

\usepackage{dsfont,url,bbm,algorithm}

\usepackage{mathtools}

\newcommand\BibTeX{{\rmfamily B\kern-.05em \textsc{i\kern-.025em b}\kern-.08em
T\kern-.1667em\lower.7ex\hbox{E}\kern-.125emX}}

\def\volumeyear{2019}

\usepackage{graphicx} 
\usepackage{xcolor}
\usepackage{amsfonts}
\usepackage{amssymb}

\graphicspath{{./figures/}}

\def\PP{{{\rm l}\kern - .15em {\rm P} }}
\def\PN2{{\PP_{N}-\PP_{N-2}}}

\newcommand{\bphi}{\boldsymbol{\varphi}}

\newcommand{\bu}{\boldsymbol{u}}

\newcommand{\bur}{{\boldsymbol{u}}_r}

\newcommand{\bv}{\boldsymbol{v}}

\newcommand{\bx}{\boldsymbol{x}}

\newcommand{\bXr}{{\bf X}^r}

\newcommand{\deleted}[1]{{}}

\usepackage{siunitx}
\definecolor{vargreen}{rgb}{0.0, 0.5, 0.0}

\begin{document}


\title{Lagrangian Reduced Order Modeling of Finite Time Lyapunov Exponents
\thanks{{National Science Foundation }{DMS-1821145}}
}

\titlerunning{Lagrangian ROM of FTLE}

\author{
	Xuping Xie\and
	Peter J. Nolan\and
	Shane D. Ross\and
	Changhong Mou\and 
	Traian Iliescu}

\institute{
    X. Xie \at
    Courant Institute of Mathematical Sciences, New York University,\\
    New York, NY, 10012, USA;\\
    \email{xxie@nyu.edu}  
\and
        P. J. Nolan  \at
        Engineering Mechanics Program, Virginia Tech,\\
        Blacksburg, VA, 24061, USA;\\
        \email{pnolan86@vt.edu}  
\and
            S. D. Ross \at
                Department of Aerospace and Ocean Engineering, Virginia Tech,\\
                Blacksburg, VA, 24061, USA;\\
                \email{sdross@vt.edu}  
\and
                C. Mou \at 
                Department of Mathematics, Virginia Tech,\\
                Blacksburg, VA, 24061, USA;\\
                \email{cmou@vt.edu}  
\and
                    T. Iliescu \at 
                    Department of Mathematics, Virginia Tech,\\
                    Blacksburg, VA, 24061, USA;\\
                    \email{iliescu@vt.edu}  
}

	
	
	
\date{Received: date / Accepted: date}

\maketitle

\begin{abstract}
There are two main strategies for improving the projection-based reduced order model (ROM) accuracy:
(i) improving the ROM, i.e., adding new terms to the standard ROM; and 
(ii) improving the ROM basis, i.e., constructing ROM bases that yield more accurate ROMs.
In this paper, we use the latter.
We propose new Lagrangian inner products that we use together with Eulerian and Lagrangian data to construct new Lagrangian ROMs.
We show that the new Lagrangian ROMs are orders of magnitude more accurate than the standard Eulerian ROMs, i.e., ROMs that use standard Eulerian inner product and data to construct the ROM basis.
Specifically, for the quasi-geostrophic equations, we show that the new Lagrangian ROMs are more accurate than the standard Eulerian ROMs in approximating not only Lagrangian fields (e.g., the finite time Lyapunov exponent (FTLE)), but also Eulerian fields (e.g., the streamfunction).
We emphasize that the new Lagrangian ROMs do not employ any closure modeling to model the effect of discarded modes (which is standard procedure for low-dimensional ROMs of complex nonlinear systems).
Thus, the dramatic increase in the new Lagrangian ROMs' accuracy is entirely due to the novel Lagrangian inner products used to build the Lagrangian ROM basis.

\end{abstract}




\section{Introduction}
	\label{sec:introduction}

Projection-based reduced order models (ROMs) have been successful in the numerical simulation of fluid flows~\cite{brunton2019data,hesthaven2015certified,HLB96,quarteroni2015reduced,noack2011reduced,taira2019modal}.
To approximate the dynamics of a given flow variable $\bu$, the ROM strategy proceeds as follows:
(i) Choose modes $\{ \bphi_1, \ldots, \bphi_R \}$, which represent the recurrent spatial structures in the flow.
(ii) Choose the dominant modes $\{ \bphi_1, \ldots, \bphi_r \}$, $r \leq R$, as  basis functions for the ROM.
(iii) Use a Galerkin truncation $\bur = \sum_{j=1}^{r} a_j \, \bphi_j$.
(iv) Replace $\bu$ with $\bur$ in the underlying equations.
(v) Use a Galerkin projection of the PDE obtained in step (iv) onto the ROM space $\bXr := \text{span} \{ \bphi_1, \ldots, \bphi_r \}$ to obtain a low-dimensional dynamical system, which represents the ROM.
(vi) In an offline stage, compute the ROM operators. 
(vii) In an online stage, repeatedly use the ROM (for various parameter settings and/or longer time intervals).
The low-dimensional ROMs can decrease the computational cost of traditional full order models (FOMs) by orders of magnitude.
ROMs, however, can be inaccurate in the numerical simulation of complex flows~\cite{HLB96,noack2011reduced}.
There are two main approaches to increasing ROM accuracy:

The first approach is to {\it improve the model}, i.e., to add new terms to the standard projection-based ROM.
Classical examples are ROM closure (see, e.g.,~\cite{chekroun2019variational,hijazi2019data,parish2020adjoint,reyes2019projection,xie2018data,mou2020dd-vms-rom}) and ROM stabilization (see, e.g.,~\cite{azaiez2017streamline,grimberg2020stability,gunzburger2019evolve,xie2018numerical}).
We will not follow this approach in this paper.

The second approach to improving the ROM accuracy is to {\it improve the ROM basis}, i.e., to construct ROM bases that yield more accurate ROMs.
One of the earliest examples in this class is the $H^1$-basis proposed in~\cite{iollo2000stability}, in which the $H^1$ inner product is used instead of the standard $L^2$ inner product to construct the ROM basis in order to increase the ROM stability.
Similarly, an enstrophy-based ROM for rotational flows was proposed in~\cite{sengupta2015enstrophy}, in which the inner product is defined for vorticity instead of velocity.
Other examples in this class are the ROM bases proposed for compressible flows~\cite{barone2009stable,iollo2000stability,kalashnikova2014reduced,rowley2004model}, which use new inner products and different flow variables to construct the ROM basis (see~\cite{kaptanoglu2020physics} for recent work on magnetohydrodynamics).
Improved ROM bases were also proposed for data assimilation~\cite{daescu2008dual}.
The inner products used to define these improved ROM bases are Eulerian inner products, i.e., they are defined only for Eulerian data.
To our knowledge, there are only a few 
Lagrangian inner products, i.e., inner products that are defined on both Eulerian and Lagrangian data, that have been recently proposed.
In~\cite{lu2020lagrangian,mojgani2017lagrangian}, the authors proposed inner products that are defined for velocity (which is an Eulerian variable) and the Lagrangian mesh coordinates (which are Lagrangian variables).

In this paper, we use the second strategy to improve the ROM accuracy, i.e., we propose improved ROM bases.
Specifically, we propose new Lagrangian inner products that utilize both Eulerian and Lagrangian data.
In the new Lagrangian inner products, Lagrangian data steers the resulting Lagrangian ROM basis toward an accurate approximation of Lagrangian quantities, whereas Eulerian data helps the Lagrangian ROM basis yield an accurate approximation of Eulerian quantities.
We emphasize that the Lagrangian inner products that we propose are different from the Lagrangian inner products in~\cite{lu2020lagrangian,mojgani2017lagrangian}.
As Lagrangian data, we use the finite time Lyapunov exponents (FTLE) field, whereas~\cite{lu2020lagrangian,mojgani2017lagrangian} use the Lagrangian mesh coordinates. 
To construct the new Lagrangian ROMs, we utilize the new Lagrangian inner products, the resulting Lagrangian ROM bases, and the Galerkin projection.
In the numerical simulation of the quasi-geostrophic equations~\cite{cushman2011introduction,Maj03,MW06,vallis2006atmospheric}  (which model large scale ocean circulation), the new Lagrangian ROMs are orders of magnitude more accurate than standard Eulerian ROMs (i.e., ROMs that use standard Eulerian data and inner products to build the ROM bases).
Furthermore, the new Lagrangian ROMs are more accurate than the standard Eulerian ROMs in approximating not only Lagrangian fields (e.g., the finite time Lyapunov exponent (FTLE)), but also Eulerian fields (e.g., the streamfunction).

For complex nonlinear systems, it is well known that the low-dimensional ROMs generally need to be equipped with a closure model (see, e.g.,~\cite{chekroun2019variational,chen2018conditional,hijazi2019data,majda2018model,majda2018strategies,parish2020adjoint,reyes2019projection,xie2018data,mou2020dd-vms-rom}) or a stabilization mechanism (see, e.g.,~\cite{azaiez2017streamline,grimberg2020stability,gunzburger2019evolve}) to model the effect of the discarded ROM modes. 
We emphasize, however, that we investigate the new Lagrangian ROMs without any closure or stabilization (a challenging test) in order to separate the ROM closure problem from the ROM basis generation, which is the main focus of our paper.
This allows us to conclude that the orders of magnitude increase in the new Lagrangian ROMs' accuracy over the standard Eulerian ROMs' accuracy is entirely due to the new Lagrangian inner products used to build the Lagrangian ROMs' bases.
Of course, we envision that using closure modeling in addition to the novel Lagrangian inner product will increase even further the Lagrangian ROMs' accuracy.

\bigskip
	
The rest of the paper is organized as follows:	
In Section~\ref{sec:lagrangian-roms}, we propose the novel Lagrangian inner products and construct the new Lagrangian ROMs.
In Section~\ref{sec:numerical-results}, for the quasi-geostrophic equations, we show that the new Lagrangian ROMs increase the numerical accuracy of standard Eulerian ROMs by orders of magnitude.
Finally, in Section~\ref{sec:conclusions}, we present conclusions and outline future research directions.

\section{Lagrangian Reduced Order Models}
	\label{sec:lagrangian-roms}

In this section, we propose two new {\it Lagrangian ROMs}, which we build as follows:
First, we use the quasi-geostrophic equations (QGE) as a mathematical model (although general models, e.g., the Boussinesq and Navier-Stokes equations, could be used instead). 
Next, we perform numerical simulations to generate the QGE velocity field (which is Eulerian data) and the QGE finite time Lyapunov exponent (FTLE) field (which is Lagrangian data).
(We note that other Lagrangian fields could be used instead of the FTLE field.) 
Finally, we propose two new {\it Lagrangian inner products} that use both Eulerian and Lagrangian data to construct new Lagrangian ROM bases, which yield the new Lagrangian ROMs.
For comparison purposes, we also outline  standard Eulerian ROMs~\cite{crommelin2004strategies,mou2020data,san2015stabilized,selten1995efficient,strazzullo2018model}, which use only Eulerian data (i.e., the velocity field) to generate the ROM basis.
In Section~\ref{sec:numerical-results}, we compare the new Lagrangian ROMs with the standard Eulerian ROM in the numerical simulation of the QGE.

\bigskip

The QGE~\cite{cushman2011introduction,Maj03,MW06,vallis2006atmospheric} are written as the following PDE:
\begin{subequations}
	\begin{align}
  		\frac{\partial \omega}{\partial t} 
		+ J(\omega, \psi) 
		- Ro^{-1} \frac{\partial \psi}{\partial x}
		&= Re^{-1} \, \Delta \omega 
		+ Ro^{-1} F \, ,
		\label{eqn:qge-1}\\
  		\omega 
		&= - \Delta \psi \, ,
		\label{eqn:qge-2}
	\end{align} 
	\label{eqn:qge}
\end{subequations}
where $\omega$ is the vorticity, $\psi$ is the streamfunction, $Re$ is the Reynolds number, and $Ro$ is the Rossby number, $J(\omega,\psi) =\omega_x\psi_y-\omega_y\psi_x$ is the Jacobian term, and $F$ is the forcing term.
The velocity can be computed from the streamfunction according to the following formula:
\begin{eqnarray}
	\mathbf{v}
	= \biggl( \frac{\partial \psi}{\partial y} , - \frac{\partial \psi}{\partial x} \biggr) \, .
	\label{eqn:streamfunction2velocity}
\end{eqnarray}
Details regarding the parameters and nondimensionalization of the QGE~\eqref{eqn:qge} are given in, e.g.,~\cite{foster2013finite,monteiro2015numerical,mou2020data,san2015stabilized,san2011approximate}.

\subsection{Finite Time Lyapunov Exponents (FTLE) Computation}
	\label{sec:ftle}

Next, we briefly describe the calculation of the FTLE field (see, e.g.,~\cite{haller2015lagrangian} for  details).
Given a velocity field $\bv(\bx,t)$ (e.g., the QGE velocity field~\eqref{eqn:streamfunction2velocity}), the trajectories are obtained from the solutions of the ODE system $\dot{\bx}=\bv(\bx,t)$. 
Each trajectory $\bx(t;t_0,\bx_0)$ is a function of time, but it also depends on the initial position $\bx_0$ and the initial time $t_0$. 
For a given initial time $t_0$ and a given final time $t$, the flow map is the function
\begin{equation}
\bx_0\mapsto\phi_{t_0}^t(\bx_0)=\bx(t;t_0,\bx_0)\; .
\end{equation} 

Consider two particles, simultaneously released at time $t_0$; one at location $\bx$, the other at location $\bx+\delta \bx $.
Under the effect of the flow map,  
the small displacement vector between  two particles, $\delta \bx$, changes. 
After an elapsed time $T = t - t_{0}$, the new vector between the two particles is
\begin{equation*} \label{eq:max_separation_fundamental_eq_Jacobi}
	\delta \bx\left ( t_{0}+T \right )
	=\phi _{t_{0}}^{t_{0}+T}\left ( \bx + \delta \bx \right )-\phi _{t_{0}}^{t_{0}+T}\left ( \bx \right )
	= D\phi _{t_{0}}^{t_{0}+T}\left ( \bx \right ) \, \delta \bx + \mathcal{O}\left ( \left \| \delta \bx \left ( t_{0} \right ) \right \|^{2} \right ),
\end{equation*}
where 
$D\phi _{t_{0}}^{t_{0}+T} = {d\phi_{t_{0}}^{t_{0}+T}\left ( {\bx} \right )}/{d\bx}$ is the Jacobian of the flow map, and  $\left \| \cdot  \right \|$ is the usual Euclidean norm.
Consider the right Cauchy-Green strain tensor,
\begin{equation}
C\left ( \bx,t_{0},T \right ) =D\phi _{t_{0}}^{t_{0}+T}\left ( \bx \right )^{\intercal} D\phi _{t_{0}}^{t_{0}+T}\left ( \bx \right )  \, .
	\label{eqn:ftle-background-1}
\end{equation}
The maximum possible separation between the released particles after a time interval $T$, assuming a sufficiently small initial distance $\left \| \delta \bx \left ( t_{0} \right ) \right \|$, is
\begin{equation} \label{eq:max_separation_CauchyGreen}
	\max \left \| \delta \bx \left ( t_{0}+T \right ) \right \|=\sqrt{\mu_{\max}\left ( C\left ( \bx ,t_{0},T \right )\right )}\left \| \delta \bx \left ( t_{0} \right ) \right \| \, ,
\end{equation}    
where $\mu_{\max}$ the largest eigenvalue of the right Cauchy-Green strain tensor $C\left ( \bx,t_{0},T \right )$.
The FTLE, with $t_{0}$ and $T$ fixed, is considered a scalar field of the Lyapunov exponent as a function of initial position, $\bx$,
\begin{equation}
	\lambda _{t_{0}}^{T}\left ( \bx \right ) = \frac{1}{\left | T \right |}\ln \sqrt{\mu_{\max}\left ( C\left ( \bx ,t_{0},T \right ) \right )}.
	 \label{eqn:ftle-definition}
\end{equation}

\subsection{Eulerian Reduced Order Model (E-ROM)}
	\label{sec:e-rom}

To generate the ROM basis for the standard Eulerian ROM, we use the proper orthogonal decomposition (POD)~\cite{HLB96,noack2011reduced}. 
We emphasize, however, that the novel Lagrangian ROMs can be used with other ROM bases~\cite{brunton2019data,crommelin2004strategies,hesthaven2015certified,perotto2017higamod,quarteroni2015reduced,taira2019modal}.
The POD starts by collecting the snapshots $\{\omega^1_h, \ldots, \omega^{M}_h\}$, which are, e.g., finite element (FE) approximations of the vorticity in the QGE~\eqref{eqn:qge} at $M$ different time instances.
The POD seeks a low-dimensional basis that approximates the snapshots optimally with respect to a certain norm. 
Probably the most popular inner product is the $L^2$ inner product :
\begin{eqnarray}
	\biggl( \omega_1, \omega_2 \biggr)
	= \int\limits_{\Omega} \omega_1(\bx) \, \omega_2(\bx) \, d\bx \, .
	\label{eqn:l2-inner-product}
\end{eqnarray}
The solution of the resulting minimization problem is equivalent to the solution of the eigenvalue problem
\begin{equation}
	Y^T M_{h} Y \widetilde{\varphi}_j 
	= \widetilde{\lambda}_j \widetilde{\varphi}_j,
     \quad j=1,\ldots, N,
     \label{eqn:pod-1}
\end{equation}
where 
$Y$ denotes the snapshot matrix, whose columns correspond to the FE coefficients of the snapshots, $M_{h}$ denotes the FE mass matrix, and $N$ is the dimension of the FE space.
The eigenvalues are real and non-negative, so they can be ordered as follows: $\widetilde{\lambda}_1 \ge \widetilde{\lambda}_2 \ge \ldots \ge \widetilde{\lambda}_R \ge \widetilde{\lambda}_{R + 1} = \ldots = \lambda_N = 0$.
The POD vorticity basis $\{ \varphi_{j}\}_{j=1}^{r}$ are obtained from the eigenfunctions in~\eqref{eqn:pod-1} that  correspond to the first $r\le R$ largest eigenvalues. 
Thus, the ROM vorticity space is defined as $X^r := \text{span} \{ \varphi_1, \ldots, \varphi_r \}$.
We follow~\cite{mou2020data,san2015stabilized} and define the POD streamfunction basis as the normalized functions $\{ \phi_{j} \}_{j=1}^{r}$, which are chosen such that
\begin{eqnarray}
	- \Delta \phi_j
	= \varphi_j \, ,
	\quad 
	j = 1, \ldots, r \, .
	\label{eqn:pod-2}
\end{eqnarray}
The ROM approximations of the vorticity and streamfunction are 
\begin{eqnarray}
	\omega_r(\bx,t)
	= \sum_{j=1}^{r} a_j(t) \, \varphi_j(\bx) \, ,
    \qquad 
	\psi_r(\bx,t)
	= \sum_{j=1}^{r} a_j(t) \, \phi_j(\bx) \, ,
	\label{eqn:g-rom-2}
\end{eqnarray}
where $\{a_{j}(t)\}_{j=1}^{r}$ are the sought time-varying ROM coefficients. 
We emphasize that, with the choices in~\eqref{eqn:pod-2}--\eqref{eqn:g-rom-2}, once the coefficients $a_j$ are determined from~\eqref{eqn:qge-1}, equation~\eqref{eqn:qge-2} is automatically satisfied.
Replacing the vorticity $\omega$ by $\omega_r$ in the QGE~\eqref{eqn:qge-1} and then using a Galerkin projection onto $X^r$, we obtain the {\it Eulerian ROM (E-ROM)} for the QGE: $\forall \, i = 1, \ldots, r,$
\begin{eqnarray}
	\biggl(
		\frac{\partial \omega_r}{\partial t} , \varphi_{i}
      \biggr)
     + 
	\biggl(
		J(\omega , \psi) , \varphi_{i}
      \biggr)
     - Ro^{-1} \, 
	\biggl(
		\frac{\partial \psi}{\partial x} , \varphi_{i}
      \biggr)
      &\,& + Re^{-1} \, 
	\biggl(
		\nabla \omega_r , \nabla \varphi_{i}
      \biggr)
    \nonumber  \\
	&\,&  
	= Ro^{-1} \, 
      \biggl(
      	F , \varphi_{i}
      \biggr) \, .
    \label{eqn:e-rom}
\end{eqnarray}
The E-ROM~\eqref{eqn:e-rom} yields the following autonomous dynamical system for the vector of time coefficients, ${\bf a}(t)$:
\begin{equation}
  	\dot{\bf a} = {\bf b} + {\bf A} \, {\bf a}   + {\bf a}^\top \, {\bf B} \, {\bf a} ,
	\label{eqn:g-rom-3}
\end{equation}
\noindent 
where $\bf{b}$, $\bf{A}$, and $\bf{B}$ correspond to the constant, linear, and quadratic terms in the numerical discretization of the QGE~\eqref{eqn:qge}, respectively. 
The finite dimensional system \eqref{eqn:g-rom-3} can be written componentwise  as follows:
For all $i = 1, \ldots, r$,
\begin{eqnarray}
	\dot{a}_i(t) 
	=  b_i
	+ \sum_{m=1}^{r} A_{im}a_m(t) 
	+ \sum_{m=1}^r \sum_{n=1}^r B_{imn} \, a_m(t) \, a_n(t),
	\label{eqn:g-rom-5}
\end{eqnarray}
where
\begin{eqnarray}
	&& \hspace*{-1.3cm} 
	b_i 
	=  
	Ro^{-1} \, 
	\biggl(
      	F , \varphi_{i}
      \biggr) \, ,
	\label{eqn:g-rom-6} \\
	&& \hspace*{-1.3cm} 
	A_{im} 
	= 
     Ro^{-1} \, 
	\biggl(
		\frac{\partial \phi_m}{\partial x} , \varphi_{i}
      \biggr)
     - Re^{-1} \, 
	\biggl(
		\nabla \varphi_{m} , \nabla \varphi_{i}
      \biggr) \, ,
	\label{eqn:g-rom-7} \\
	&& \hspace*{-1.3cm} 
	B_{imn} 
	= 
	- \biggl(
		J(\varphi_m , \phi_n) , \varphi_{i}
      \biggr) \, .
	\label{eqn:g-rom-8}
\end{eqnarray}

The E-ROM~\eqref{eqn:e-rom} has been investigated in the numerical simulation of the QGE~\eqref{eqn:qge} (see, e.g.,~\cite{mou2020data,san2015stabilized,selten1995efficient,strazzullo2018model}), where it was shown that it can decrease the computational cost of standard algorithms by orders of magnitude.  However, the numerical simulations in~\cite{mou2020data,san2015stabilized} have also shown that a low-dimensional E-ROM is not able to produce accurate approximations of standard Eulerian quantities, such as the streamfunction and the velocity fields.
(We will also show in Section~\ref{sec:numerical-results} that the standard E-ROM~\eqref{eqn:e-rom} produces inaccurate approximations of Lagrangian quantities, such as the FTLE field.)
The E-ROM's numerical inaccuracy in~\cite{mou2020data,san2015stabilized} is due to the lack of a closure model~\cite{mou2020dd-vms-rom,xie2018data}, i.e., a model for the effect of the discarded ROM modes. 
Thus, to alleviate its numerical inaccuracy, in~\cite{mou2020data,san2015stabilized} the standard E-ROM~\eqref{eqn:e-rom} was supplemented with a stabilizing mechanism that yielded relatively accurate results.

In the next section, we pursue a fundamentally different research avenue to improve the standard E-ROM's numerical accuracy.
Instead of modifying the ROM (i.e., adding a closure model, as done in~\cite{san2015stabilized}), we propose using a novel set of basis functions that combine Lagrangian and Eulerian data.

\subsection{Lagrangian ROMs}
	\label{sec:Lagrangian ROM}

In this section, we put forth two Lagrangian ROMs, in which both the snapshots and the inner product use Lagrangian data (i.e., the FTLE field, $\lambda$) in addition to the Eulerian data (i.e., the vorticity field, $\omega$).
The Eulerian data helps the resulting ROM basis yield an accurate approximation of the Eulerian output.
On the other hand, the Lagrangian data ``steers" the ROM basis toward an accurate approximation of the Lagrangian output.

The main tools that we use to construct the new Lagrangian ROMs for FTLE computation are two novel Lagrangian inner products, which are fundamentally different from the standard $L^{2}$ inner product~\eqref{eqn:l2-inner-product} used to develop the Eulerian ROM (i.e., the E-ROM~\eqref{eqn:e-rom}).	
These new inner products are {\it Lagrangian inner products $( \cdot , \cdot )_{FTLE}$}, 
which aim at including both Eulerian data (i.e., the vorticity field) and Lagrangian data (i.e., the FTLE field) in the ROM basis generation.
We emphasize that adding FTLE data to the snapshots is not redundant, since this changes the relative ordering of the eigenpairs of the POD eigenproblem and, therefore, yields a significantly different ROM basis (see snapshot difference quotients used for E-ROM~\eqref{eqn:e-rom} in~\cite{iliescu2014are} for similar behavior in an Eulerian setting).
The two new Lagrangian inner products generate ROM basis functions that are different from the standard E-ROM modes, which are built with the standard $L^{2}$ inner product (see Fig.~\ref{fig:basis-functions}). 
These two new bases yield two new Lagrangian ROMs, which we present in Sections~\ref{sec:alpha-rom} and \ref{sec:lambda-rom}.
In Section~\ref{sec:numerical-results}, in the numerical simulation of the QGE, we show that these two novel Lagrangian ROMs are orders of magnitude more accurate than the standard E-ROM~\eqref{eqn:e-rom}.

\subsubsection{$\alpha$-ROM}
	\label{sec:alpha-rom}

The first Lagrangian inner product that we propose is
\begin{eqnarray}
\biggl ( (\omega_1 , \lambda_1) \, , \, (\omega_2 , \lambda_2) \biggr )_{FTLE}
= \int\limits_{\Omega} \omega_1(\bx) \, \omega_2(\bx) + \alpha \, \lambda_1(\bx) \, \lambda_2(\bx) \, d \bx \, ,
\label{eqn:ftle-inner-product}
\end{eqnarray}
where $\omega_1$ and $\omega_2$ are vorticity fields and $\lambda_1$ and $\lambda_2$ are FTLE fields. 
Thus, the Lagrangian inner product~\eqref{eqn:ftle-inner-product} combines Lagrangian data ($\lambda_1, \lambda_2$) with Eulerian data ($\omega_1, \omega_2$).
The parameter $\alpha$ in~\eqref{eqn:ftle-inner-product} is a {\it weighting parameter} that measures the Lagrangian data's contribution to the inner product:  
When $\alpha = 0$, the Lagrangian data does not play any role, so the inner product~\eqref{eqn:ftle-inner-product} is the standard $L^2$ inner product~\eqref{eqn:l2-inner-product} used to build the standard E-ROM~\eqref{eqn:e-rom}.
When $\alpha > 0$, the Lagrangian data plays a significant role: 
The higher the $\alpha$ value, the more important  the Lagrangian data contribution to the inner product~\eqref{eqn:ftle-inner-product}.

\begin{remark}[Nondimensional inner product]
The Lagrangian inner product~\eqref{eqn:ftle-inner-product} combines data (the vorticity field and the FTLE field) that has the same  dimensional units (i.e., inverse time). 
Thus, the two different types of variables ($\omega$ and $\lambda$) in~\eqref{eqn:ftle-inner-product} can be added together. 
Furthermore, the QGE~\eqref{eqn:qge} used to compute $\omega$ and $\lambda$ are {\it nondimensionalized}, so the two types of variables could be added even if they did not have the same dimensional units. 
Finally, if the QGE were left in their original dimensional form, we would need to scale the variables $\omega$ and $\lambda$ appropriately. 
\label{remark:non-dimensional}
\end{remark}

We use the new Lagrangian inner product~\eqref{eqn:ftle-inner-product} to generate the ROM basis for a new Lagrangian ROM.
First, we collect snapshots that consist of both vorticity and FTLE approximations.
(Note that this is different from the standard E-ROM~\eqref{eqn:e-rom} basis generation, where only vorticity snapshots were collected.)
Then, we construct the new Lagrangian ROM basis that approximates the snapshots optimally with respect to the Lagrangian norm
\begin{equation}
	\| \omega \| 
	+ \sqrt{\alpha} \, \| \lambda \| \, . 
	\label{eqn:ftle-norm}
\end{equation}
(Again, we note that this is different from the approach used for the standard E-ROM~\eqref{eqn:e-rom}, which utilizes the norm $\| \omega \|$.)
Finally, from the resulting ROM basis functions, we only use their vorticity components in the ROM~\eqref{eqn:e-rom}.

The novel Lagrangian ROM for the FTLE computation is the ROM~\eqref{eqn:e-rom} in which the ROM basis is generated by using the new Lagrangian inner product~\eqref{eqn:ftle-inner-product} instead of the standard $L^2$ inner product~\eqref{eqn:l2-inner-product} used to build the E-ROM~\eqref{eqn:e-rom}.
In what follows, we will denote by {\it $\alpha$-ROM} the resulting new Lagrangian ROM. 
Since the new $\alpha$-ROM includes FTLE data (through both the snapshots and the Lagrangian inner product~\eqref{eqn:ftle-inner-product}), we expect it to yield a more accurate FTLE approximation than the standard E-ROM~\eqref{eqn:e-rom}, which does not explicitly include FTLE data.

\subsubsection{$\lambda$-ROM}
	\label{sec:lambda-rom}

The second Lagrangian inner product that we propose is
	\begin{equation}
		\bigl ( \omega_1 \, , \, \omega_2 \bigr )_{FTLE}
		= \int \limits_{\Omega} \langle \lambda \rangle(\bx) \, \omega_1(\bx) \, \omega_2(\bx) \, d \bx \, ,
		\label{eqn:ftle-rom-subsec-1}
\end{equation}
where $\langle \lambda \rangle$ is the {\it time average} of the FTLE field, $\lambda$.
The Lagrangian inner product~\eqref{eqn:ftle-rom-subsec-1} is similar to the Lagrangian inner product~\eqref{eqn:ftle-inner-product} in that both use Lagrangian data (i.e., the FTLE field).
We note, however, that the way in which Lagrangian and Eulerian data is combined in the two inner products is different: the Lagrangian data is added to the Eulerian data in~\eqref{eqn:ftle-inner-product}, whereas in~\eqref{eqn:ftle-rom-subsec-1} the Lagrangian data is first time averaged and then it is used as a scaling factor for the Eulerian data.
We also note that in the numerical investigation in Section~\ref{sec:numerical-results}, we use a QGE setting in which the time averages of the streamfunction and FTLE fields play an important role.
Thus, we expect the Lagrangian inner product~\eqref{eqn:ftle-rom-subsec-1} to produce accurate results in that setting.
Finally, as noted in Remark~\ref{remark:non-dimensional}, the data used in the Lagrangian inner product~\eqref{eqn:ftle-rom-subsec-1} is nondimensional, so the definition~\eqref{eqn:ftle-rom-subsec-1} is appropriate.

When we use the Lagrangian inner product~\eqref{eqn:ftle-rom-subsec-1} to generate the Lagrangian ROM basis, these basis functions approximate the snapshots optimally with respect to the norm 
\begin{equation}
    \left(
	\int \limits_{\Omega} \langle \lambda \rangle(\bx) \, \omega^2(\bx) \, d \bx 
	\right)^{\frac{1}{2}} \, . 
	\label{eqn:ftle-rom-subsec-2}
\end{equation}
Note that, by definition, the FTLE field~\eqref{eqn:ftle-definition} is always positive.
Thus, the Lagrangian inner product~\eqref{eqn:ftle-rom-subsec-1} and the associated norm~\eqref{eqn:ftle-rom-subsec-2} are well defined.

The second new Lagrangian ROM for the FTLE computation is the ROM \eqref{eqn:e-rom} in which the ROM basis is generated by using the new Lagrangian inner product~\eqref{eqn:ftle-rom-subsec-1} instead of the standard $L^2$ inner product~\eqref{eqn:l2-inner-product} used to build the E-ROM~\eqref{eqn:e-rom}.
In what follows, we will denote by {\it $\lambda$-ROM} the resulting new Lagrangian ROM. 
Again, since the new $\lambda$-ROM includes FTLE information, (through both the snapshots and the Lagrangian inner product~\eqref{eqn:ftle-inner-product}), we expect it to yield a more accurate FTLE approximation than the E-ROM~\eqref{eqn:e-rom}, which does not explicitly include FTLE information.

\subsection{Previous Relevant Work}
 	\label{sec:previous-relevant-work}

To our knowledge, there is only little work on reduced order modeling for the FTLE calculation~\cite{babaee2017reduced,kourentis2012uncovering,surana2008reduced}.
We emphasize that the Lagrangian ROMs proposed in this paper are fundamentally different from the ROMs used in~\cite{babaee2017reduced,kourentis2012uncovering}, which are Eulerian ROMs.
The Lagrangian ROMs are also different from the ROM used in~\cite{surana2008reduced}, since the FTLE field is used in~\cite{surana2008reduced} only to choose the number, not the actual form of ROM modes, whereas we explicitly use the FTLE field to define the FTLE inner product~\eqref{eqn:ftle-inner-product} and, thus, to construct the ROM basis.

The Lagrangian ROMs proposed in this paper are related to ROMs that aim at  tackling the challenges posed by transport-dominated problems, e.g., wave-like phenomena, moving interfaces, and moving shocks. 
The ROMs for transport-dominated problems are surveyed in~\cite{lu2020lagrangian,mojgani2017lagrangian,nonino2019overcoming} and include development of, e.g., local bases~\cite{amsallem2012nonlinear,san2015principal}, domain decomposition~\cite{lucia2001reduced}, adaptivity~\cite{carlberg2015adaptive,peherstorfer2018model}, symmetry and self similarity transformations~\cite{rowley2003reduction,rowley2000reconstruction}, approximated Lax pairs~\cite{gerbeau2014approximated}, and 
transport maps~\cite{bernard2018reduced,cagniart2019model,nair2019transported,nonino2019overcoming,ohlberger2013nonlinear,reiss2018shifted,rim2018transport,welper2017h}.

There are also connections between the new Lagrangian ROMs and the ROMs that preserve Lagrangian structure~\cite{carlberg2015preserving,lall2003structure} (see also~\cite{afkham2017structure,breiten2012interpolation,chaturantabut2016structure,gong2017structure,peng2016symplectic} for ROMs that preserve Hamiltonian structure)  as  well as the energy-conserving ROMs for the Navier-Stokes equations~\cite{chan2019entropy,farhat2014dimensional,farhat2015structure,loiseau2018constrained,mohebujjaman2017energy,mohebujjaman2019physically,mou2020data,yano2019discontinuous}.

We also note that including Lagrangian information to build the ROM basis is similar to the difference quotients used in~\cite{iliescu2014are,KV01} and collecting snapshots for the nonlinear terms in the Empirical Interpolation Method (EIM)~\cite{barrault2004eim} and its discrete version, the Discrete Empirical Interpolation Method (DEIM)~\cite{chaturantabut2010nonlinear}.
Indeed, in all these methods, one collects linear combinations of the snapshots. 
Of course, this does not change the rank of the snapshot matrix, but can change the ordering of its singular values and, thus, yield different ROM bases.
Adding Lagrangian information to the set of snapshots is similar in spirit:  We do not necessarily add new information, but we ``steer" the ROM basis in a certain direction.

\section{Numerical Results}
	\label{sec:numerical-results}

In Section~\ref{sec:Lagrangian ROM}, we proposed two new Lagrangian ROMs (i.e., the $\alpha$-ROM and the $\lambda$-ROM) for the  numerical simulation of the FTLE field.
For clarity, Table~\ref{table:ftle-rom-summary} summarizes the inner products used to build the basis functions of the two new Lagrangian ROMs, as well as the standard Eulerian ROM (i.e., the E-ROM~\eqref{eqn:e-rom}).
In this section, we perform a numerical investigation of the two new Lagrangian ROMs.
To separate the ROM closure modeling from the ROM basis generation, we investigate the two new Lagrangian ROMs without any closure model or stabilization mechanism.
\begin{table}[H]
\centering
	\caption{
		The new Lagrangian ROMs ($\alpha$-ROM and $\lambda$-ROM), the standard Eulerian ROM (E-ROM), and the inner products used to construct their bases.	
	\label{table:ftle-rom-summary}
	}
\begin{tabular}{|l|c|c|}
\hline
inner product & ROM & Type \\ 
\hline
\hline
equation~\eqref{eqn:l2-inner-product} & E-ROM & Eulerian \\ \hline
equation~\eqref{eqn:ftle-inner-product} & $\alpha$-ROM & Lagrangian \\ \hline
equation~\eqref{eqn:ftle-rom-subsec-1} & $\lambda$-ROM & Lagrangian \\ \hline
\end{tabular}
\end{table}



In this section, we investigate the Lagrangian $\alpha$-ROM and $\lambda$-ROM in the numerical simulation of the velocity and FTLE fields of the QGE~\eqref{eqn:qge}.
For comparison purposes, we also test the standard Eulerian ROM (i.e., the E-ROM~\eqref{eqn:e-rom}).
As a benchmark, we use the full order model (FOM), which is outlined in Algorithm~\ref{alg:fom}:
\begin{algorithm}[H]
	\caption{Full Order Model (FOM)}
	\label{alg:fom}
  	\begin{enumerate}
    		\item[(1)] Compute high resolution streamfunction $\psi^{FOM}$ on $[T_{min},T_{max}]$. 
		\item[(2)] Use $\psi^{FOM}$ in (1) and formula~\eqref{eqn:streamfunction2velocity} to compute high resolution velocity $\bv^\text{\it FOM}$ on $[T_{min},T_{max}]$.
		\item[(3)] Use $\bv^\text{\it FOM}$ in (2) to calculate (see \S~\ref{sec:ftle}) high resolution FTLE field $\lambda^{FOM}$ on $[T_{min},T_{max}]$.
	\end{enumerate}
\end{algorithm}

In the numerical investigation of the three ROMs (i.e., $\alpha$-ROM, $\lambda$-ROM, and E-ROM), we use Algorithm~\ref{alg:rom}.
We also use two types of regimes: (i) the {\it reconstructive} regime; and (ii) the {\it predictive} regime.
The two regimes have fundamentally different goals:
The reconstructive regime is an easier test, 
in which the ROM is validated on the same time interval as the time interval used to train the ROM.
The predictive regime is a  harder test case, in which the ROM is trained on a short time interval, e.g., $\displaystyle \left[ T_{min},\frac{T_{max}}{2} \right]$ and validated on a longer time interval $[T_{min},T_{max}]$. 

\begin{algorithm}[H]
	\caption{Reduced Order Model (ROM)}
	\label{alg:rom}
  	\begin{enumerate}
    		\item[(1)] Compute high resolution streamfunction $\psi^{FOM}$ on $[T_{min},T_{max}]$.
		\item[(2)] Use $\psi^{FOM}$ in (1) and formula~\eqref{eqn:streamfunction2velocity} to compute high resolution velocity $\bv^\text{\it FOM}$ on $[T_{min},T_{max}]$.
		\item[(3)] Use $\bv^\text{\it FOM}$ in (2)  on $[T_{min},T_{max}]$ to construct Lagrangian ROMs ($\alpha$-ROM and $\lambda$-ROM) and Eulerian ROM (E-ROM).
		\item[(4)] Use ROMs in (3) to compute low resolution ROM streamfunction $\psi^{ROM}$ on $[T_{min},T_{max}]$.
		\item[(5)] Use low resolution streamfunction $\psi^{ROM}$ in (4) and formula~\eqref{eqn:streamfunction2velocity} to compute low resolution velocity $\bv^\text{\it ROM}$ on $[T_{min},T_{max}]$.
		\item[(6)] Use low resolution velocity $\bv^\text{\it ROM}$ in (5) to calculate low resolution ROM-FTLE field $\lambda^{ROM}$ on $[T_{min},T_{max}]$.
	\end{enumerate}
\end{algorithm}


\subsection{Test Problem Setup}
	\label{sec:test-problem}

As a test problem in our numerical investigation, we consider the QGE~\eqref{eqn:qge} with a symmetric double-gyre wind forcing given by $F = \sin \bigl(\pi \, (y-1) \bigr)$, which yields a four-gyre circulation in the time mean.
This test problem has been used in numerous studies (see,  e.g.,~\cite{cummins1992inertial,greatbatch2000four,holm2003modeling,monteiro2015numerical,mou2020data,nadiga2001dispersive,rahman2019dynamic,san2011approximate,san2015stabilized,san2018extreme}) as a simplified model for more realistic ocean dynamics.
\begin{figure}[htp]
\centering
\centering
\vspace{0.3cm}
\includegraphics[width=0.3\linewidth]{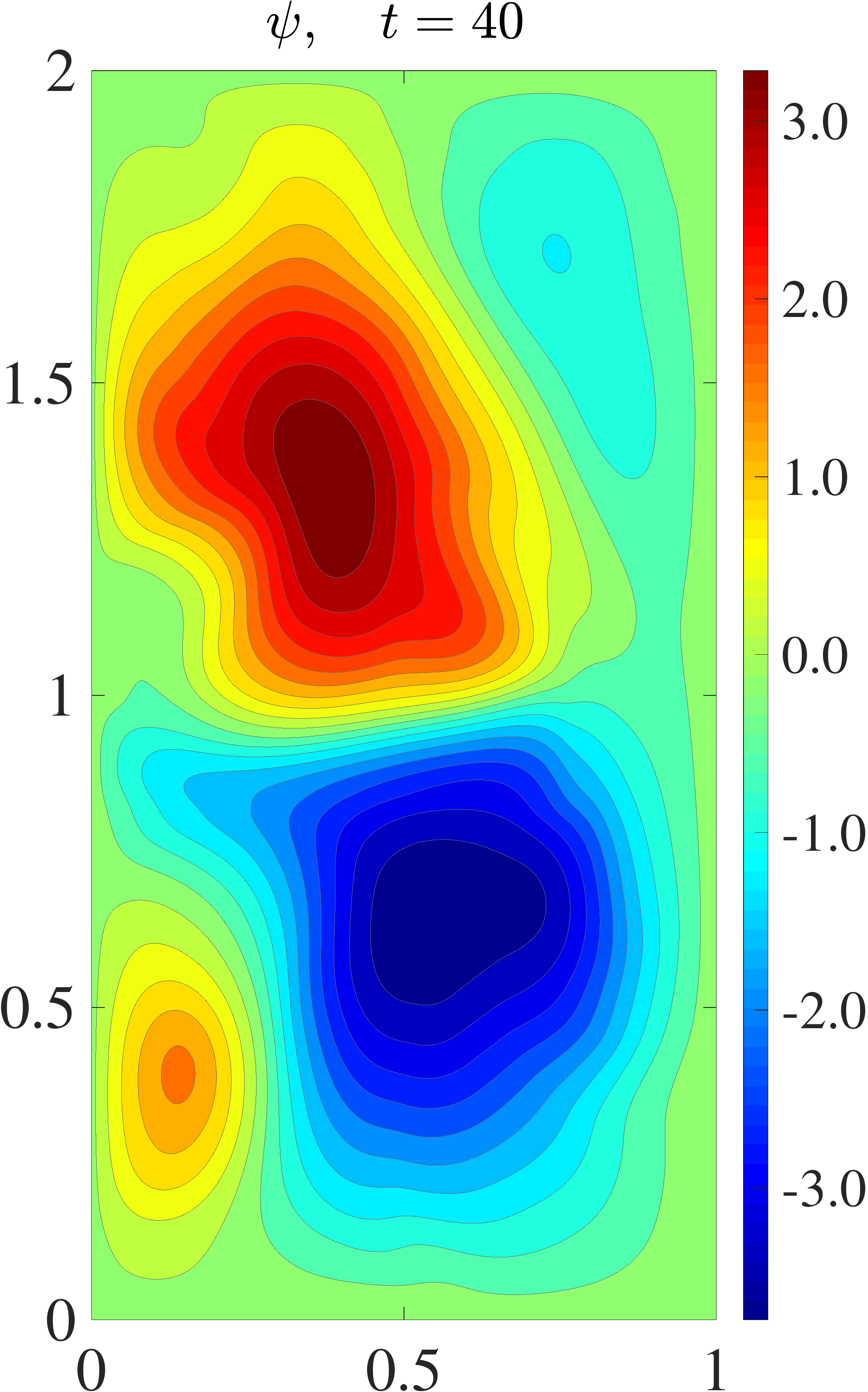}
\hspace{0.2cm}
\includegraphics[width=0.3\linewidth]{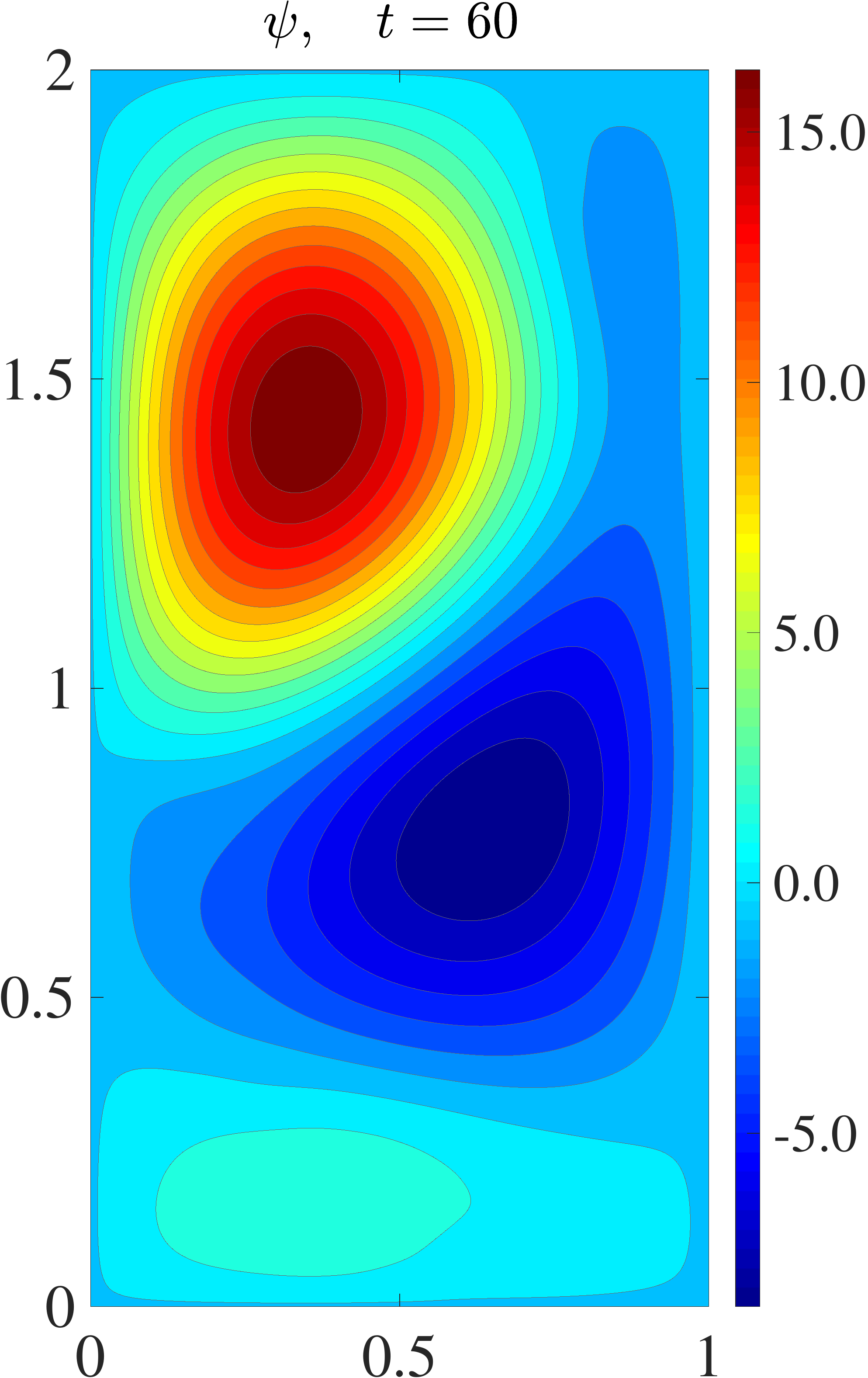}
\hspace{0.2cm}
\includegraphics[width=0.3\linewidth]{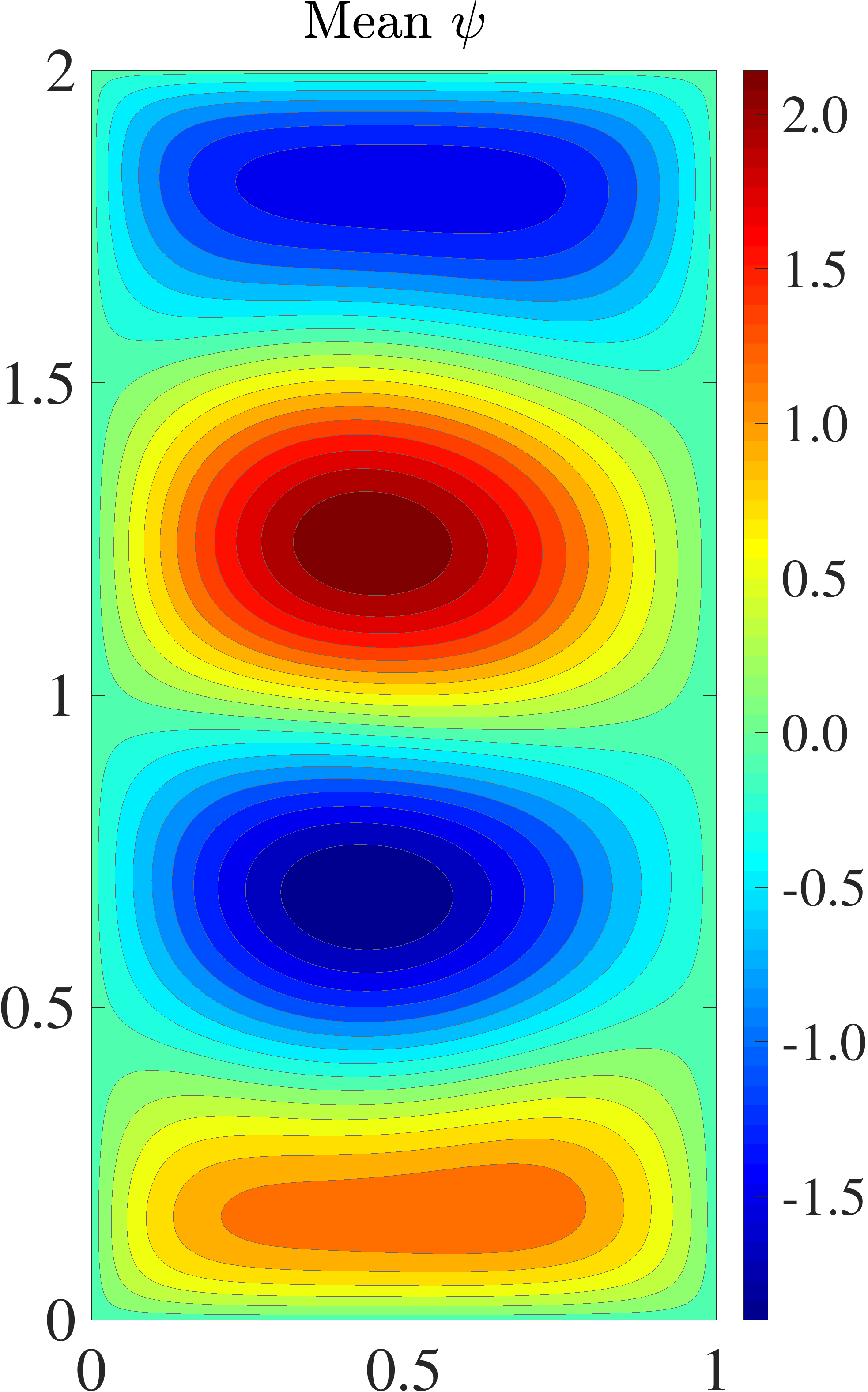}
\vspace{0.3cm}

\includegraphics[width=0.3\linewidth]{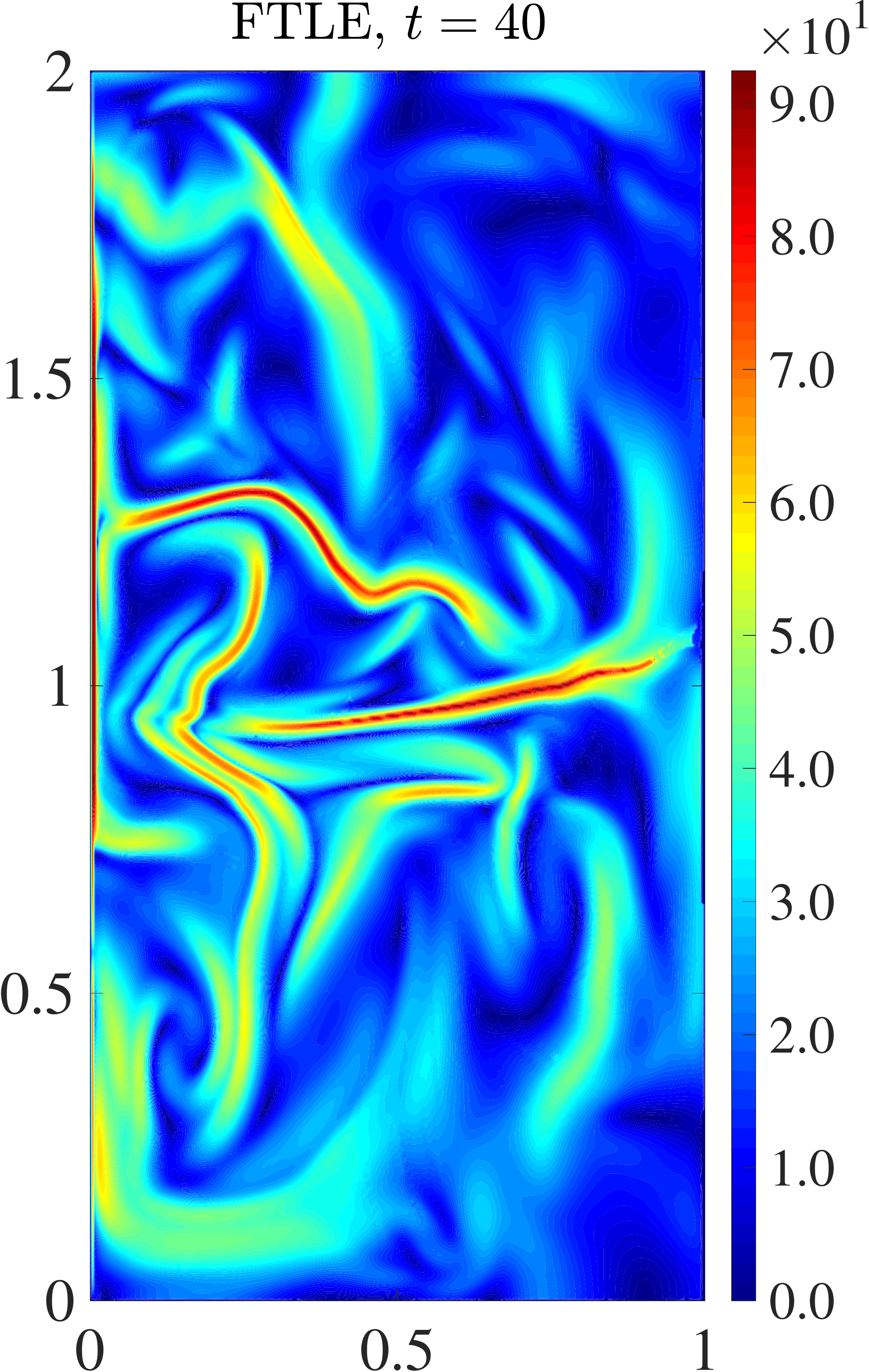}
\hspace{0.2cm}
\includegraphics[width=0.3\linewidth]{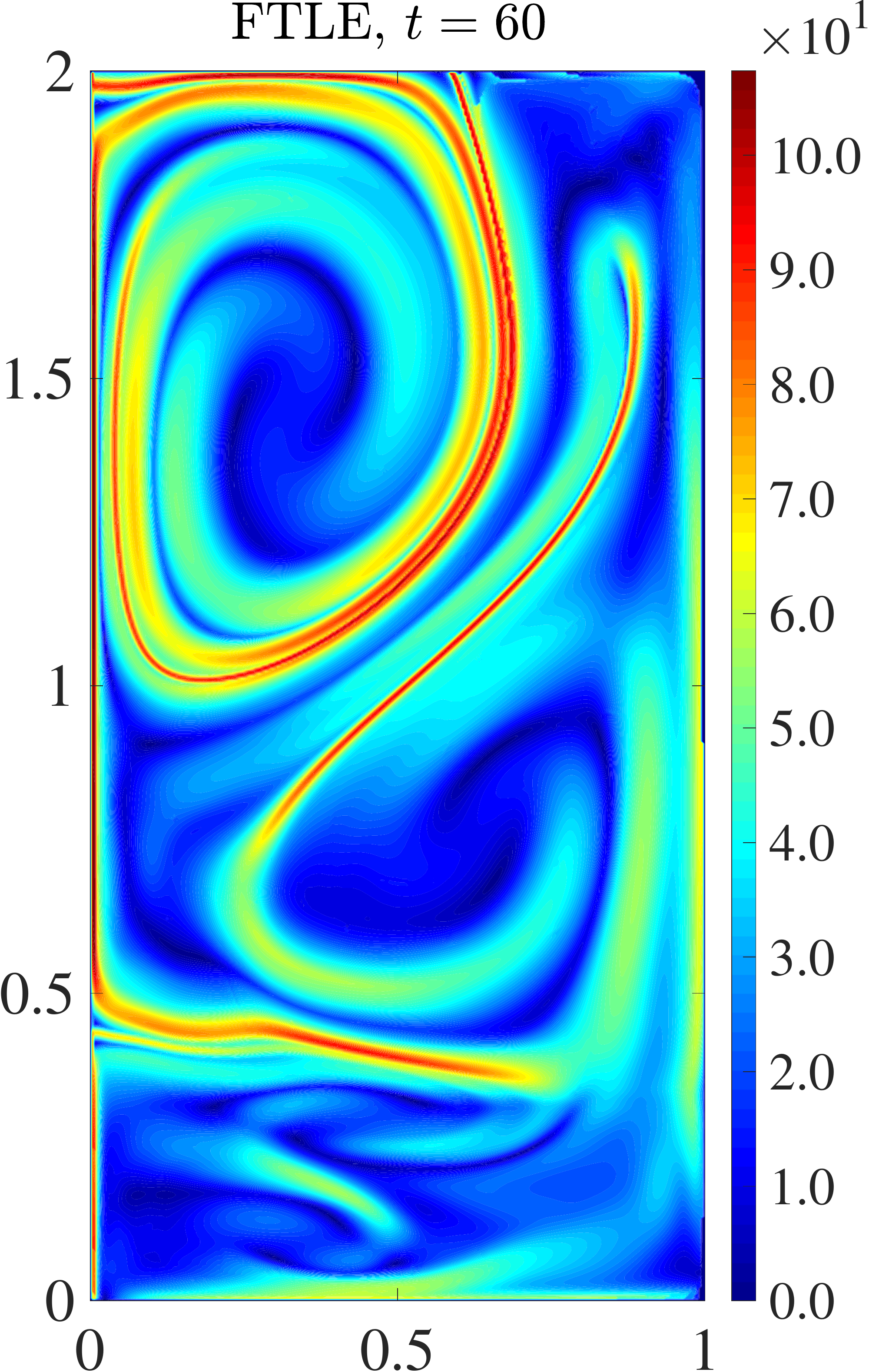}
\hspace{0.2cm}
\includegraphics[width=0.3\linewidth]{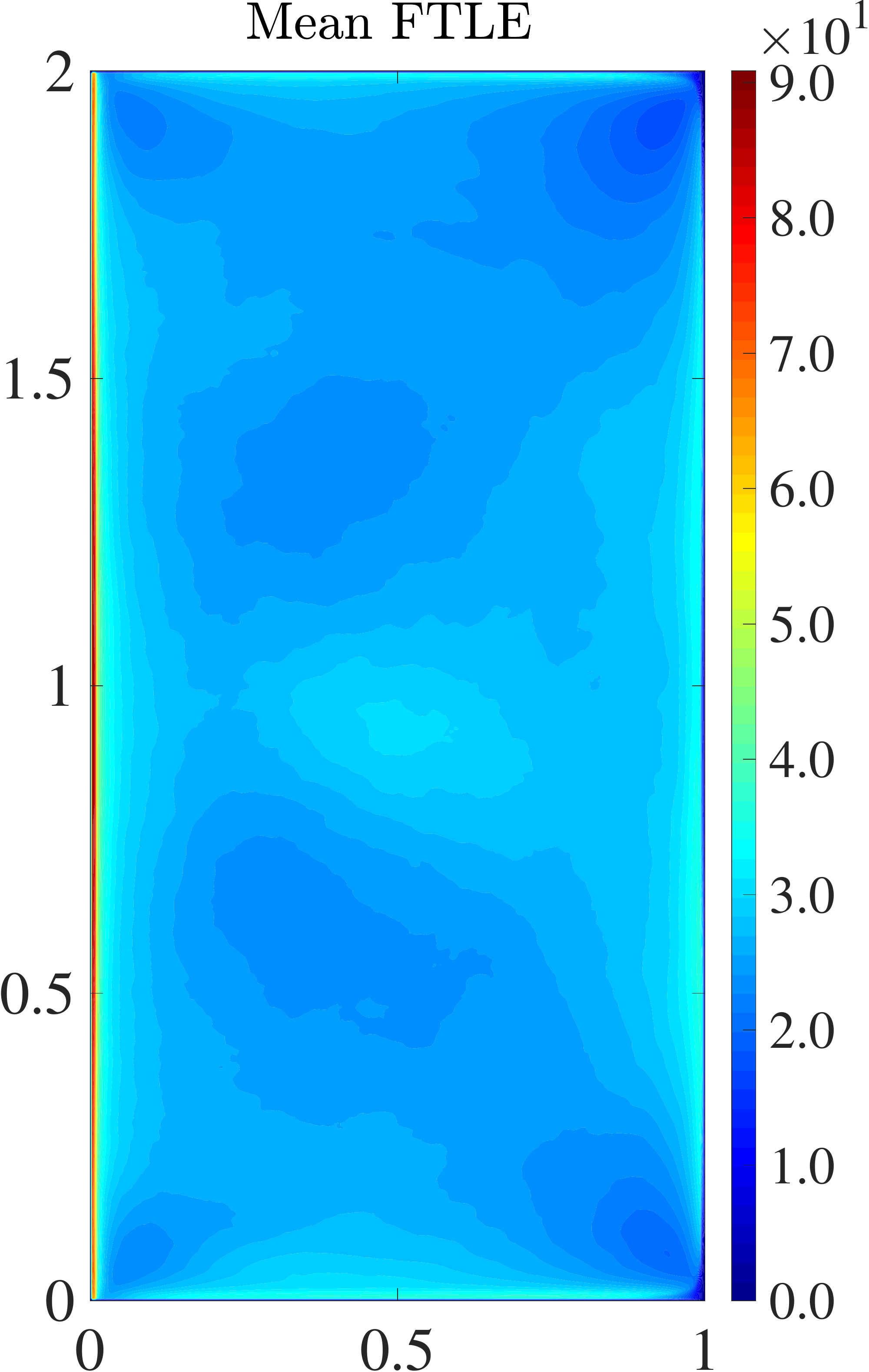}

	\caption{
		Streamfunction contour plots at $t = 40$ (top, left), $t = 60$ (top, middle), and time-averaged (top, right).
		FTLE contour plots at $t = 40$ (bottom, left), $t = 60$ (bottom, middle), and time-averaged (bottom, right).
		An FTLE movie is available at 
		\protect\url{https://youtu.be/JXqdcBVfhMw}.
	}
	\label{fig:experiment-2-2} 
 \end{figure}
We emphasize that the four-gyre QGE test problem represents a significant challenge for standard numerical methods: 
Indeed, as shown in~\cite{greatbatch2000four}, although a double-gyre wind forcing is used, the long term time-average yields a {\it four-gyre} pattern (see Fig.~\ref{fig:experiment-2-2}).
On realistic coarse meshes, classical numerical methods (e.g., finite element and finite volume methods) generally produce inaccurate approximations to this test problem.
In particular, standard numerical discretizations fail to recover the correct four-gyre pattern (see, e.g.,~\cite{san2015stabilized,san2011approximate}). 

In the QGE~\eqref{eqn:qge}, we use the same parameters as those used in~\cite{holm2003modeling,mou2020data,san2015stabilized,san2011approximate}: $Re=450$ and $Ro = 0.0036$.
The spatial domain of the QGE is $[0, 1]\times [0, 2]$. 
In the FTLE field computation~\eqref{eqn:ftle-definition}, we use 
$T=0.05$.

\subsection{Criteria}
	\label{sec:criteria}

To investigate the numerical accuracy of the three ROMs (i.e., $\alpha$-ROM, $\lambda$-ROM, and E-ROM), we compare the ROM results with the FOM results.
To this end, we use two fundamentally different types of criteria: 

The first type of criteria are {\it Eulerian criteria}. 
Specifically, we compute the $L^2$ norm of the the time-averaged streamfunction errors between $\psi^{FOM}$ obtained in Step (2) of Algorithm~\ref{alg:fom} and $\psi^{ROM}$ obtained in Step (5) of Algorithm~\ref{alg:rom}:
	\begin{eqnarray}
	\left \| \frac{1}{M}\sum_{j=1}^M\psi^{FOM}(t_j)-\frac{1}{M}\sum_{j=1}^M\psi^{ROM}(t_j) \right \|_{L^2}^2 \, .
	\label{eqn:norm-time-averaged-streamfunction}
\end{eqnarray}
In addition to the quantitative criterion~\eqref{eqn:norm-time-averaged-streamfunction}, we are also using the following qualitative Eulerian criterion: 
We investigate whether the three ROMs can  recover the four-gyre pattern of the time average of the streamfunction in Fig.~\ref{fig:experiment-2-2}, which represents a challenging test for standard numerical methods at realistic low resolutions (see, e.g.,~\cite{san2015stabilized,san2011approximate}).

The second type of criterion we use in our numerical investigation is a {\it Lagrangian criterion}. 
Specifically, we compute the $L^2$ norm of the the time-averaged FTLE errors between $\lambda^{FOM}$ obtained in Step (3) of Algorithm~\ref{alg:fom} and $\lambda^{ROM}$ obtained in Step (6) of Algorithm~\ref{alg:rom}:
\begin{eqnarray}
        \left \| \frac{1}{M} \sum_{j=1}^{M} \lambda^{FOM}(t_j) - \frac{1}{M} \sum_{j=1}^{M} \lambda^{ROM}(t_j) \right \|_{L^2}^2 \, .
	\label{eqn:norm-time-averaged-ftle}
\end{eqnarray}


\subsection{ROM Snapshot Generation}

\begin{figure}[H]
	\begin{center}
		\includegraphics[width=0.6\linewidth]{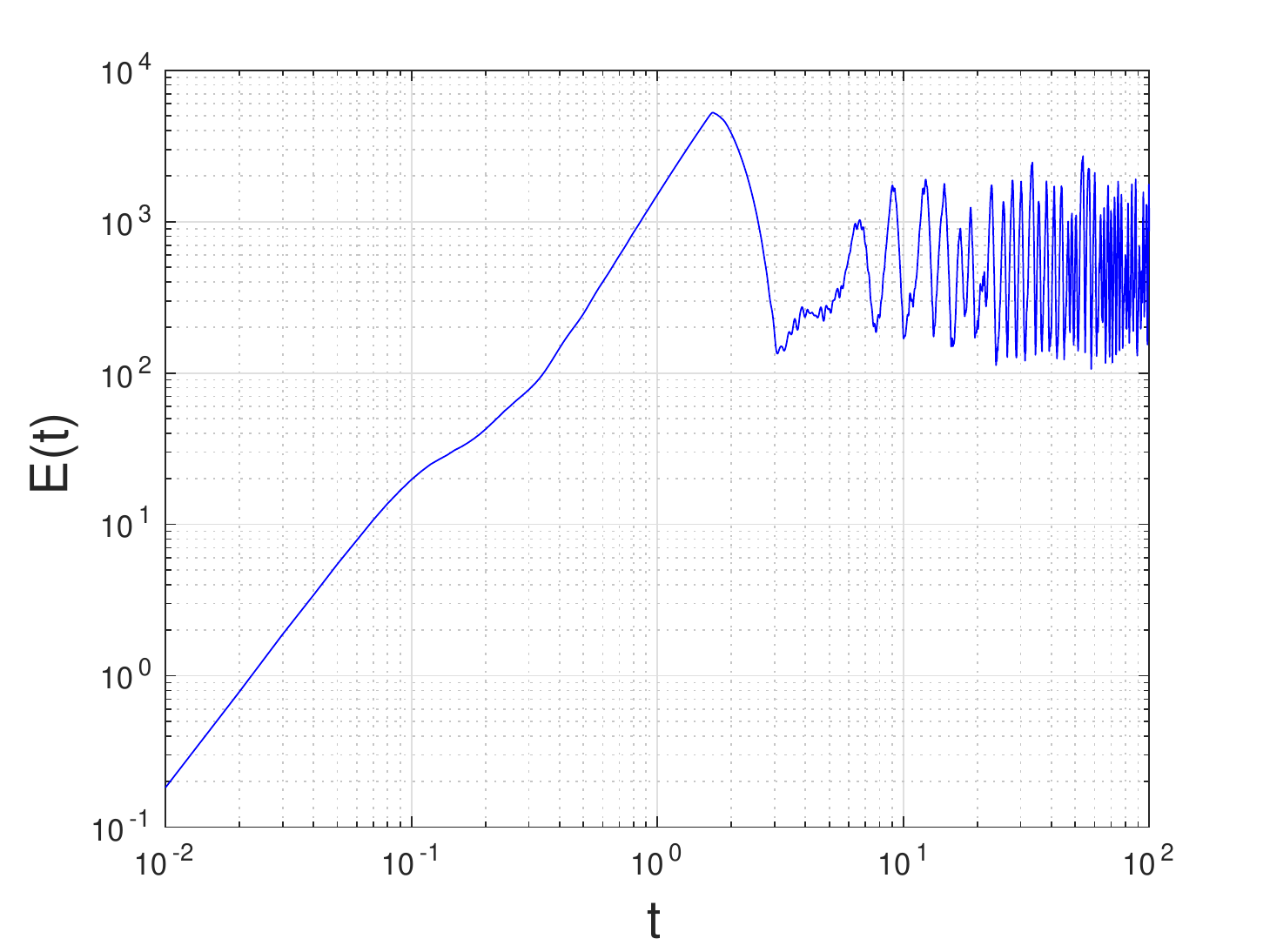}
	\end{center}
	\caption{
		Time evolution of the spatially averaged kinetic energy of the FOM.
	}
	\label{fig:fom-ke} 
 \end{figure}

For the FOM (see Algorithm~\ref{alg:fom}) spatial discretization, we use a spectral method with a $257 \times 513$ spatial resolution~\cite{mou2020data}. 
For the FOM time discretization, we utilize a time step $\Delta t = 0.01$ and an explicit Runge-Kutta method (Tanaka-Yamashita, an order $7$ method with an embedded order $6$ method for error control) and an error tolerance of \num{1.0e-8} in time with adaptive time refinement and coarsening~\cite{mou2020data}.
These spatial and temporal discretizations yield numerical results that are similar to the fine resolution numerical results obtained in~\cite{san2015stabilized,san2011approximate}. 
In Fig.~\ref{fig:fom-ke}, we plot the time evolution of the spatially averaged kinetic energy, $E(t)$.
Figure~\ref{fig:fom-ke} (see also Fig. 1 in~\cite{san2015stabilized}) shows that the flow converges to a statistically steady state, after a short transient interval that ends around $t = 10$.
Thus, in our numerical investigation, we follow~\cite{mou2020data,san2015stabilized,san2011approximate} and consider the FOM results only on $[T_{min},T_{max}] = [10,80]$.
In Fig.~\ref{fig:experiment-2-2}, we display the instantaneous contour plot for the streamfunction field at $t = 40$ and $t = 60$.
We emphasize that, although $t = 40$  and $t = 60$ are well within the statistically steady state regime, the flow displays a high degree of variability.
Thus, the numerical approximation of this statistically steady regime remains challenging for the low resolution ROMs that we investigate in this section. 
To generate the ROM basis (see Section~\ref{sec:e-rom}), we follow~\cite{mou2020data,san2015stabilized,san2011approximate} and collect $701$ snapshots in the time interval $[T_{min},T_{max}] = [10,80]$ (on which the statistically steady state regime is attained) at equidistant time intervals.

\subsection{ROM Basis Investigation}

The new Lagrangian ROM (i.e., $\alpha$-ROM and $\lambda$-ROM) bases are fundamentally different from the standard  E-ROM~\eqref{eqn:e-rom} basis.
Indeed, the E-ROM basis is built only from Eulerian data (i.e., the vorticity $\omega$) by using the standard $L^{2}$ inner product~\eqref{eqn:l2-inner-product}.
On the other hand, the $\alpha$-ROM and $\lambda$-ROM bases are constructed from Lagrangian data (i.e., the FTLE field $\lambda$) in addition to Eulerian data (i.e., the vorticity $\omega$) by using the new Lagrangian inner product~\eqref{eqn:ftle-inner-product} and the new Lagrangian inner product~\eqref{eqn:ftle-rom-subsec-1}, respectively.

\begin{figure}[htp]
\centering
\centering
\vspace{0.3cm}
\includegraphics[width=0.3\linewidth]{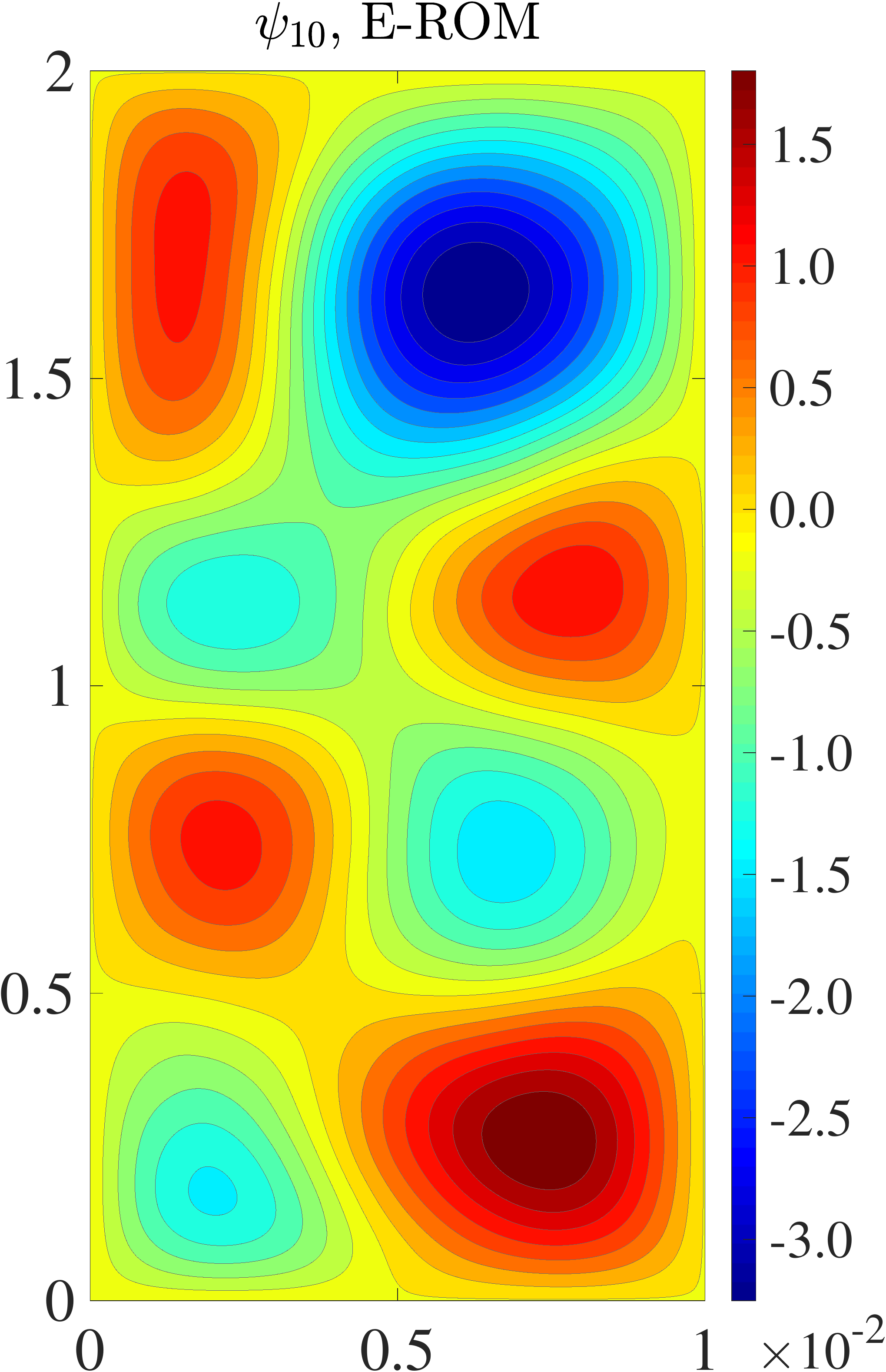}
\hspace{0.2cm}
\includegraphics[width=0.3\linewidth]{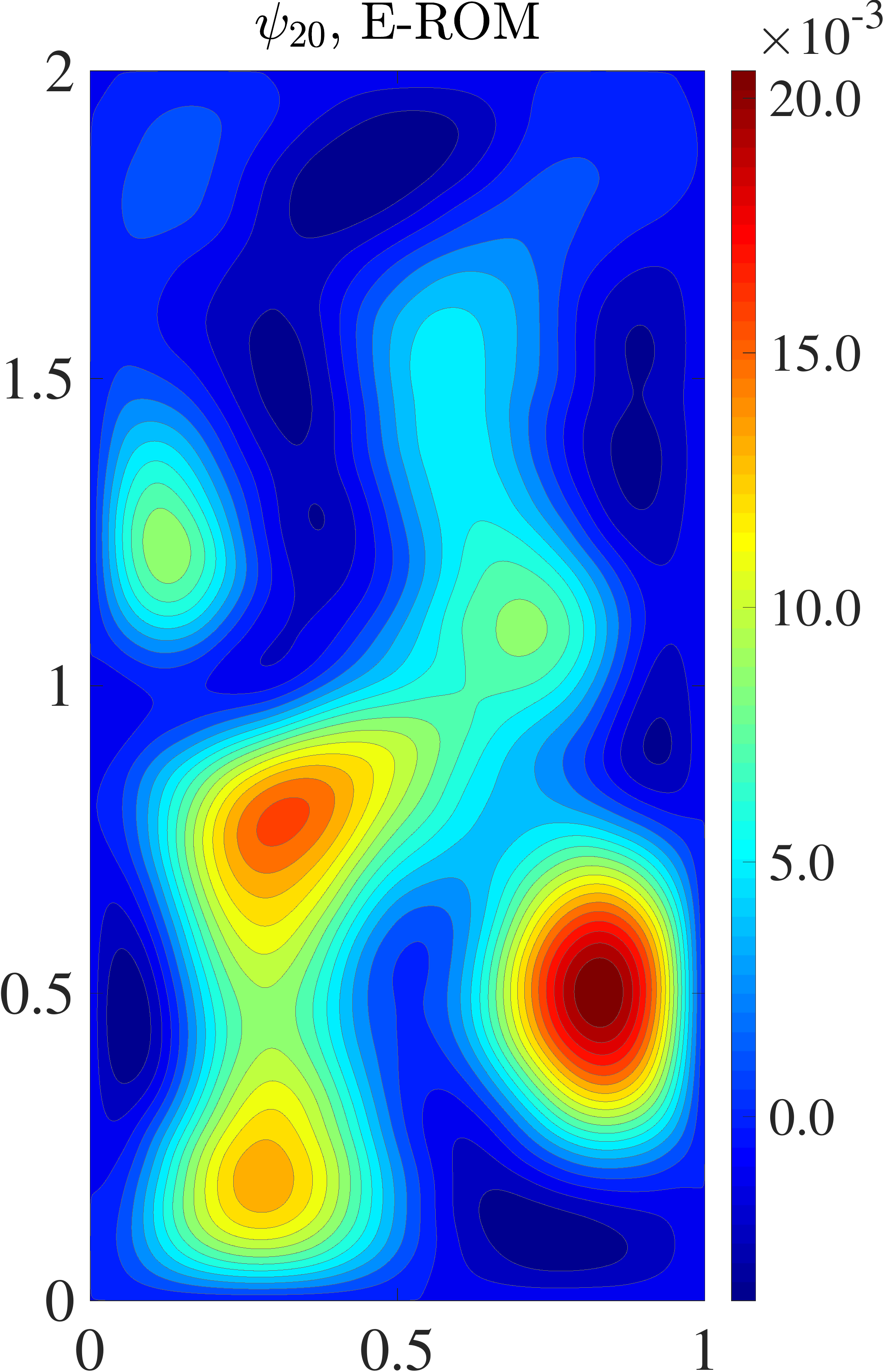}
\hspace{0.2cm}
\includegraphics[width=0.3\linewidth]{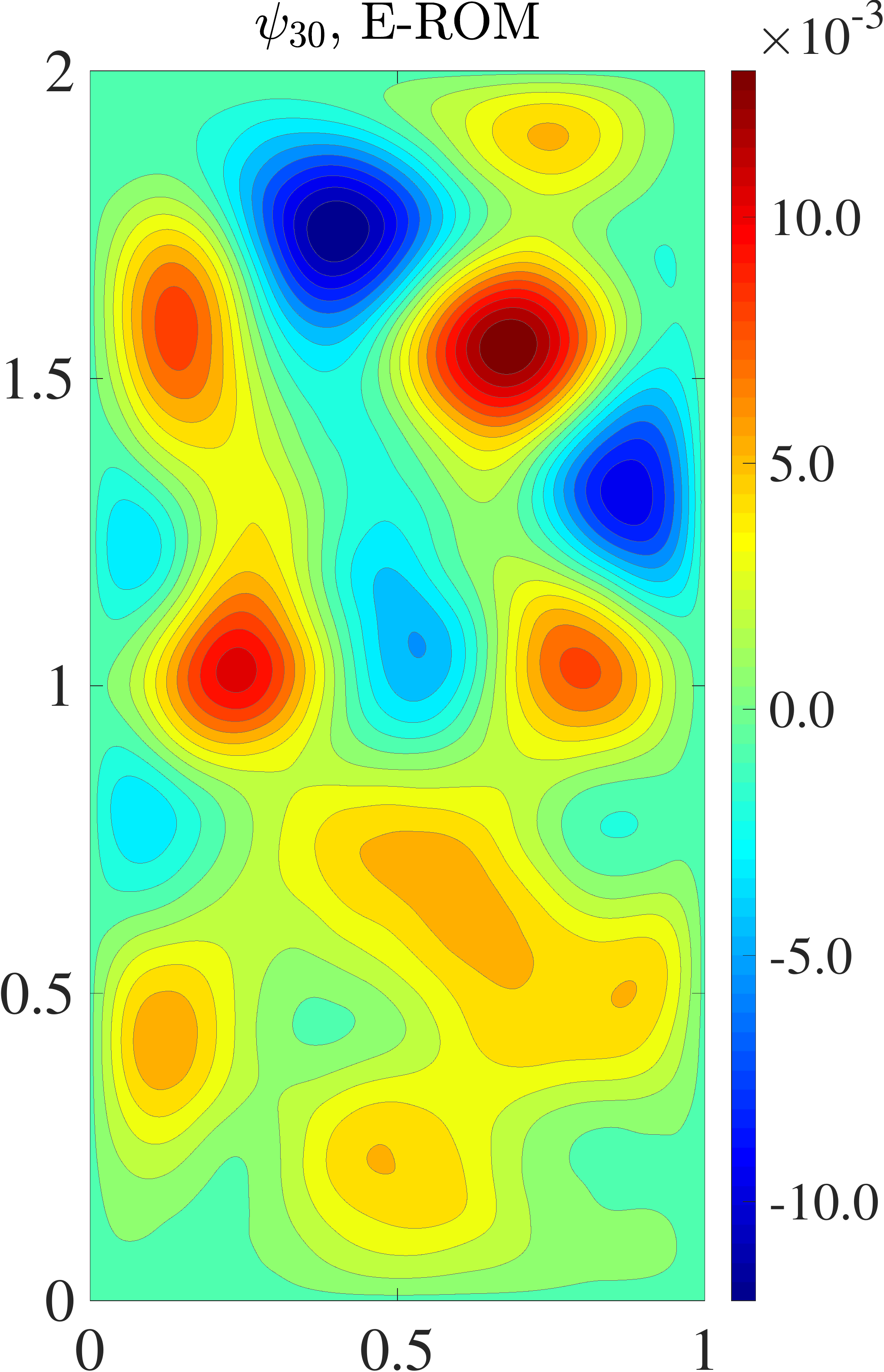}
\vspace{0.3cm}

\includegraphics[width=0.3\linewidth]{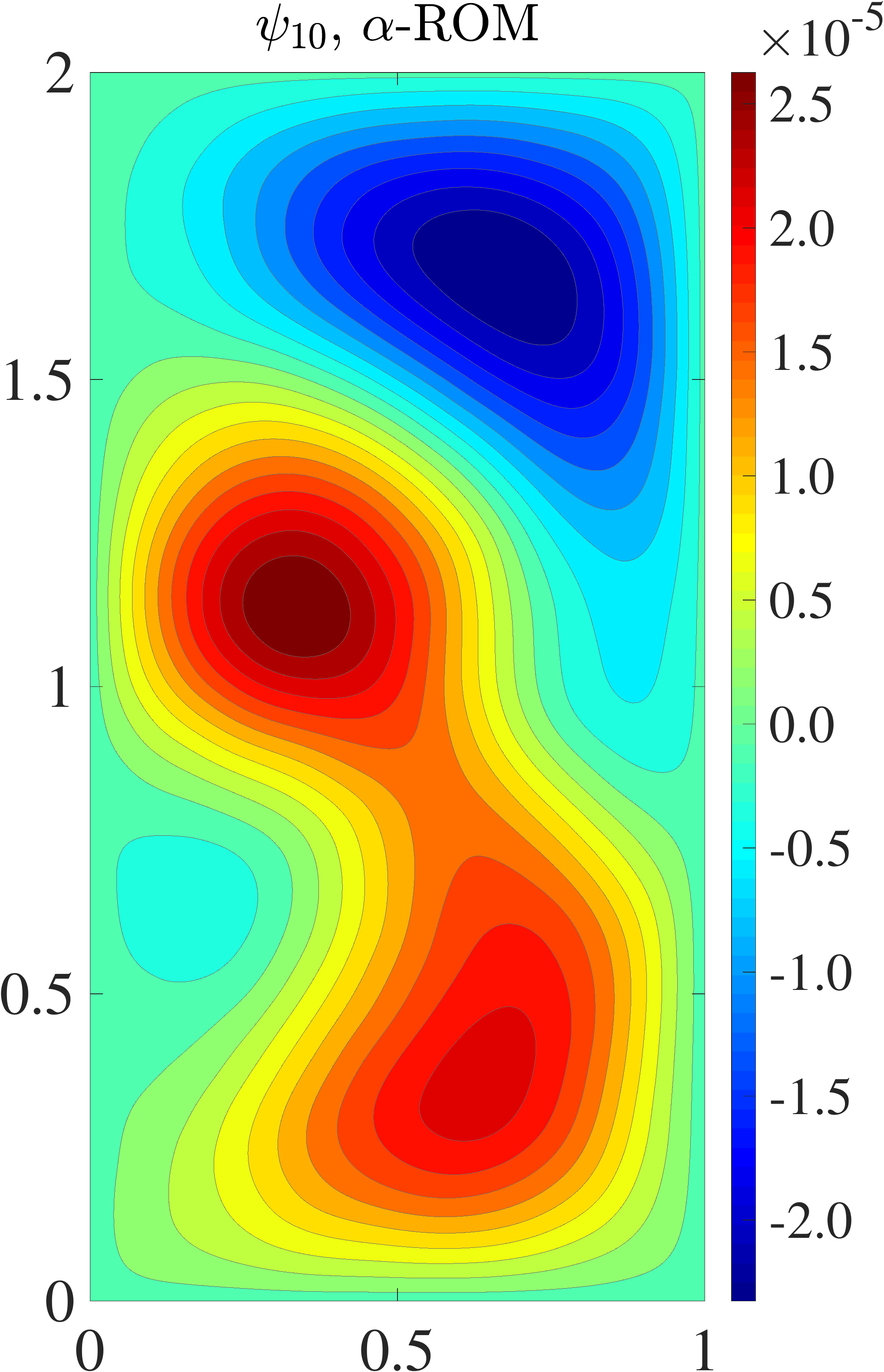}
\hspace{0.2cm}
\includegraphics[width=0.3\linewidth]{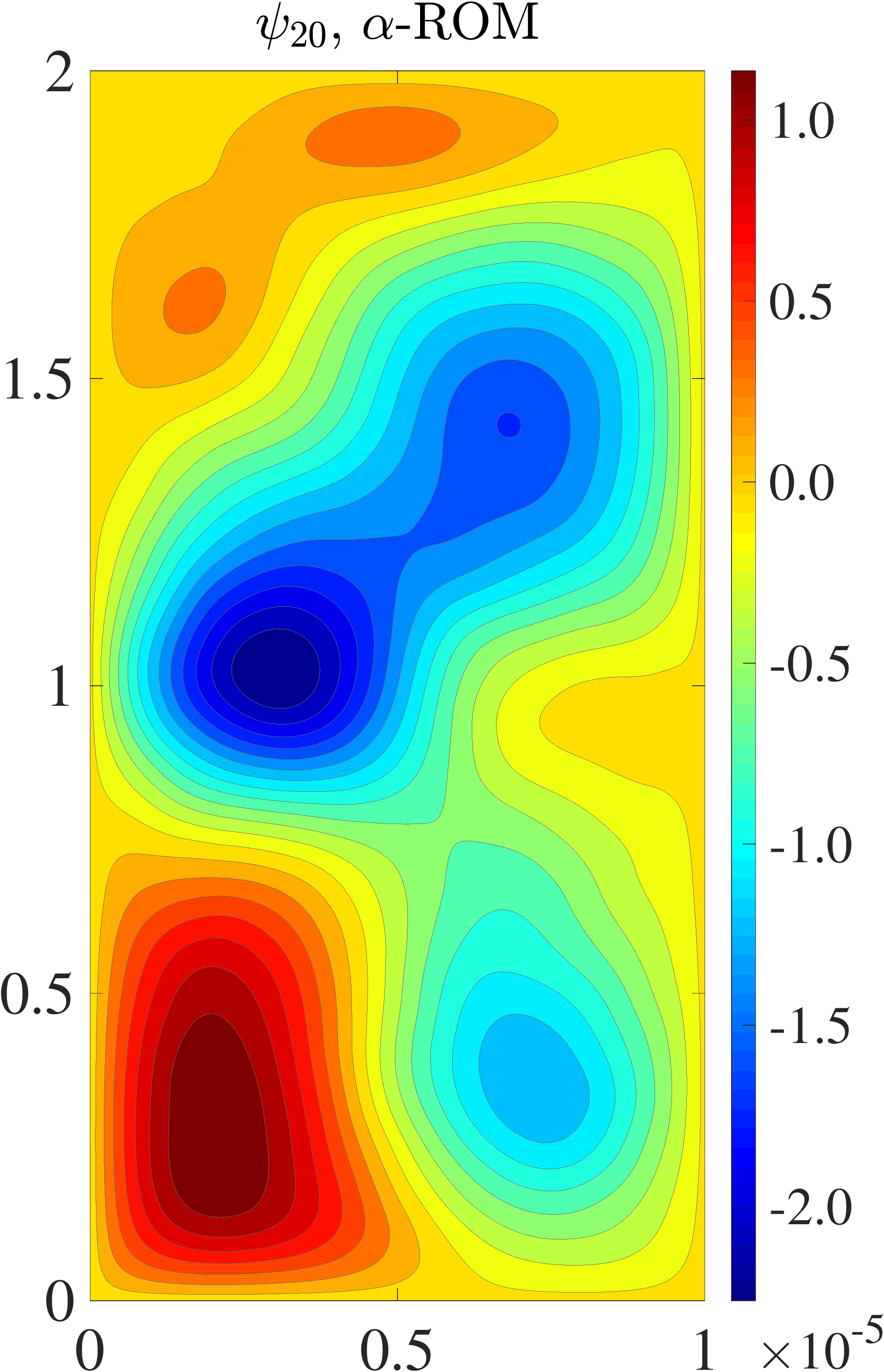}
\hspace{0.2cm}
\includegraphics[width=0.3\linewidth]{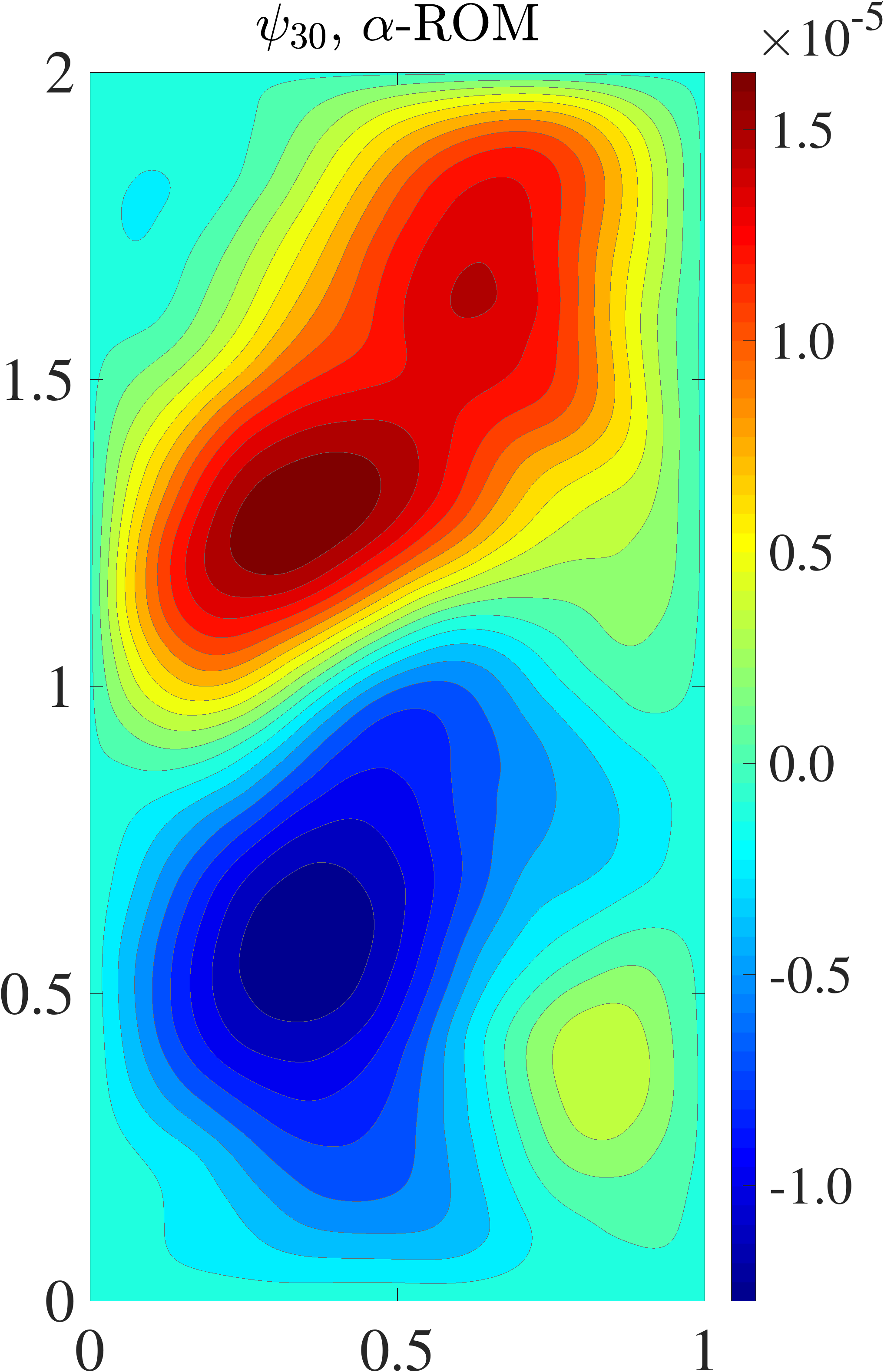}
\vspace{0.3cm}

\includegraphics[width=0.3\linewidth]{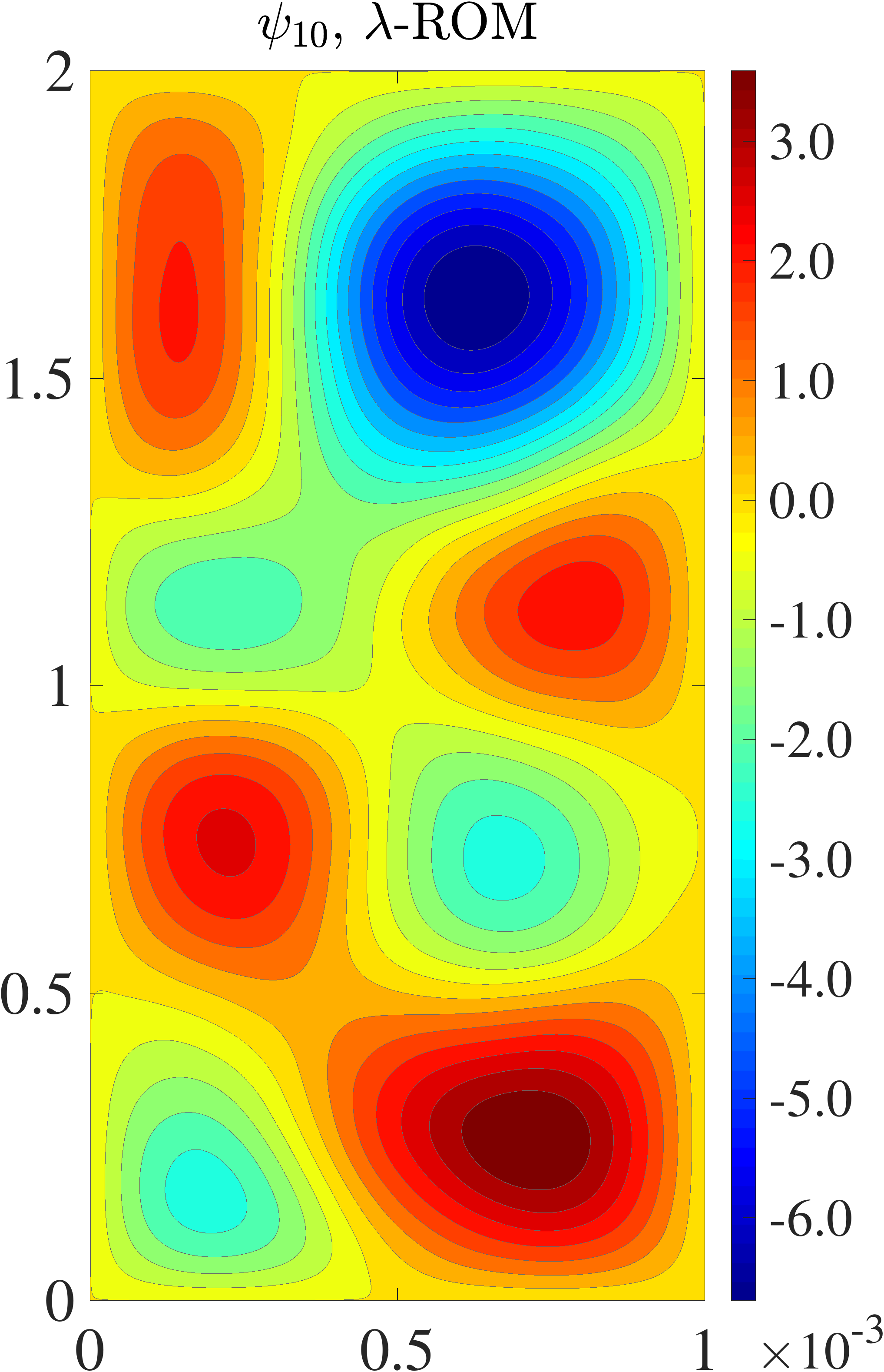}
\hspace{0.2cm}
\includegraphics[width=0.3\linewidth]{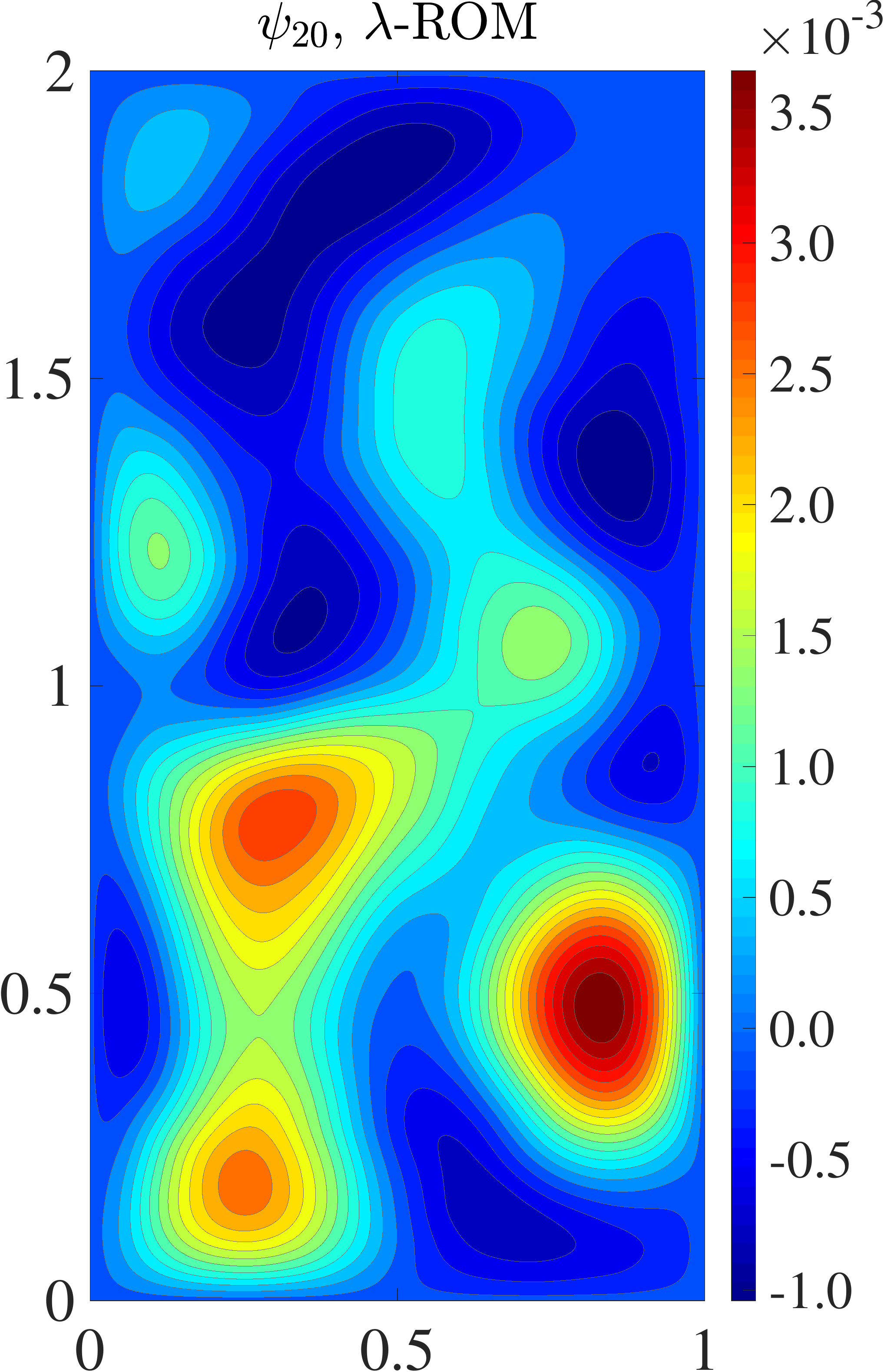}
\hspace{0.2cm}
\includegraphics[width=0.3\linewidth]{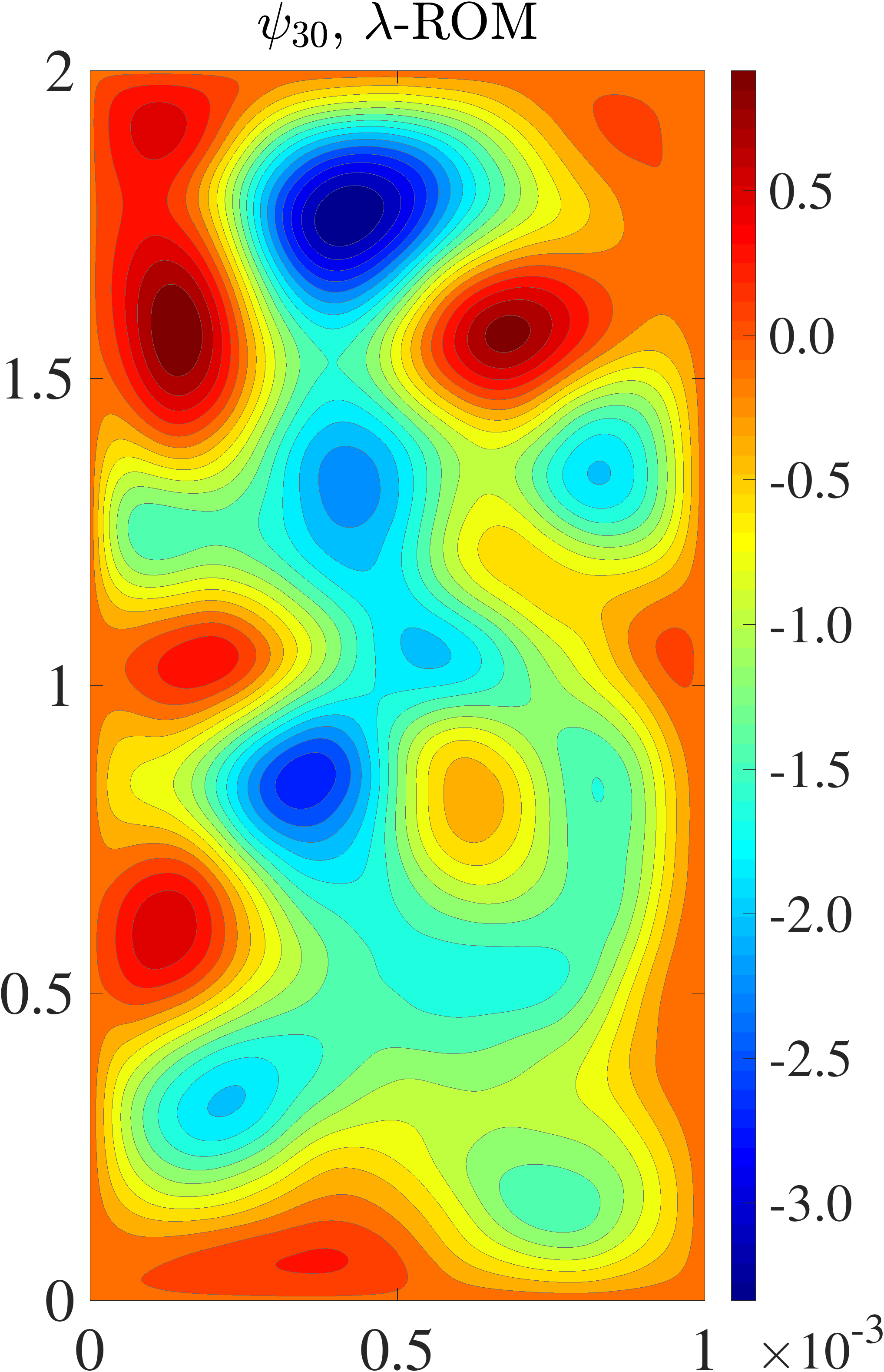}
	\caption{
		ROM basis functions $\psi_{10}$ (first column), $\psi_{20}$ (second column), and $\psi_{30}$ (third column) for the standard E-ROM (first row), new Lagrangian $\alpha$-ROM with $\alpha=10^{4}$ (second row), and new Lagrangian $\lambda$-ROM (third row).
	}
	\label{fig:basis-functions} 
 \end{figure}

To investigate whether the $\alpha$-ROM and $\lambda$-ROM bases are different from the E-ROM~\eqref{eqn:e-rom} basis, in Fig.~\ref{fig:basis-functions} we display the ROM basis functions $\psi_{10}$, $\psi_{20}$, and $\psi_{30}$ generated with the standard $L^{2}$ inner product~\eqref{eqn:l2-inner-product} (i.e., the E-ROM basis functions), the new Lagrangian inner product~\eqref{eqn:ftle-inner-product} (i.e., the new $\alpha$-ROM basis functions) with $\alpha=10^{4}$, and the new Lagrangian inner product~\eqref{eqn:ftle-rom-subsec-1} (i.e., the new $\lambda$-ROM basis functions).

The $\alpha$-ROM basis functions (second row of Fig.~\ref{fig:basis-functions}) are completely different from the E-ROM basis functions (first row of Fig.~\ref{fig:basis-functions}) for $\psi_{10}$, $\psi_{20}$, and $\psi_{30}$.
The $\alpha$-ROM basis functions are also completely different from the $\lambda$-ROM basis functions (third row of Fig.~\ref{fig:basis-functions}).
The $\lambda$-ROM basis functions (third row of Fig.~\ref{fig:basis-functions}) are also different from the E-ROM basis functions (first row of Fig.~\ref{fig:basis-functions}), although this time the differences are not as dramatic as before: there are large differences in $\psi_{30}$, moderate differences in $\psi_{20}$, and minor differences in $\psi_{10}$. 
Overall, the results in Fig.~\ref{fig:basis-functions} show that the new Lagrangian inner product~\eqref{eqn:ftle-inner-product}, the new Lagrangian inner product~\eqref{eqn:ftle-rom-subsec-1}, and the standard Eulerian $L^{2}$ inner product~\eqref{eqn:l2-inner-product} generate completely different bases for the Lagrangian $\alpha$-ROM, the Lagrangian $\lambda$-ROM, and the standard E-ROM, respectively.
In the next section, we investigate which of these bases yields more accurate ROMs in the FTLE field computation.

\subsection{ROM Numerical Accuracy}
	\label{sec:rom-numerical-accuracy}

In this section, we perform a numerical investigation of the accuracy of the two Lagrangian ROMs (i.e., $\alpha$-ROM, $\lambda$-ROM).
We only consider the effect of the basis functions on the ROM accuracy without using a ROM closure model or ROM stabilization mechanism, which is a challenging test. 

We compare the Lagrangian $\alpha$-ROM and $\lambda$-ROM accuracy with the standard  E-ROM accuracy.
As a benchmark for our comparison, we use the FOM results (Algorithm~\ref{alg:fom}). 
In Section~\ref{sec:eulerian-investigation}, we perform an Eulerian investigation of the three ROMs, i.e., we investigate the ROMs' accuracy in approximating the streamfunction (which is an Eulerian quantity).
In Section~\ref{sec:lagrangian-investigation}, we perform a Lagrangian investigation of the three ROMs, i.e., we investigate the ROMs' accuracy in approximating the FTLE field (which is a Lagrangian quantity).
For both the Eulerian and Lagrangian investigations, we consider both the reconstructive regime and the predictive regime.

In our numerical experiments, we use the following parameter values:
For the $\alpha$-ROM, we use $\alpha = 1, \alpha=10^{2}, \alpha=10^{3}$, and $\alpha = 10^{4}$.
We choose this wide range of parameter values to elucidate the effect of the Lagrangian data on the new $\alpha$-ROM.
Indeed, the parameter $\alpha$ in~\eqref{eqn:ftle-inner-product} is a weighting parameter that measures the Lagrangian data's contribution to the inner product:  
The higher the $\alpha$ value, the more important  the Lagrangian data contribution to the inner product~\eqref{eqn:ftle-inner-product}.
For all the ROM simulations, we use an RK4 time discretization with $\Delta t=10^{-3}$. 
Finally, for all the ROMs, we utilize the following $r$ values: $5, 10, 15, 20, 25, 30, 35, 40, 45$, and $50$.
We choose this wide range of values to clarify the effect of the ROM dimension on the ROM accuracy.

\subsubsection{Eulerian Investigation}
	\label{sec:eulerian-investigation}

In this section, we perform an Eulerian investigation of the accuracy of the two Lagrangian ROMs (i.e., the $\alpha$-ROM and $\lambda$-ROM).
First, we consider the reconstructive regime and then the more challenging predictive regime.
In both regimes, we use the two Eulerian criteria described in Section~\ref{sec:criteria}: 
(i) the quantitative Eulerian criterion~\eqref{eqn:norm-time-averaged-streamfunction}, i.e., the $L^2$ norm of the the time-averaged streamfunction errors between $\psi^{FOM}$ obtained in Step (2) of Algorithm~\ref{alg:fom} and $\psi^{ROM}$ obtained in Step (5) of Algorithm~\ref{alg:rom}; and
(ii) the qualitative Eulerian criterion based on the ability of the ROMs to recover the four-gyre pattern of the time average of the streamfunction in Fig.~\ref{fig:experiment-2-2} (which is a challenging test for standard numerical methods at realistic low resolutions~\cite{san2015stabilized,san2011approximate}).

\paragraph{Reconstructive Regime:}
For the reconstructive regime, we check whether the ROMs can reproduce the dynamics of the underlying system on the same time interval as that used to generate the ROM basis functions, i.e., we validate the ROMs on the same time interval as the time interval used to train the ROM.

In Table~\ref{table:error-psi-bar-reconstructive}, for different $r$ values, we list the $L^2$ norm of the errors in the time-averaged streamfunction~\eqref{eqn:norm-time-averaged-streamfunction} for E-ROM, $\lambda$-ROM, and $\alpha$-ROM with $\alpha = 1, \alpha=10^{2}, \alpha=10^{3}$, and $\alpha = 10^{4}$. 
These results yield the following conclusions:
The E-ROM yields inaccurate results for low $r$ values. 
As expected, the E-ROM results get better for large $r$ values.  
The $\lambda$-ROM results are slightly worse than or similar to the E-ROM results for low $r$ values and somewhat better for large $r$ values.
The results for $\alpha$-ROM with $\alpha = 1$ are generally worse than the E-ROM results.
The results for $\alpha$-ROM with $\alpha = 10$ and $\alpha = 10^{2}$ are better than the E-ROM results for all $r$ values except $r=10$.
The results for $\alpha$-ROM with $\alpha = 10^{3}$ and $\alpha = 10^{4}$ are dramatically better than the E-ROM results:
For example, for $r=5$, the only ROM that yields acceptable results is the $\alpha$-ROM with $\alpha = 10^{4}$; all other ROMs simply blow up (denoted by ``N/A'' in Table~\ref{table:error-psi-bar-reconstructive}).
Furthermore, for $r=10$ and $r=15$, the errors of $\alpha$-ROM with $\alpha = 10^{3}$ and $\alpha = 10^{4}$ are {\it two orders of magnitude} lower than the E-ROM error.
For the larger $r$ values, the errors of $\alpha$-ROM with $\alpha = 10^{3}$ and $\alpha = 10^{4}$ continue to be lower than the E-ROM errors, although (as expected) the differences decrease as the $r$ values increase.
As expected, when the $r$ values increase, the results in Table~\ref{table:error-psi-bar-reconstructive} show that all the ROMs' errors converge until they reach a plateau around $5.00$e-$01$ (which is probably due to the ROM truncation error).  
Overall, the results in Table~\ref{table:error-psi-bar-reconstructive} show that {\it the Lagrangian $\alpha$-ROM with high $\alpha$ values} (i.e., $\alpha=10^{3}$ and $\alpha=10^{4}$) yields {\it significantly more accurate results than the standard E-ROM and the Lagrangian $\lambda$-ROM}, especially for the small $r$ values.
\bigskip

\begin{table}[H]
\centering
\caption{
        Eulerian investigation, reconstructive regime:
		$L^2$ norm of the errors in the time-averaged streamfunction~\eqref{eqn:norm-time-averaged-streamfunction} for E-ROM (second column), $\lambda$-ROM (third column), and $\alpha$-ROM for $\alpha=1$ (fourth column), $\alpha=10$ (fifth column), $\alpha=10^{2}$ (sixth column), $\alpha=10^{3}$ (seventh column), and $\alpha=10^{4}$ (eighth column). 
	\label{table:error-psi-bar-reconstructive}
	}
\begin{tabular}{|c|c|c|c|c|c|c|c|}
\hline
$r$& E-ROM& $\lambda$-ROM& $\alpha =1$& $\alpha =10$& $\alpha =10^2$& $\alpha =10^3$& $\alpha =10^4$ \\ \hline
 $5$&   N/A&        N/A&        N/A&        N/A&        N/A&        N/A&        1.0e+01 \\ \hline
$10$&   3.6e+02&    4.0e+02&    7.9e+02&    6.2e+03&    6.5e+02&    2.9e+00&    1.2e+00 \\ \hline
$15$&   1.8e+02&    5.5e+02&    4.1e+02&    1.3e+01&    1.1e+01&    1.1e+00&    2.3e+00 \\ \hline
$20$&   1.3e+01&    8.0e+00&    1.5e+01&    4.2e+00&    2.7e+00&    1.1e+00&    1.1e+00 \\ \hline
$25$&   3.3e+00&    2.7e+00&    4.7e+00&    3.0e+00&    2.1e+00&    3.9e-01&    3.4e-01 \\ \hline
$30$&   2.8e+00&    2.0e+00&    2.5e+00&    2.5e+00&    1.1e+00&    5.9e-01&    4.2e-01 \\ \hline
$35$&   1.5e+00&    9.3e-01&    1.2e+00&    1.3e+00&    8.6e-01&    5.1e-01&    3.1e-01 \\ \hline
$40$&   1.0e+00&    6.1e-01&    9.3e-01&    9.9e-01&    5.4e-01&    4.6e-01&    5.4e-01 \\ \hline
$45$&   5.4e-01&    5.2e-01&    7.0e-01&    6.8e-01&    5.1e-01&    4.0e-01&    7.0e-01 \\ \hline
$50$&   4.5e-01&    5.0e-01&    5.0e-01&    6.2e-01&    5.9e-01&    6.0e-01&    3.8e-01 \\ \hline
\end{tabular}
\end{table}

\bigskip

\begin{figure}[htp]
\centering
\centering
\vspace{0.3cm}
\includegraphics[width=0.24\linewidth]{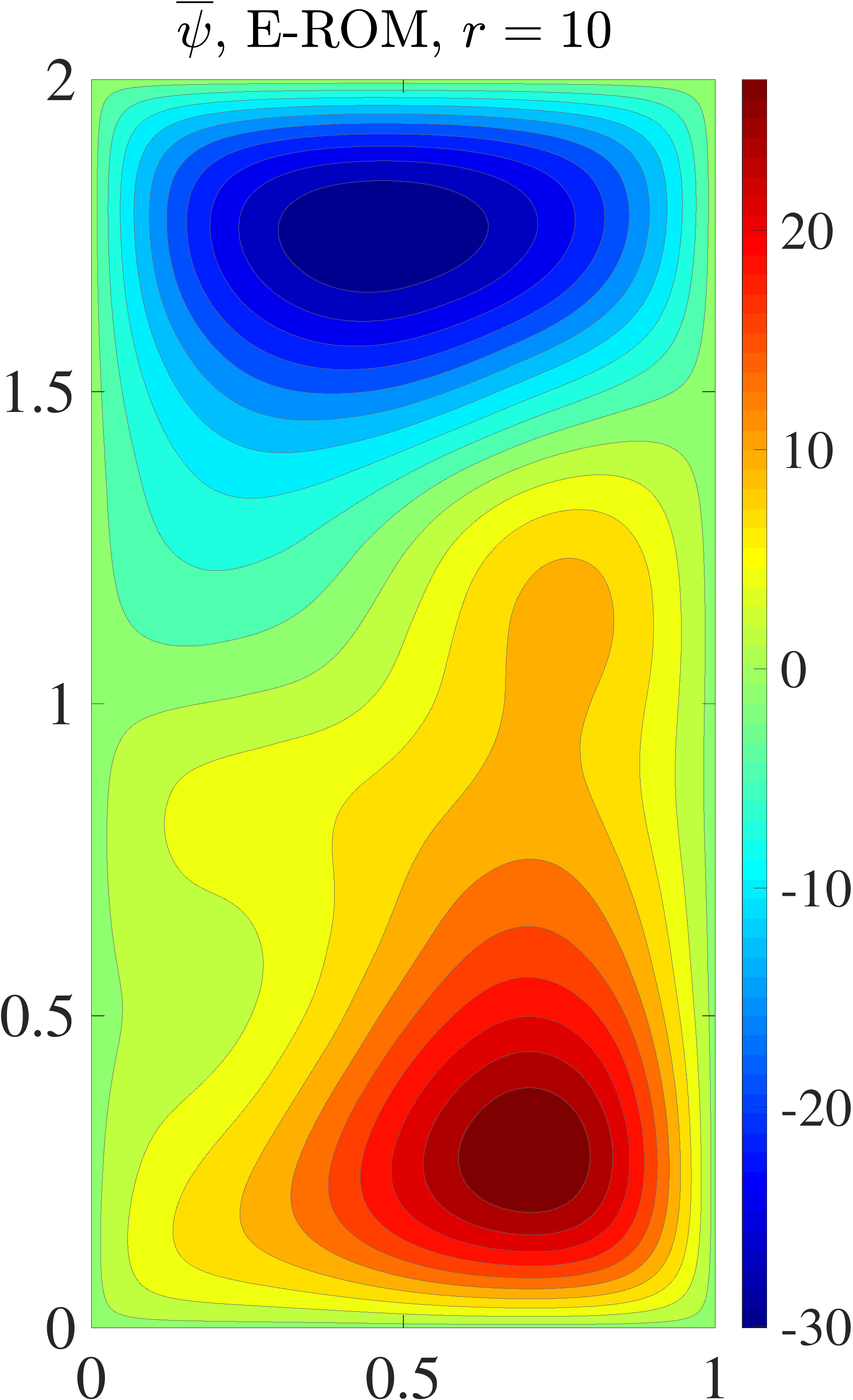}
\includegraphics[width=0.24\linewidth]{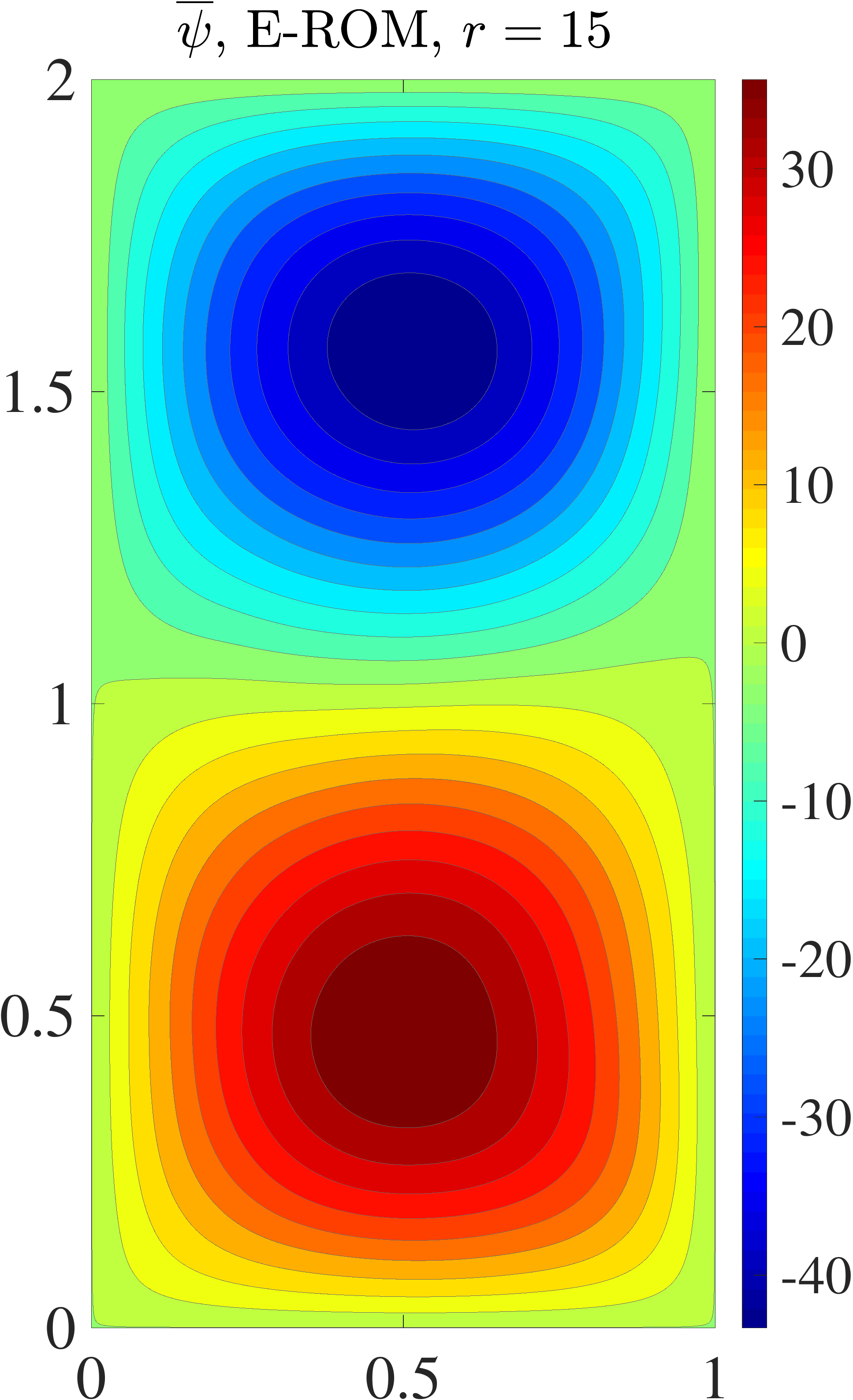}
\includegraphics[width=0.243\linewidth]{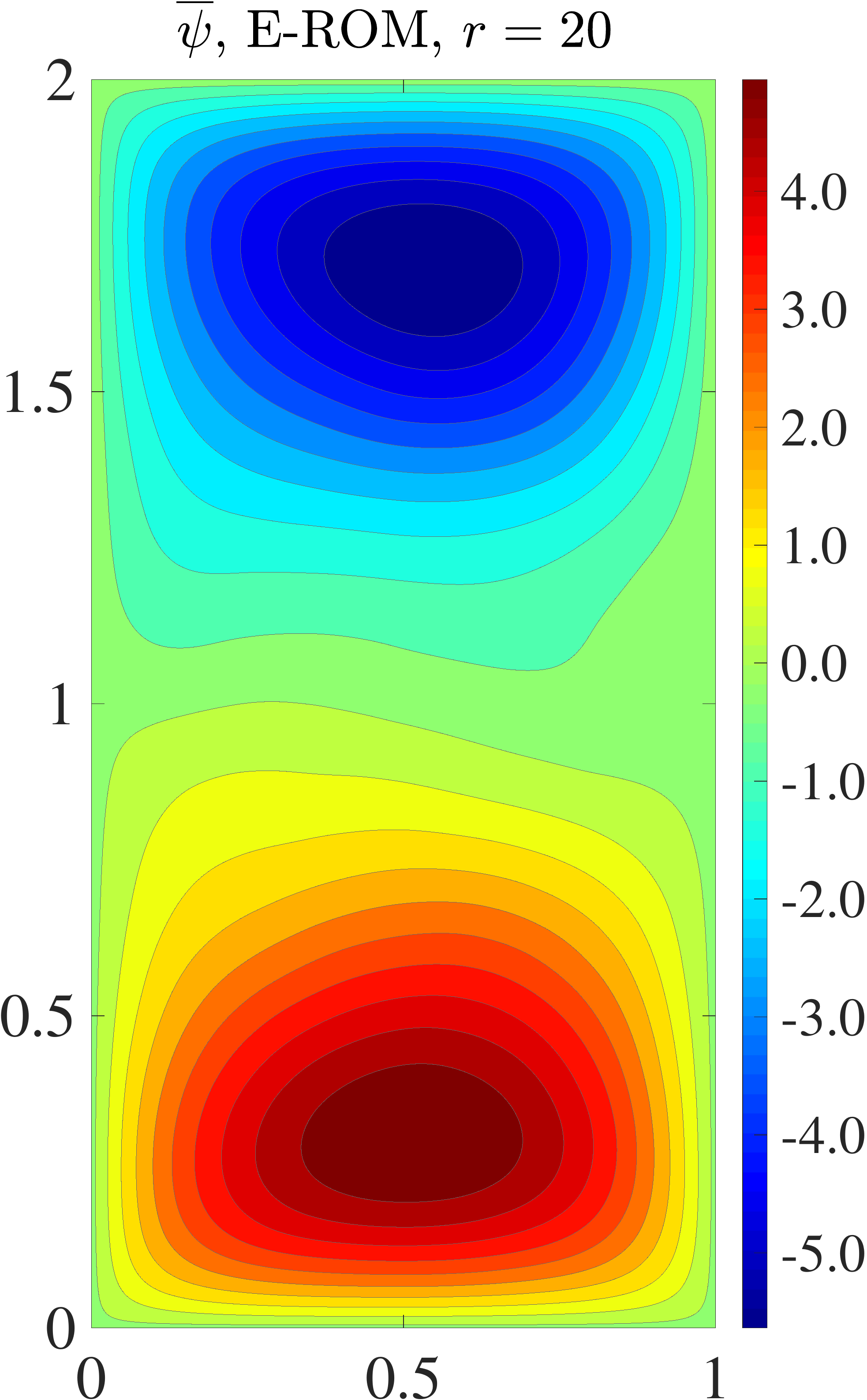}
\includegraphics[width=0.243\linewidth]{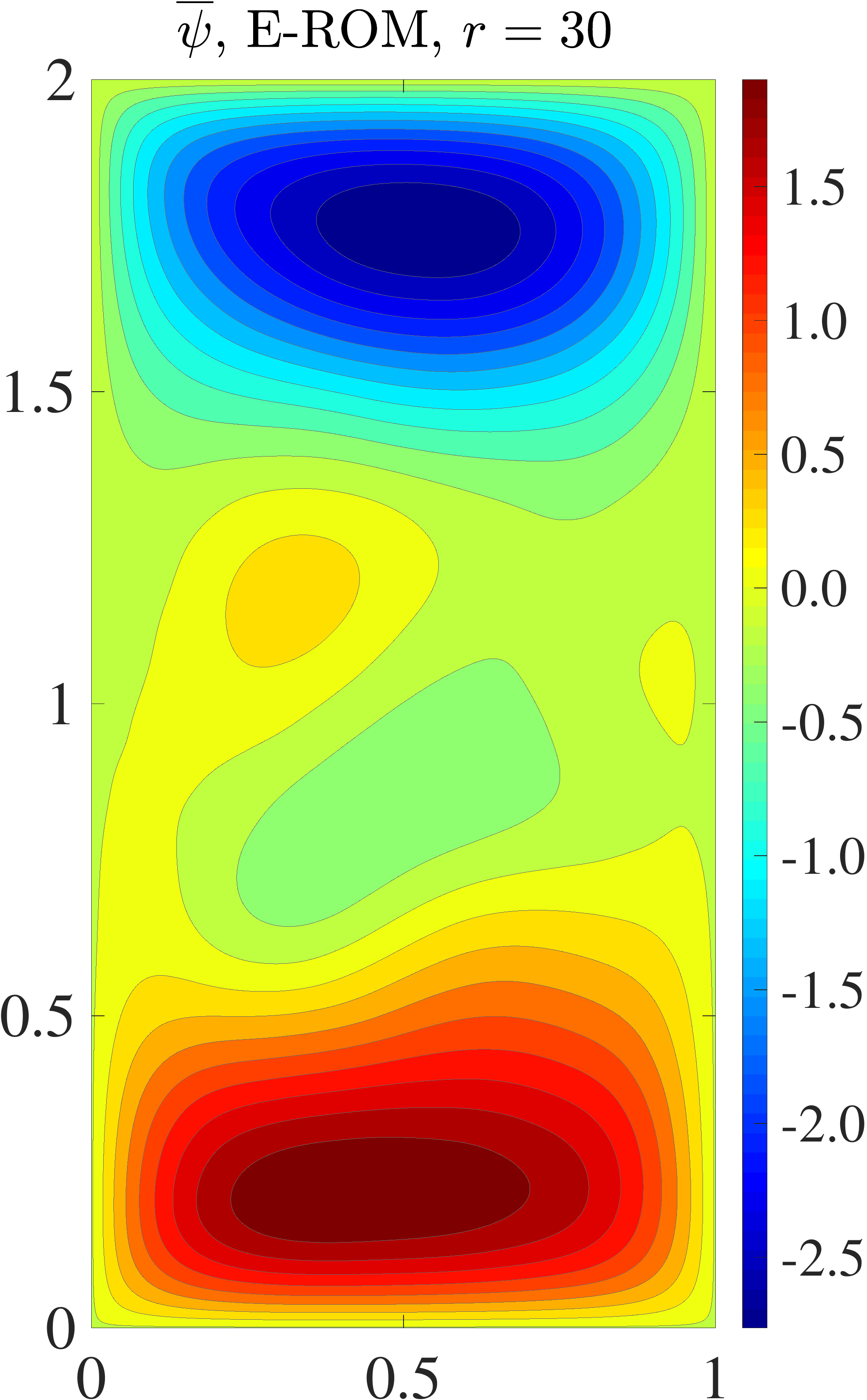}
\vspace{0.3cm}

\includegraphics[width=0.24\linewidth]{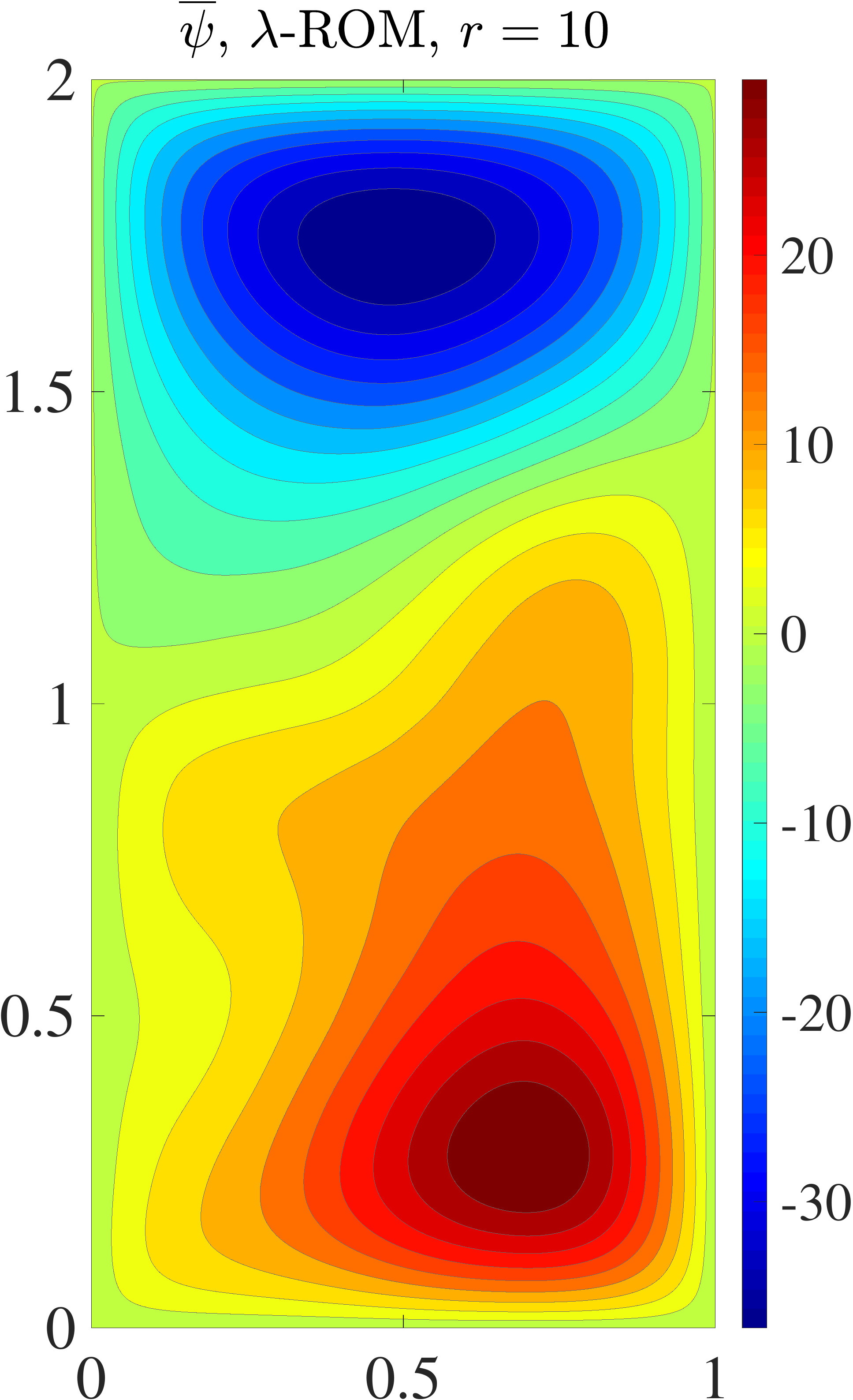}
\includegraphics[width=0.24\linewidth]{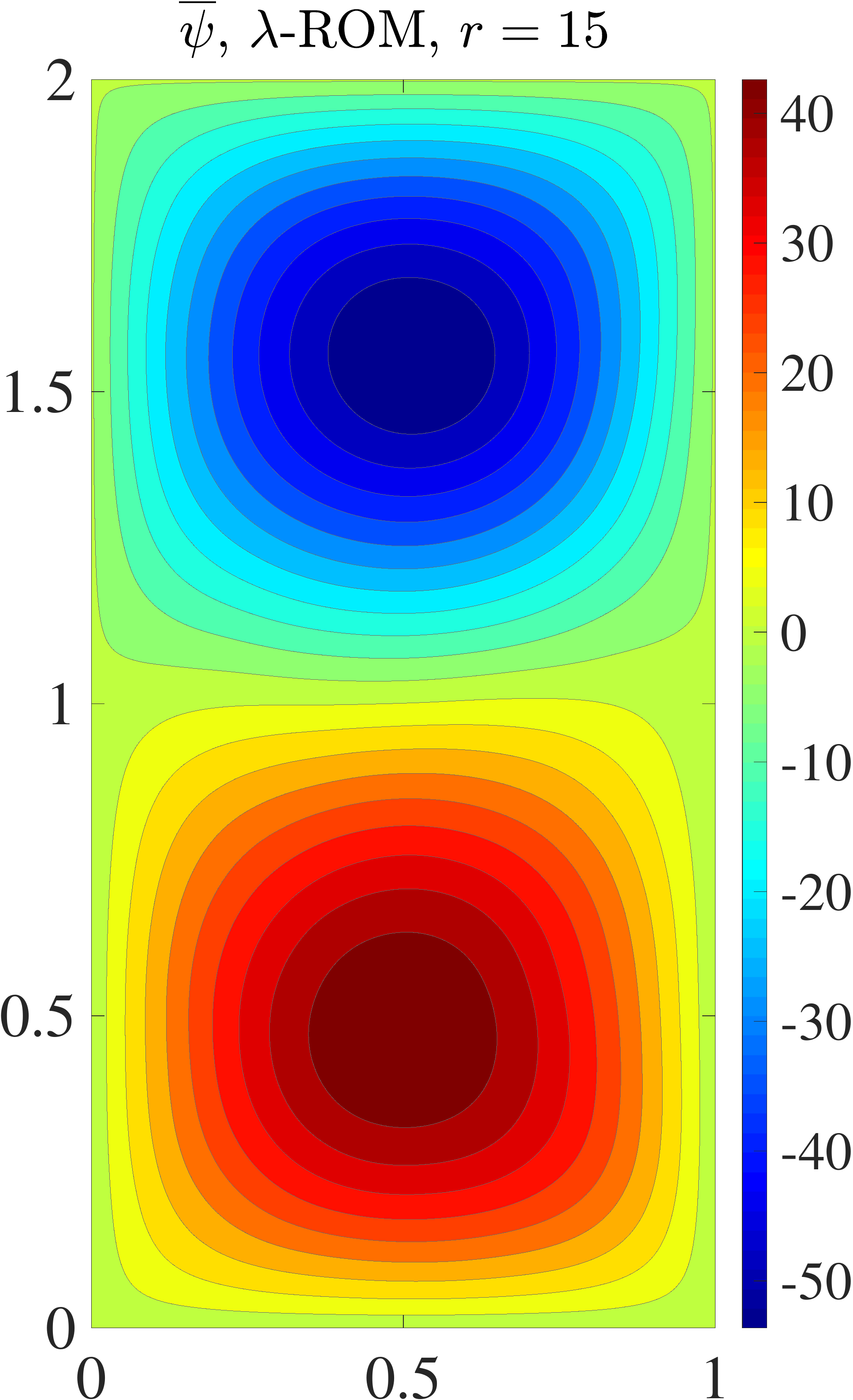}
\includegraphics[width=0.243\linewidth]{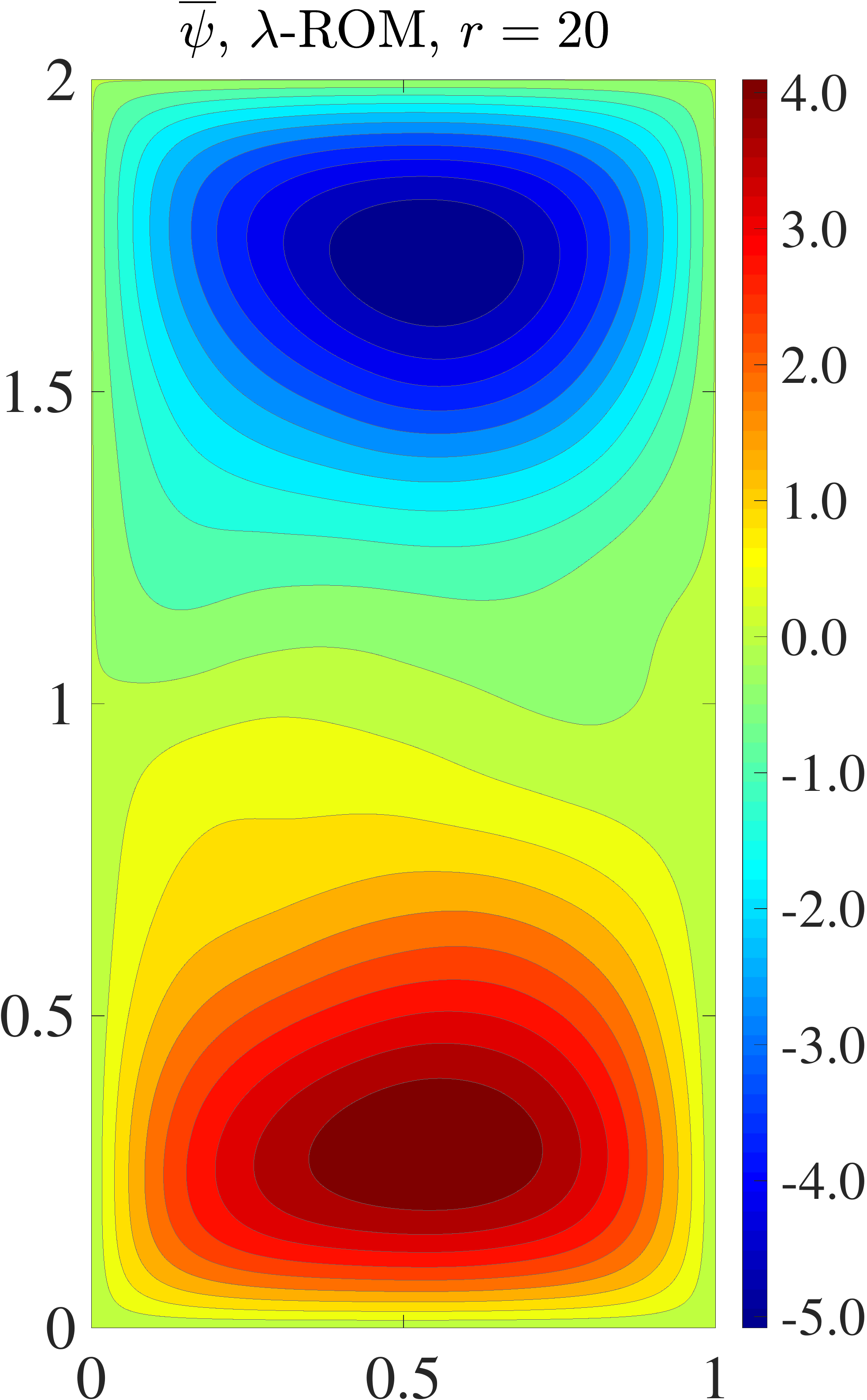}
\includegraphics[width=0.243\linewidth]{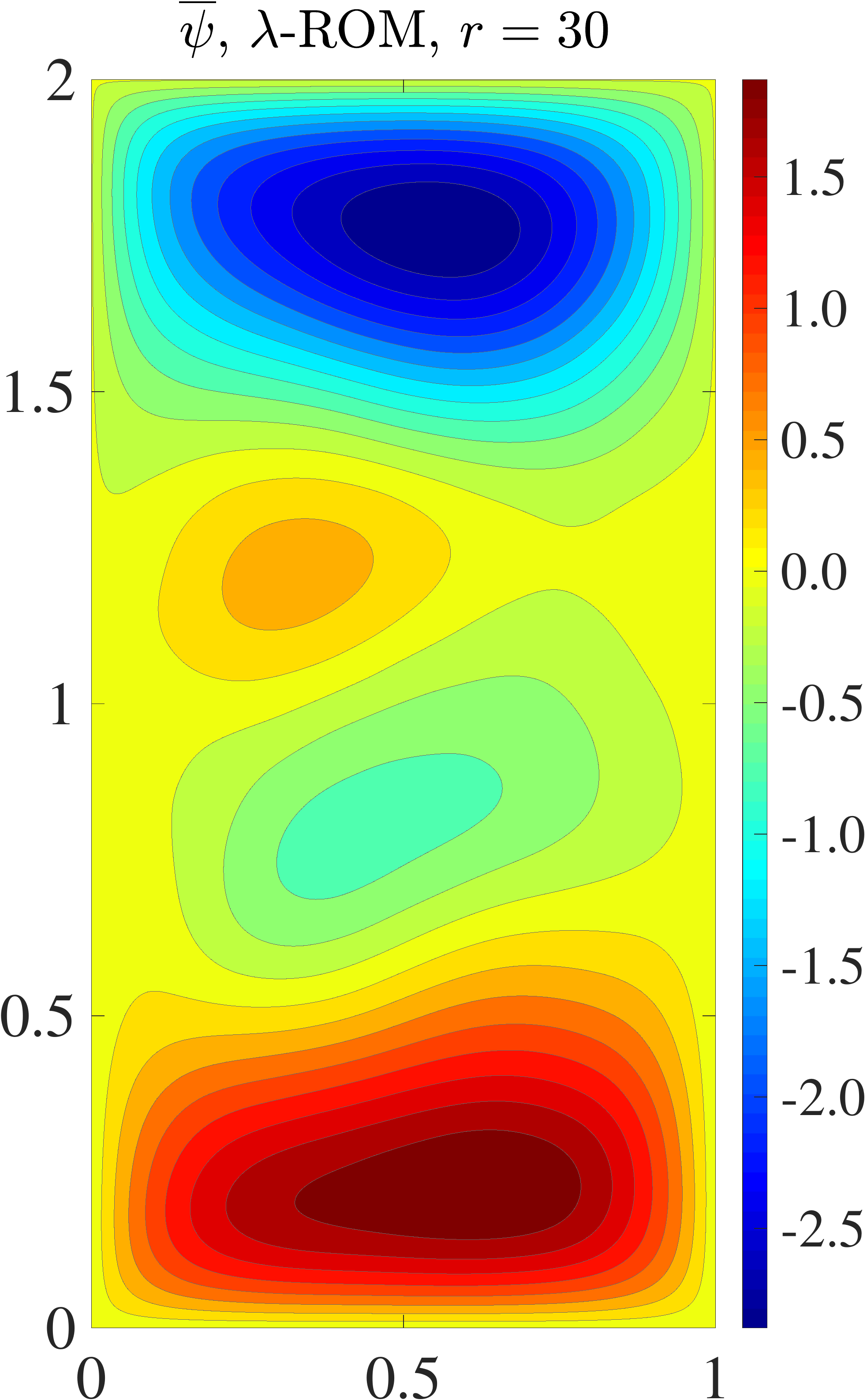}
\vspace{0.3cm}

\includegraphics[width=0.24\linewidth]{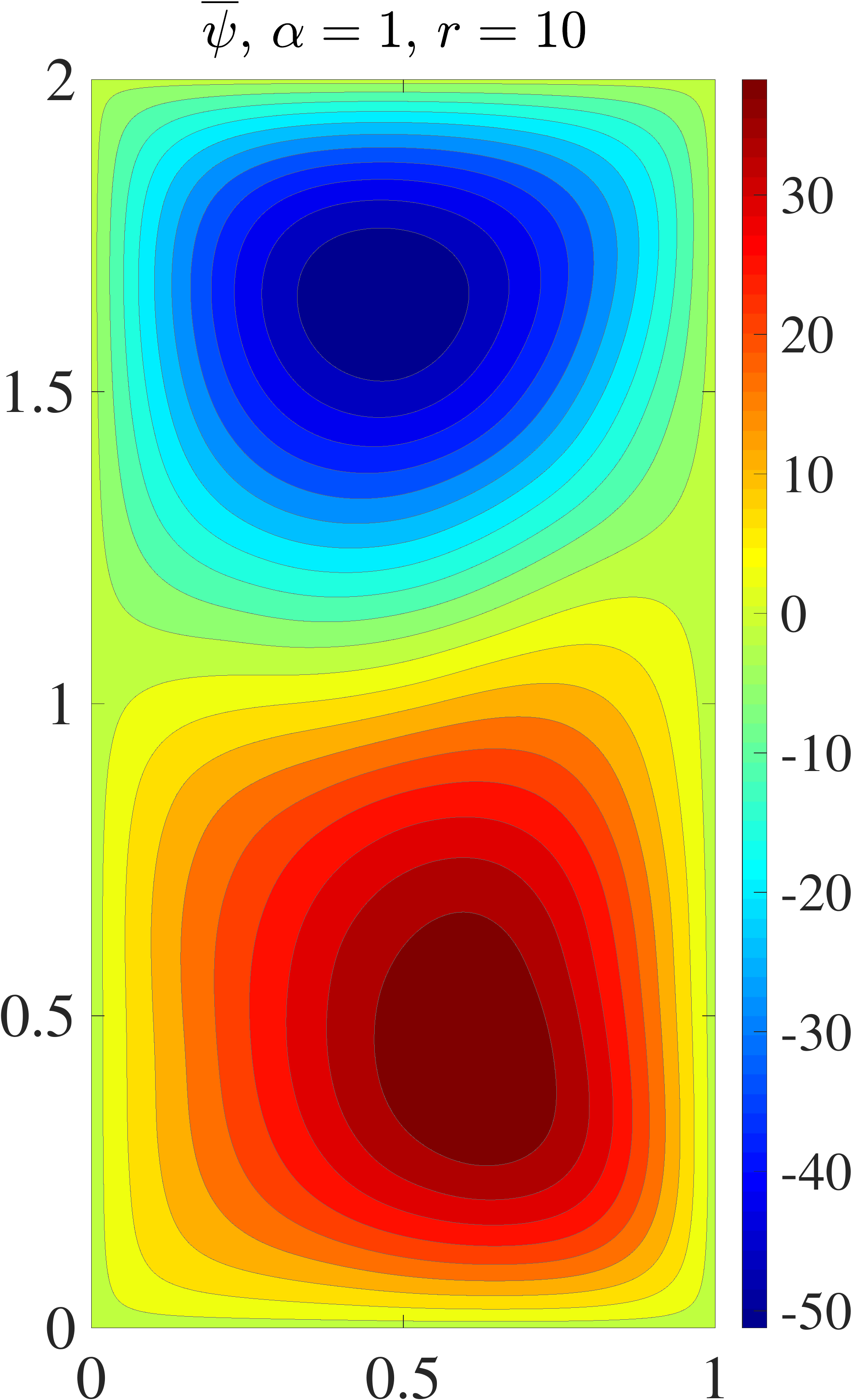}
\includegraphics[width=0.24\linewidth]{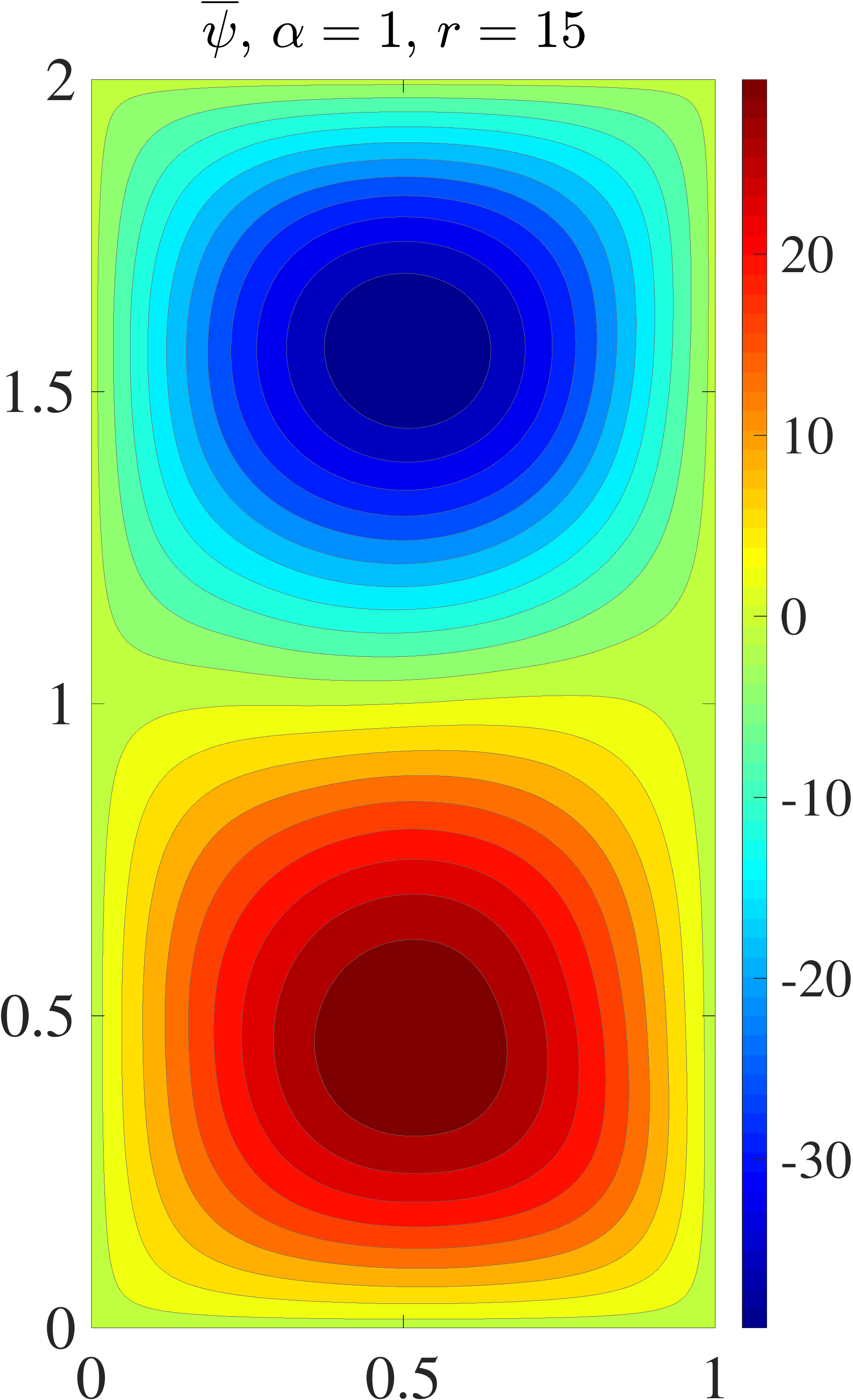}
\includegraphics[width=0.24\linewidth]{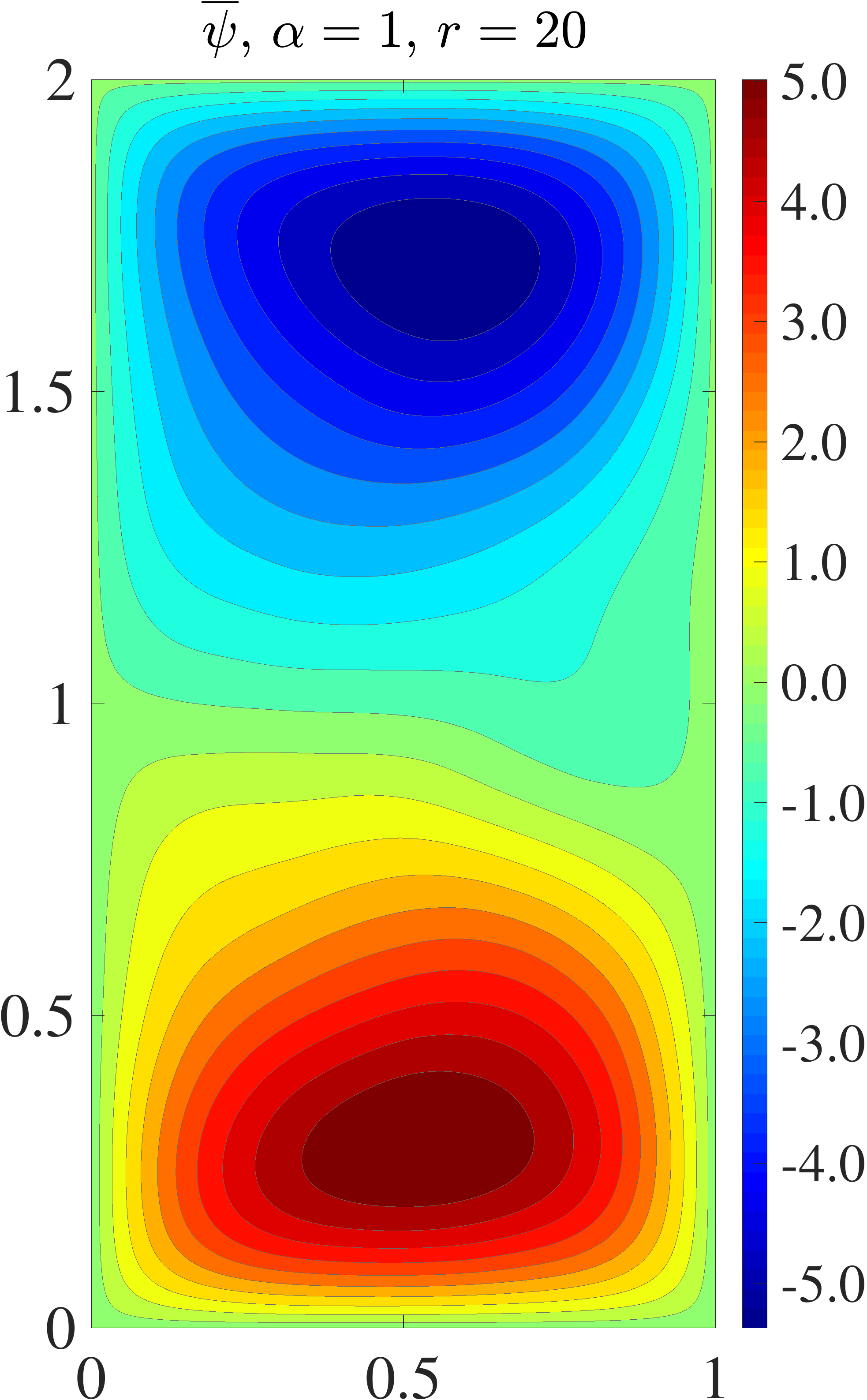}
\includegraphics[width=0.24\linewidth]{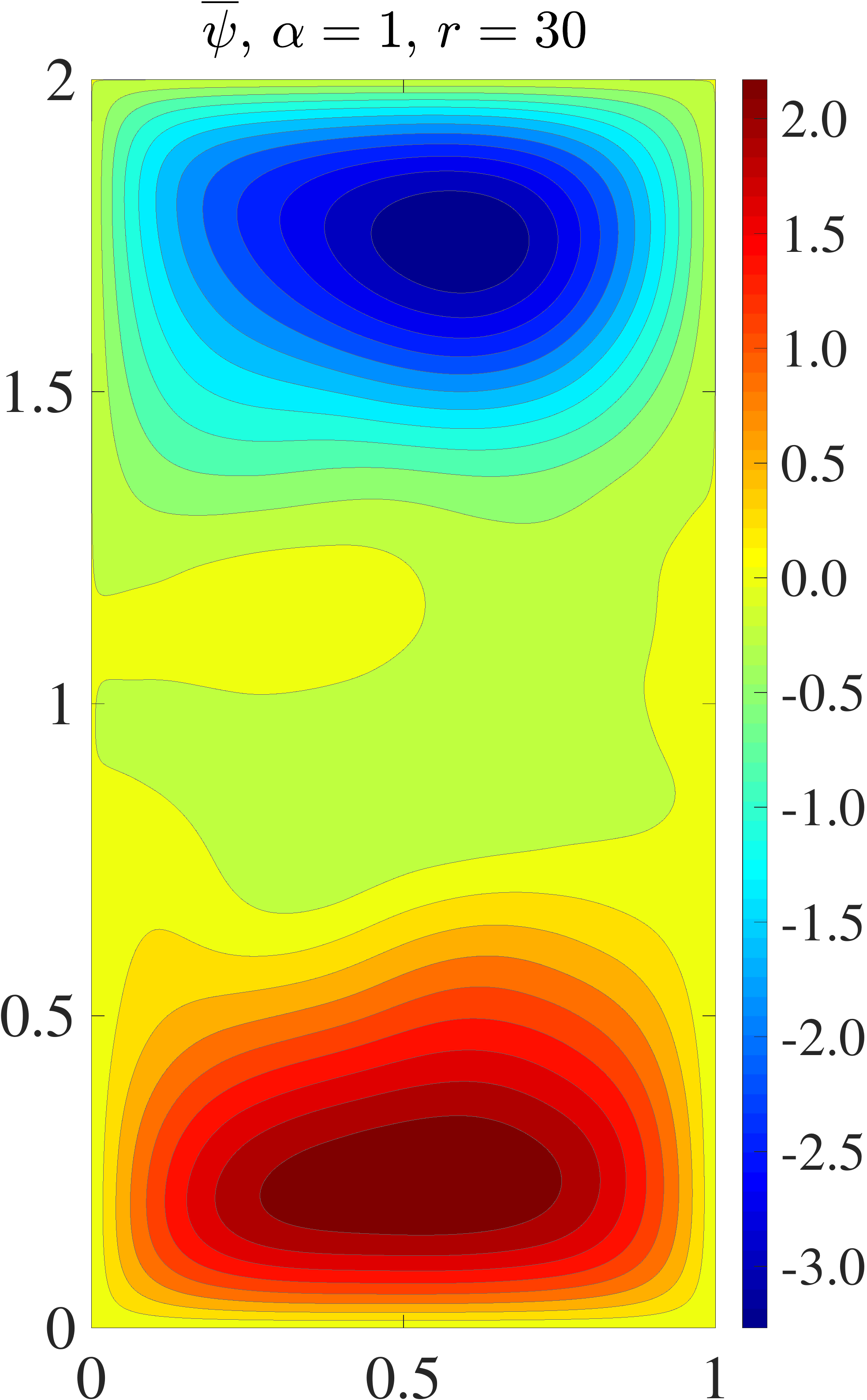}
\vspace{0.3cm}

\includegraphics[width=0.24\linewidth]{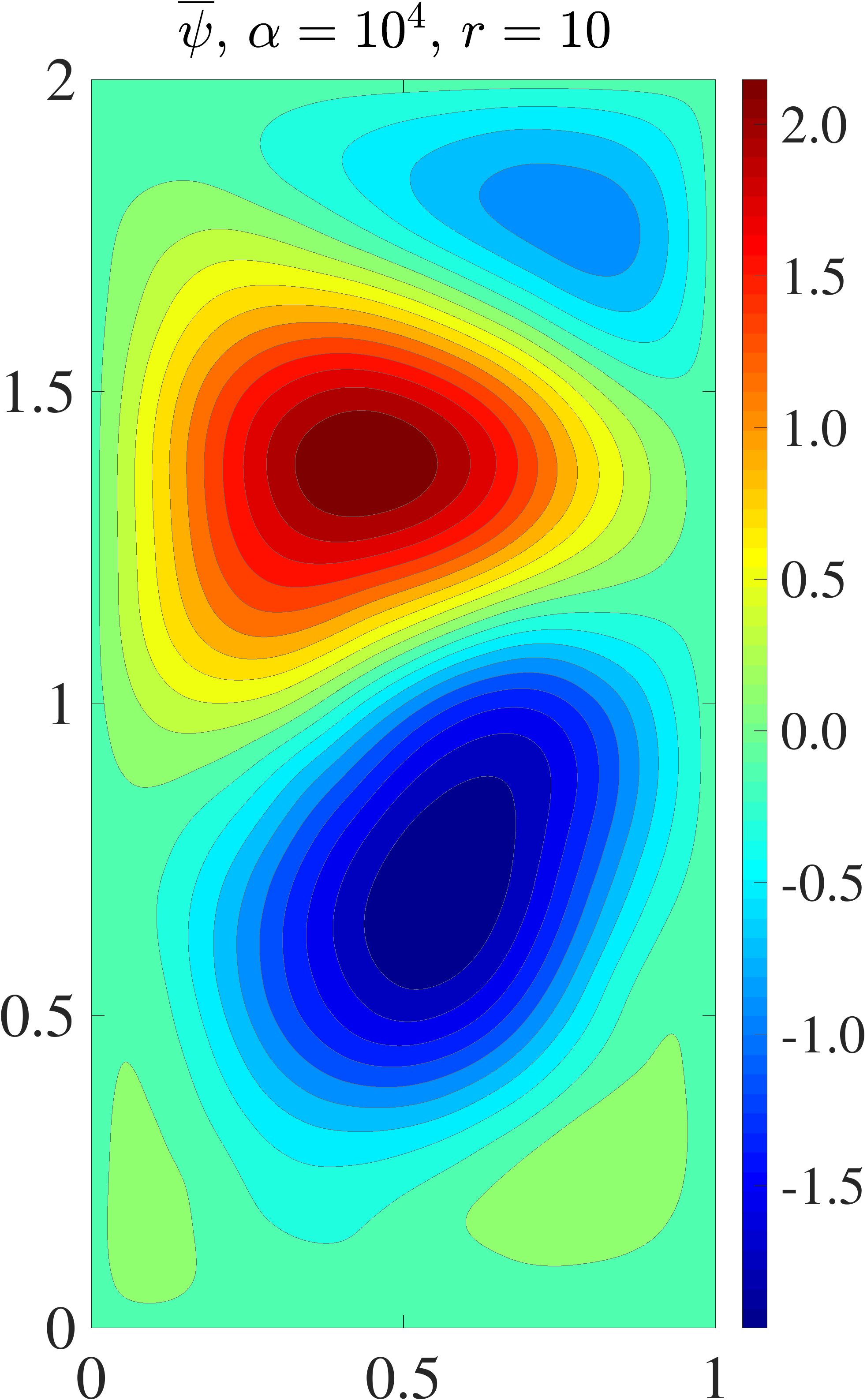}
\includegraphics[width=0.24\linewidth]{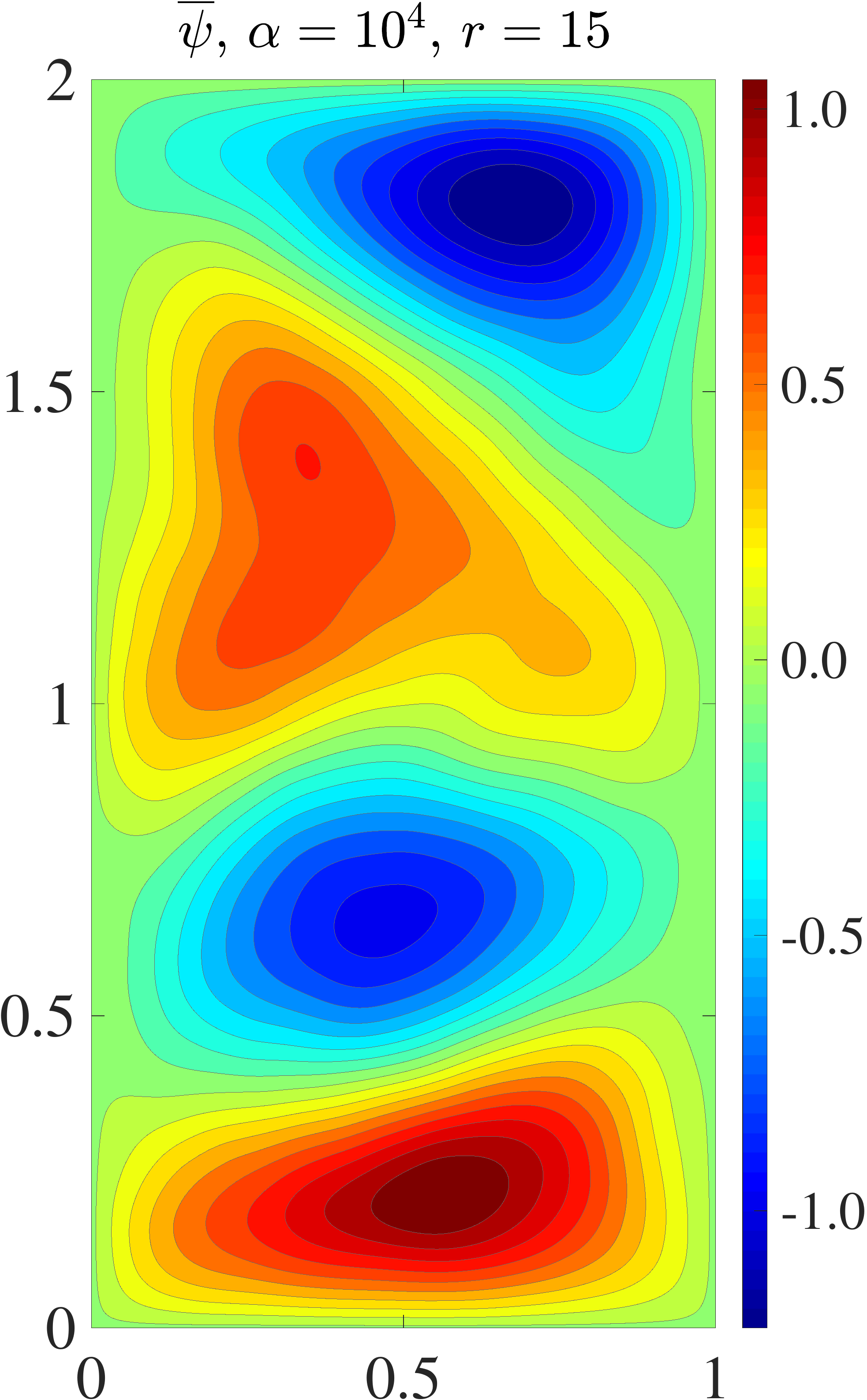}
\includegraphics[width=0.24\linewidth]{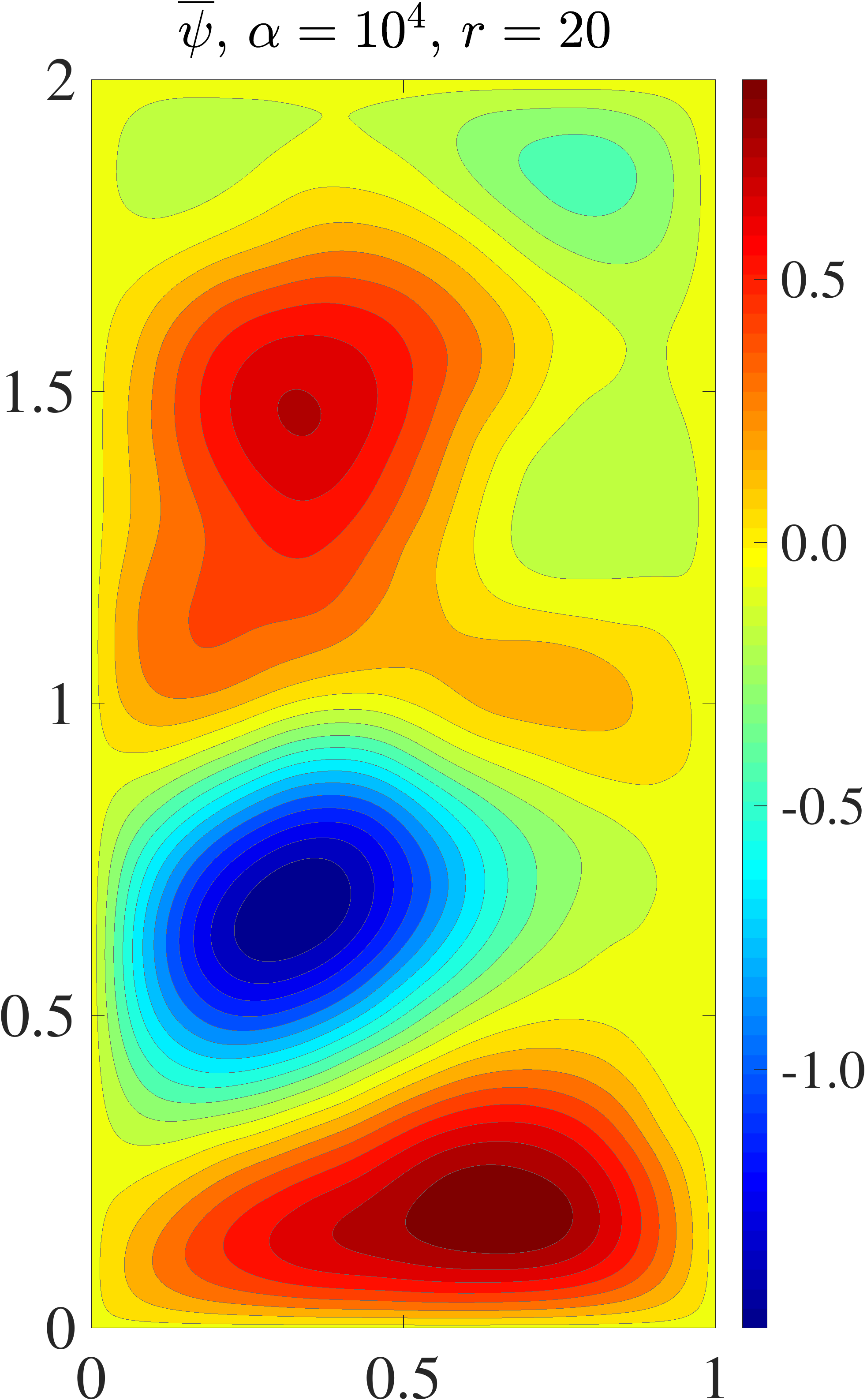}
\includegraphics[width=0.24\linewidth]{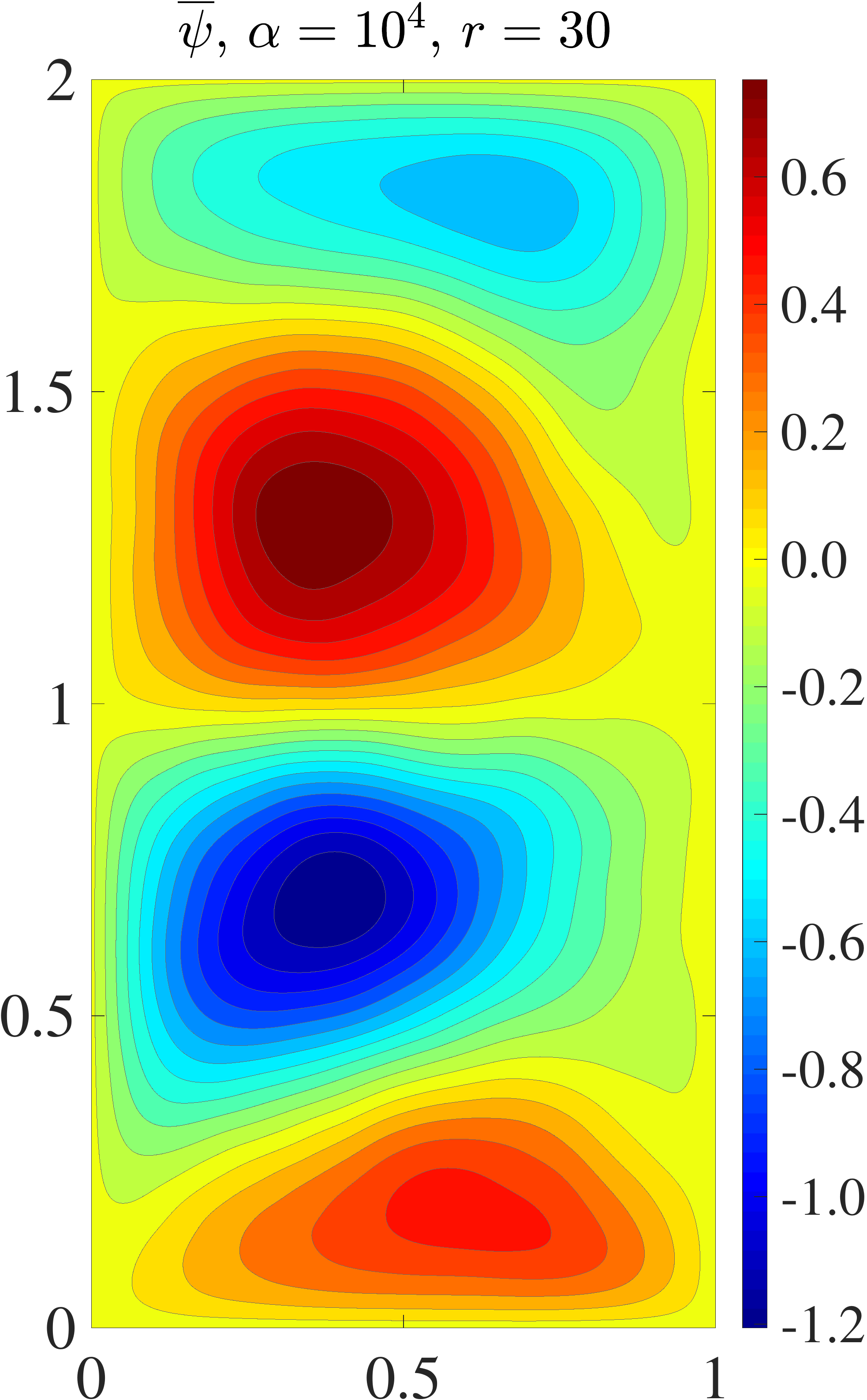}

\caption{
    Eulerian investigation, reconstructive regime:
	Mean streamfunction from 
	E-ROM (first row), 
	$\lambda$-ROM (second row)
	$\alpha$-ROM with $\alpha=1$ (third row), and
	$\alpha=10^{4}$ (fourth row), 
	for $r=10$ (first column), 
	$r=15$ (second column), 
	$r=20$ (third column), and 
	$r=30$ (fourth column).
	}
 \label{fig:mean-psi-meanrom}
 \end{figure}

Next, we use the qualitative Eulerian criterion to investigate the ability of the ROMs to  recover the four-gyre pattern of the time average of the streamfunction in Fig.~\ref{fig:experiment-2-2}.
In Fig.~\ref{fig:mean-psi-meanrom}, for $r=10, 15, 20$, and $30$, we plot the mean streamfunction for E-ROM, $\lambda$-ROM, and $\alpha$-ROM with $\alpha = 1$ and $\alpha = 10^{4}$. 
These results yield the following conclusions:
The E-ROM, the $\lambda$-ROM, and the $\alpha$-ROM with a low $\alpha$ value (i.e., $\alpha = 1$) yield similar results: These ROMs cannot recover the four-gyre pattern for any of the four $r$ values.
However, the $\alpha$-ROM with a large $\alpha$ value (i.e., $\alpha = 10^{4}$) yields dramatically better results: This ROM can clearly capture the four-gyre pattern for $r=30$; for $r=20$, the pattern is somewhat captured, although not as clearly as for $r=30$; finally, for $r=10$ and $r=15$, only hints of the four-gyre pattern are present.
Overall, the plots in Fig.~\ref{fig:mean-psi-meanrom} show that the {\it Lagrangian $\alpha$-ROM with a large $\alpha$ value} (i.e., $\alpha = 10^{4}$) {\it can capture the four-gyre pattern}, whereas the {\it standard E-ROM and the Lagrangian $\lambda$-ROM cannot}.

\bigskip 

The results in Table~\ref{table:error-psi-bar-reconstructive} and Fig.~\ref{fig:mean-psi-meanrom} consistently show that the {\it new Lagrangian $\alpha$-ROM with large $\alpha$ values outperforms the standard E-ROM and the Lagrangian $\lambda$-ROM} with respect to the two Eulerian metrics used in this section.
These results also show that the Lagrangian data used to construct the new Lagrangian $\alpha$-ROM play an important role: {\it the higher the Lagrangian data contribution} (i.e., the higher the $\alpha$ value), {\it the more accurate the results}.

\clearpage



\paragraph{Predictive Regime:}
For the predictive regime, we check whether the investigated ROMs can predict the dynamics of the underlying system. 
Specifically, we validate the ROMs on a time interval that is twice as long as the time interval used to train the ROMs.
In Table~\ref{table:error-psi-bar-predictive}, for different $r$ values, we list the $L^2$ norm of the errors in the time-averaged streamfunction~\eqref{eqn:norm-time-averaged-streamfunction} for E-ROM, $\lambda$-ROM, and $\alpha$-ROM with $\alpha = 1, \alpha=10, \alpha=10^{2}, \alpha=10^{3}$, and $\alpha = 10^{4}$. 
These results yield conclusions that are similar to those drawn in the reconstructive regime:
For low $r$ values, the E-ROM yields inaccurate results. 
The $\lambda$-ROM results are similar to or slightly better than the E-ROM results for all $r$ values.
The results for $\alpha$-ROM with $\alpha = 1, \alpha = 10$, and $\alpha = 10^{2}$ are generally better than or similar to the E-ROM results.
The results for $\alpha$-ROM with $\alpha = 10^{3}$ and $\alpha = 10^{4}$ are dramatically better than the E-ROM results:
For example, for $r=5$, the only ROM that yields acceptable results is the $\alpha$-ROM with $\alpha = 10^{4}$; all other ROMs simply blow up (this is denoted by ``N/A'' in Table~\ref{table:error-psi-bar-predictive}).
Furthermore, for $r=10$, $r=15$ and $r=25$, the errors of $\alpha$-ROM with $\alpha = 10^{3}$ and $\alpha = 10^{4}$ are {\it at least one order of magnitude} lower than the E-ROM error.
For the larger $r$ values, the errors of $\alpha$-ROM with $\alpha = 10^{3}$ and $\alpha = 10^{4}$ continue to be lower than the E-ROM errors, although (as expected) the differences decrease as the $r$ values increase.
Overall, the results in Table~\ref{table:error-psi-bar-predictive} show that, in the predictive regime, {\it the Lagrangian $\alpha$-ROM with high $\alpha$ values} (i.e., $\alpha=10^{3}$ and $\alpha=10^{4}$) yields {\it significantly more accurate results than the standard E-ROM and the Lagrangian $\lambda$-ROM}, especially for the small $r$ values. 

\bigskip

\begin{table}[H]
\centering
\caption{
        Eulerian investigation, predictive regime: 
		$L^2$ norm of the errors in the time-averaged streamfunction~\eqref{eqn:norm-time-averaged-streamfunction} for E-ROM (second column), $\lambda$-ROM (third column), and $\alpha$-ROM for $\alpha=1$ (fourth column), $\alpha=10$ (fifth column), $\alpha=10^{2}$ (sixth column), $\alpha=10^{3}$ (seventh column), and $\alpha=10^{4}$ (eighth column). }
	\label{table:error-psi-bar-predictive}
\begin{tabular}{|c|c|c|c|c|c|c|c|}
\hline
$r$& E-ROM&$\lambda$-ROM&$\alpha =1$ &$\alpha =10$ &$\alpha =10^2$&$\alpha =10^3$& $\alpha =10^4$    \\ \hline
 $5$&   N/A&        N/A&        N/A&        N/A&        N/A&        2.8e+01&    1.2e+01 \\ \hline
$10$&   2.6e+04&    8.6e+03&    1.6e+04&    5.3e+03&    4.8e+01&    8.1e+00&    7.7e+00 \\ \hline
$15$&   1.1e+01&    1.8e+01&    1.9e+01&    2.2e+01&    7.3e+00&    1.8e+00&    1.3e+00 \\ \hline
$20$&   9.0e+00&    6.3e+00&    8.8e+00&    9.5e+00&    1.7e+00&    3.5e-01&    5.8e-01 \\ \hline
$25$&   4.0e+00&    2.5e+00&    3.8e+00&    2.7e+00&    8.4e-01&    3.3e-01&    3.5e-01 \\ \hline
$30$&   1.0e+00&    7.7e-01&    8.8e-01&    6.8e-01&    6.0e-01&    1.9e-01&    6.2e-01 \\ \hline
$35$&   4.3e-01&    4.9e-01&    5.8e-01&    5.5e-01&    5.4e-01&    3.4e-01&    5.2e-01 \\ \hline
$40$&   4.8e-01&    3.8e-01&    3.8e-01&    4.9e-01&    4.4e-01&    7.6e-01&    5.7e-01 \\ \hline
$45$&   6.4e-01&    4.4e-01&    4.6e-01&    4.9e-01&    6.3e-01&    7.5e-01&    5.3e-01 \\ \hline
$50$&   4.0e-01&    4.9e-01&    3.52e-01&   3.8e-01&    6.1e-01&    4.0e-01&    4.3e-01 \\ \hline

\end{tabular}
\end{table}

\clearpage

\begin{figure}[htp]
\centering
\centering
\vspace{0.3cm}
\includegraphics[width=0.244\linewidth]{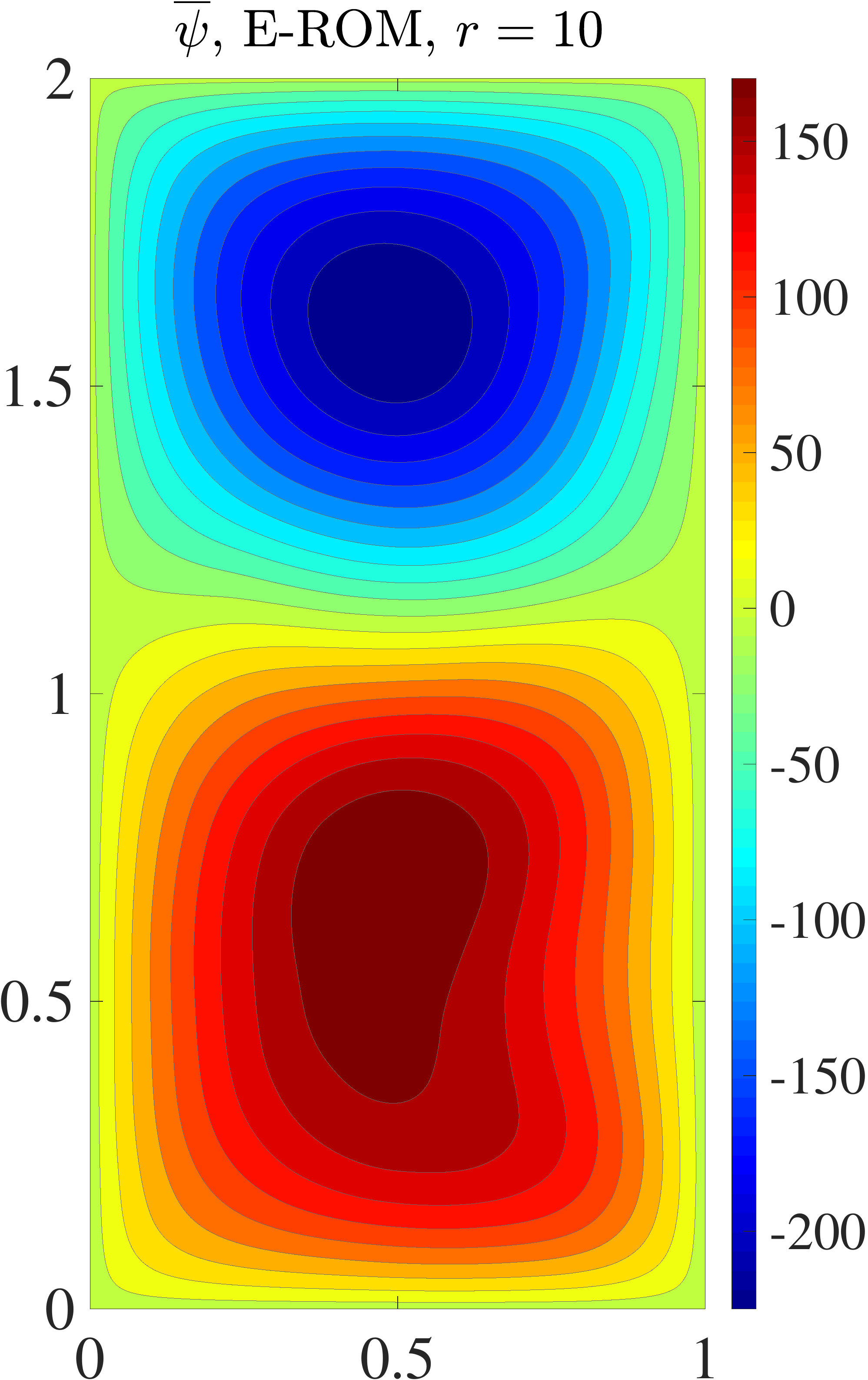}
\includegraphics[width=0.24\linewidth]{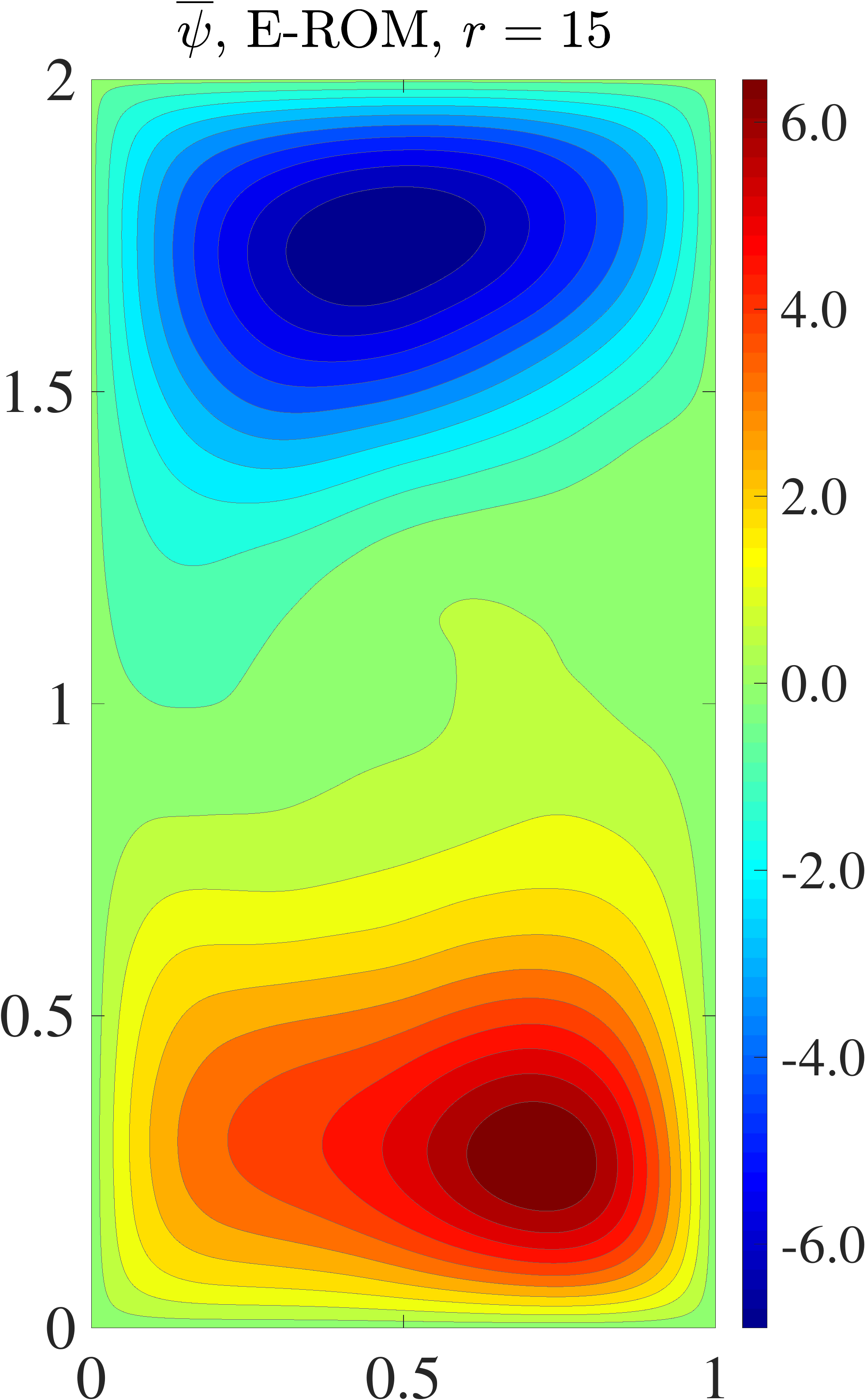}
\includegraphics[width=0.24\linewidth]{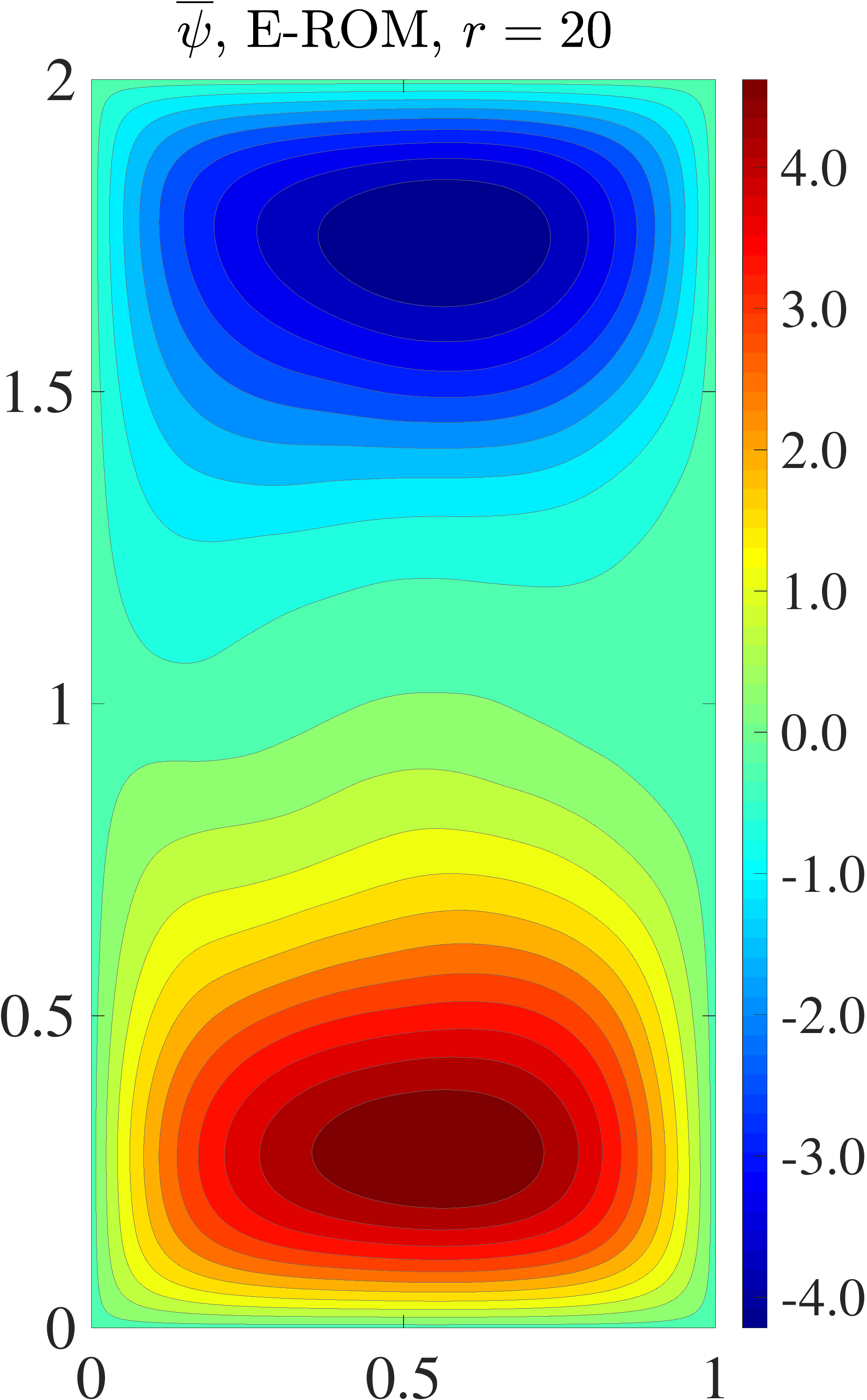}
\includegraphics[width=0.24\linewidth]{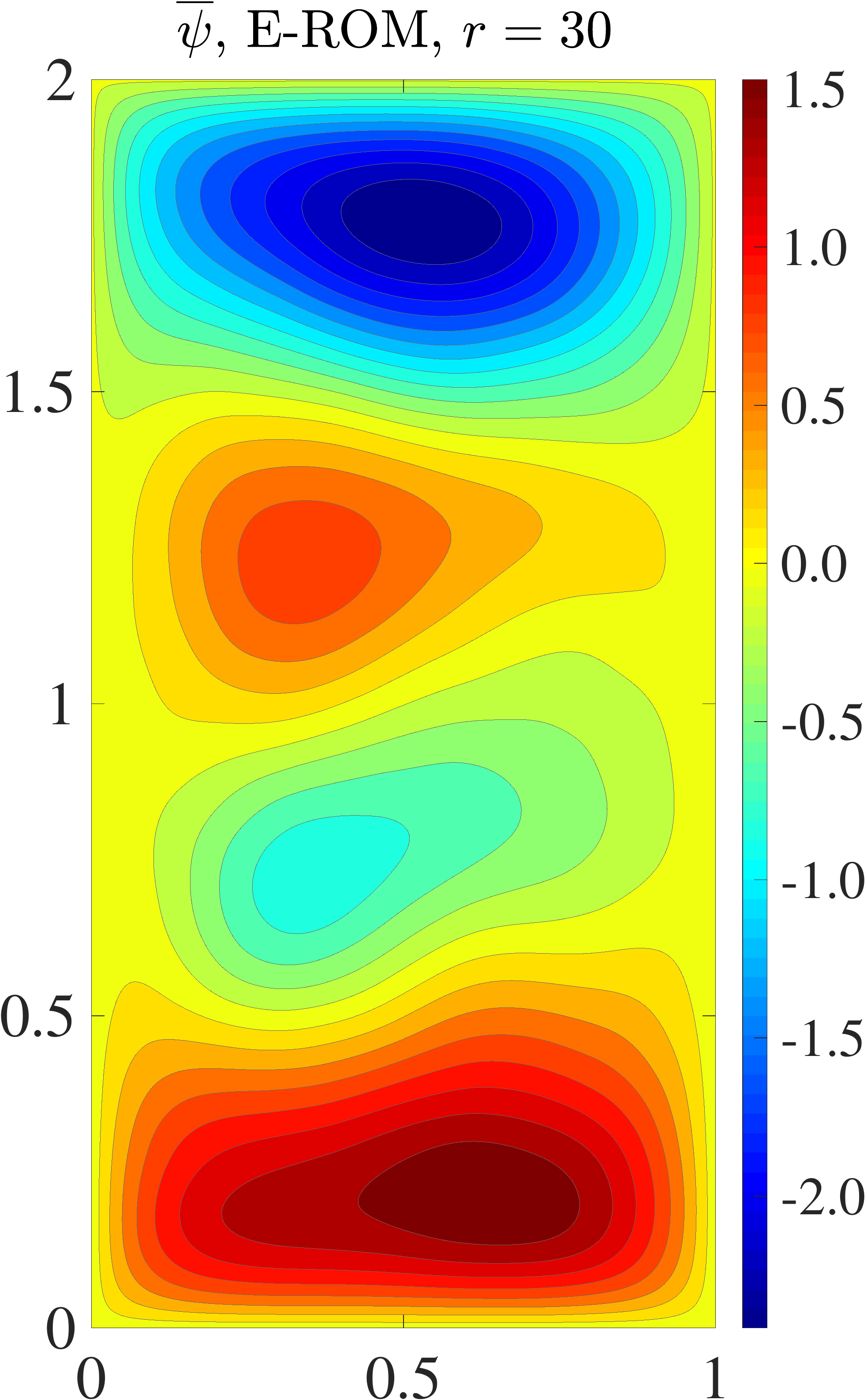}
\vspace{0.3cm}

\includegraphics[width=0.244\linewidth]{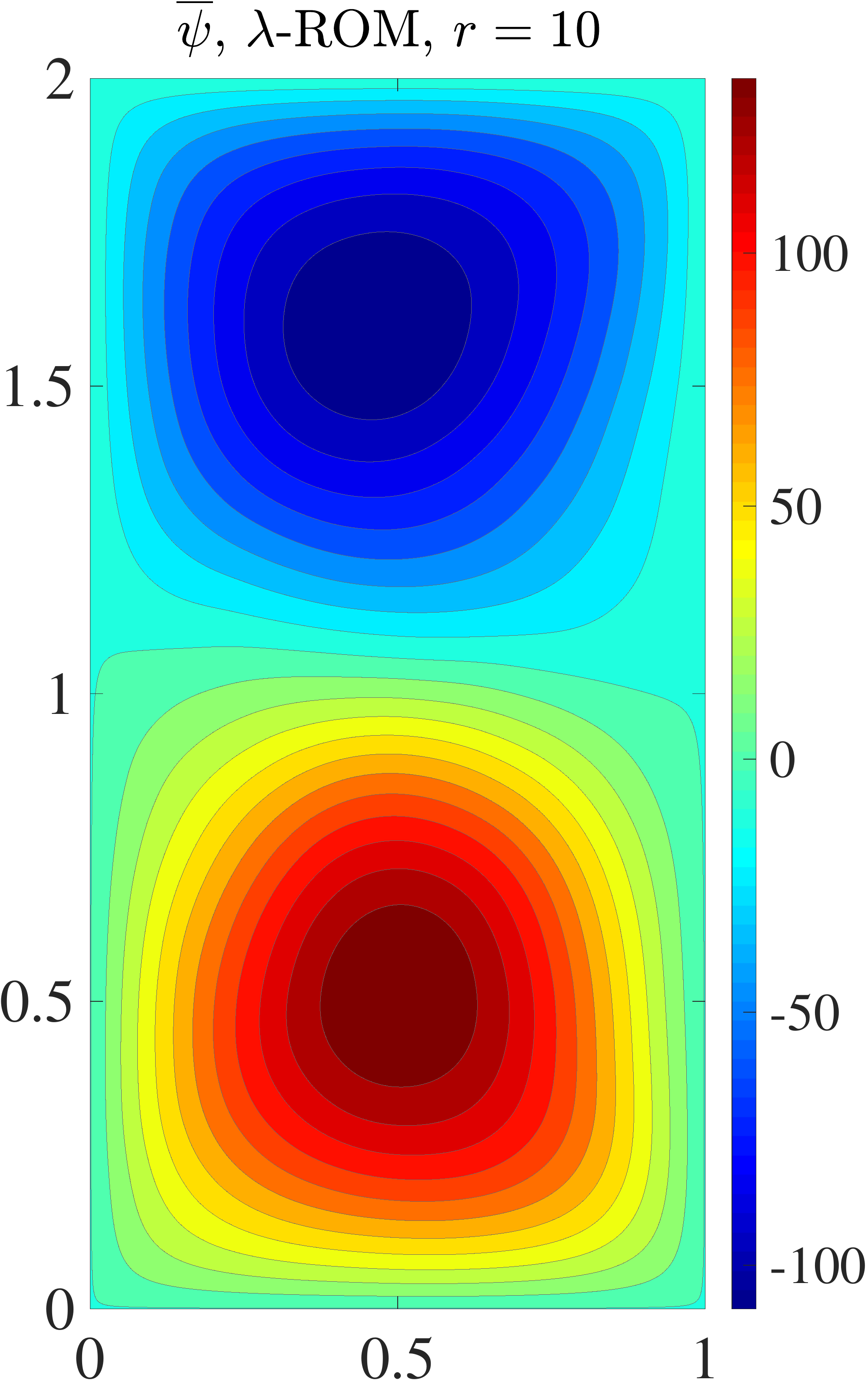}
\includegraphics[width=0.24\linewidth]{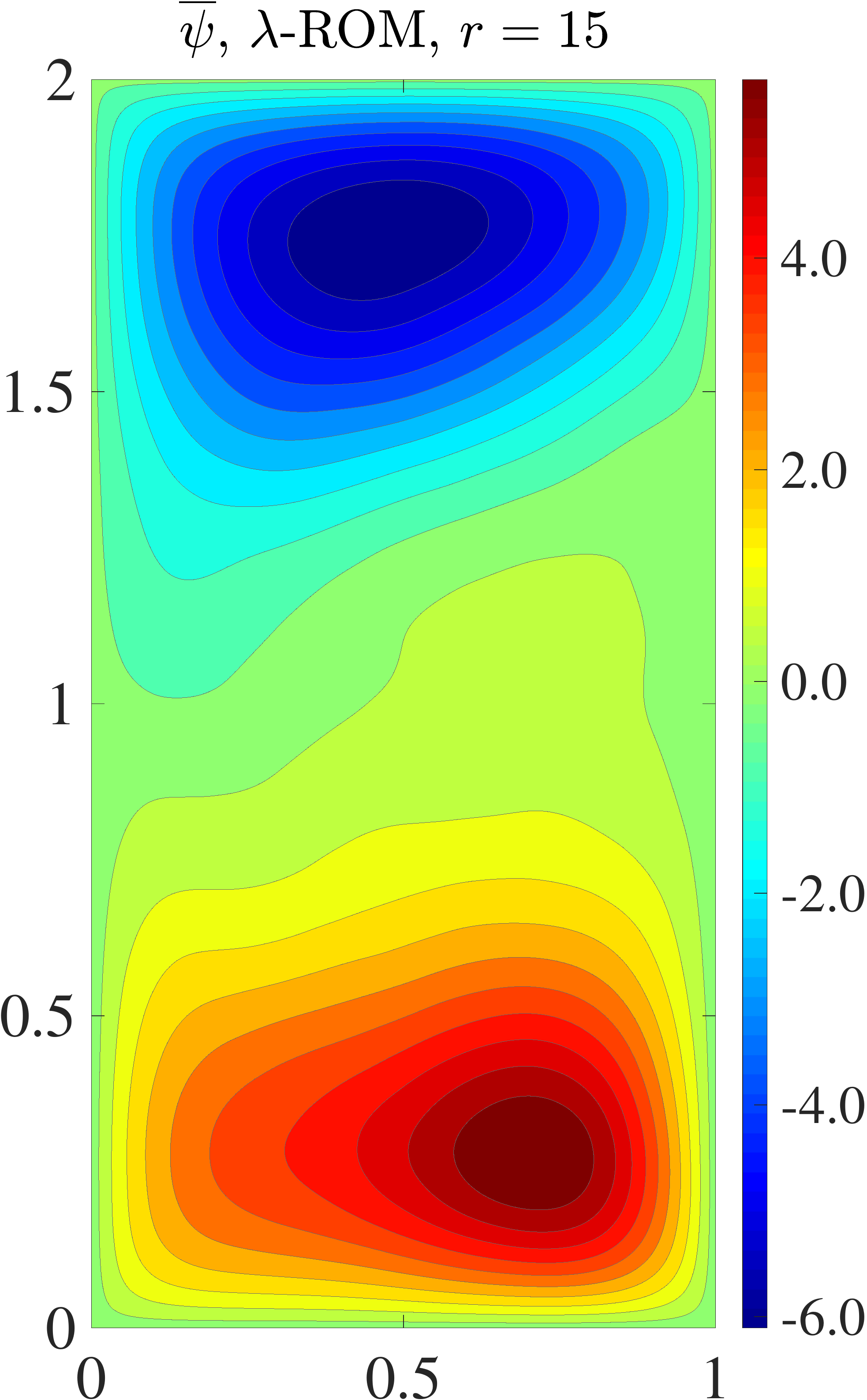}
\includegraphics[width=0.24\linewidth]{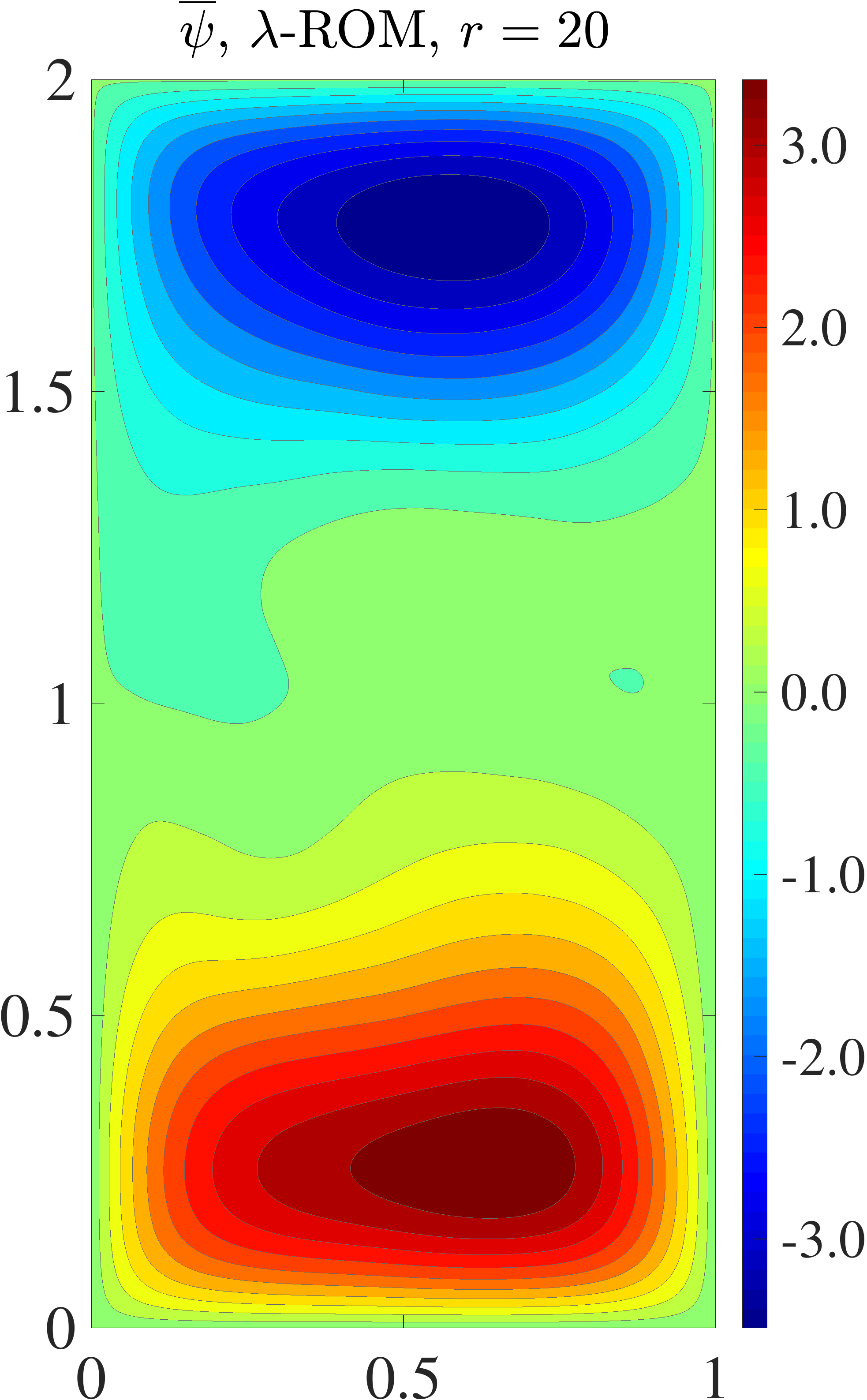}
\includegraphics[width=0.24\linewidth]{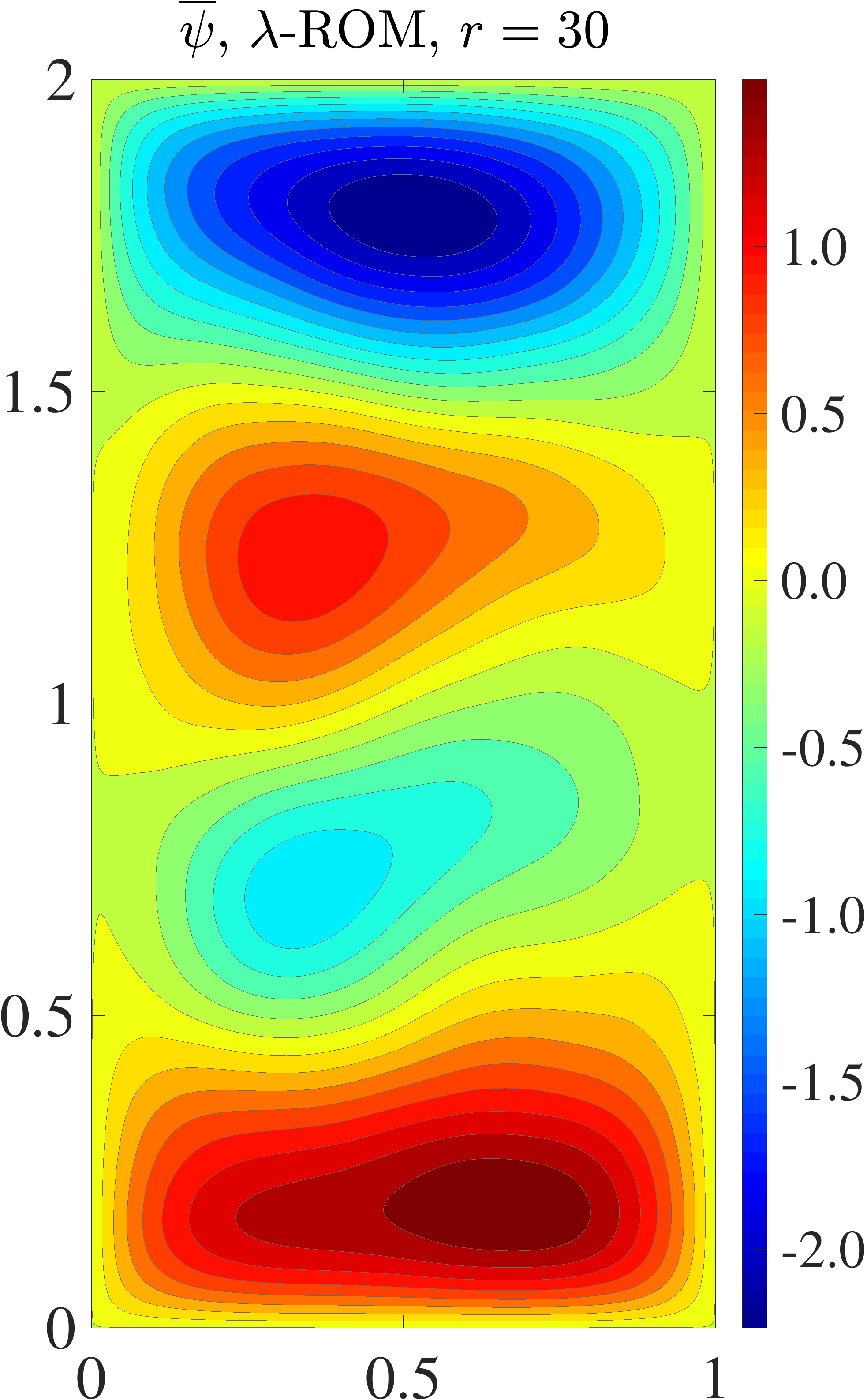}
\vspace{0.3cm}

\includegraphics[width=0.24\linewidth]{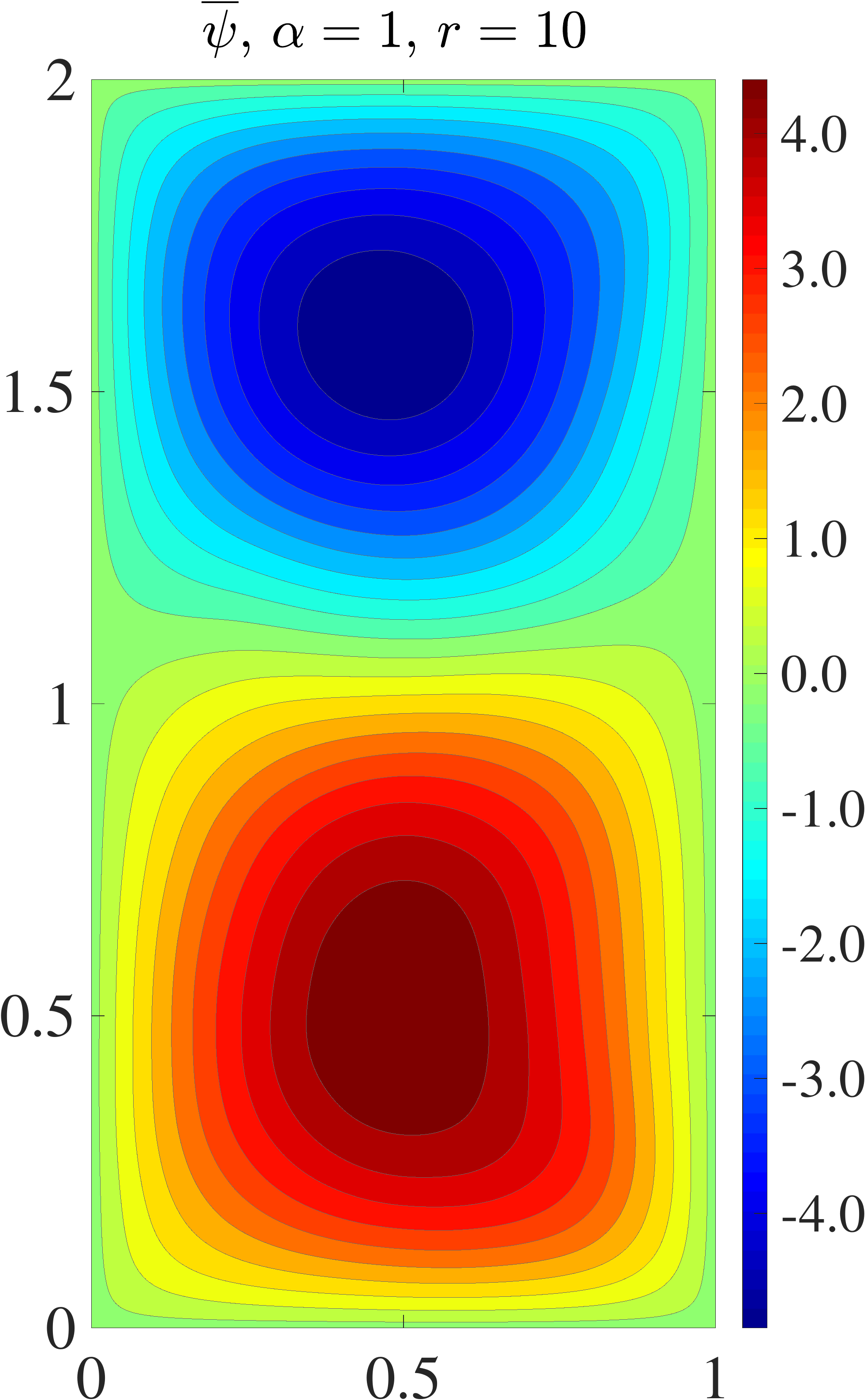}
\includegraphics[width=0.24\linewidth]{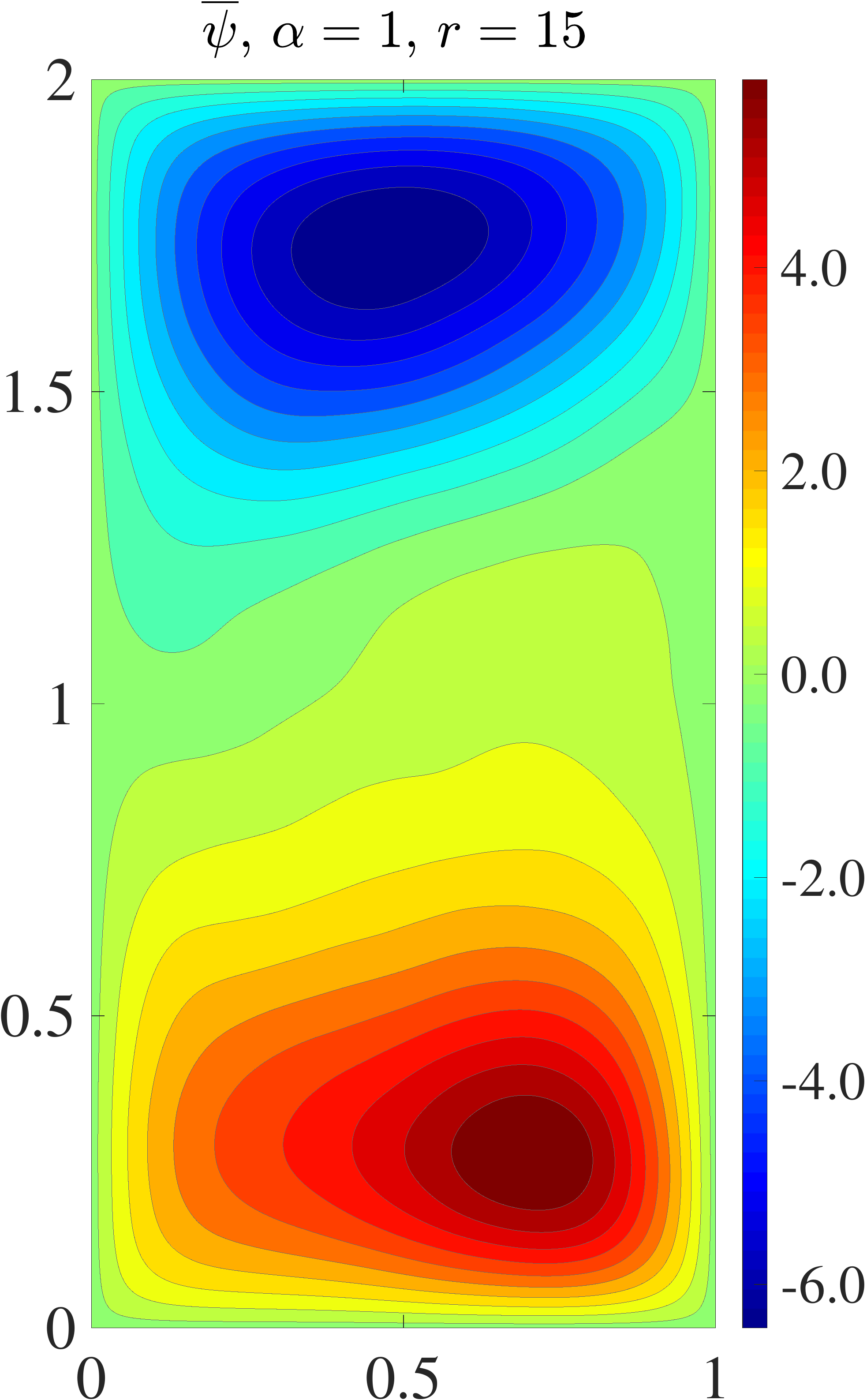}
\includegraphics[width=0.24\linewidth]{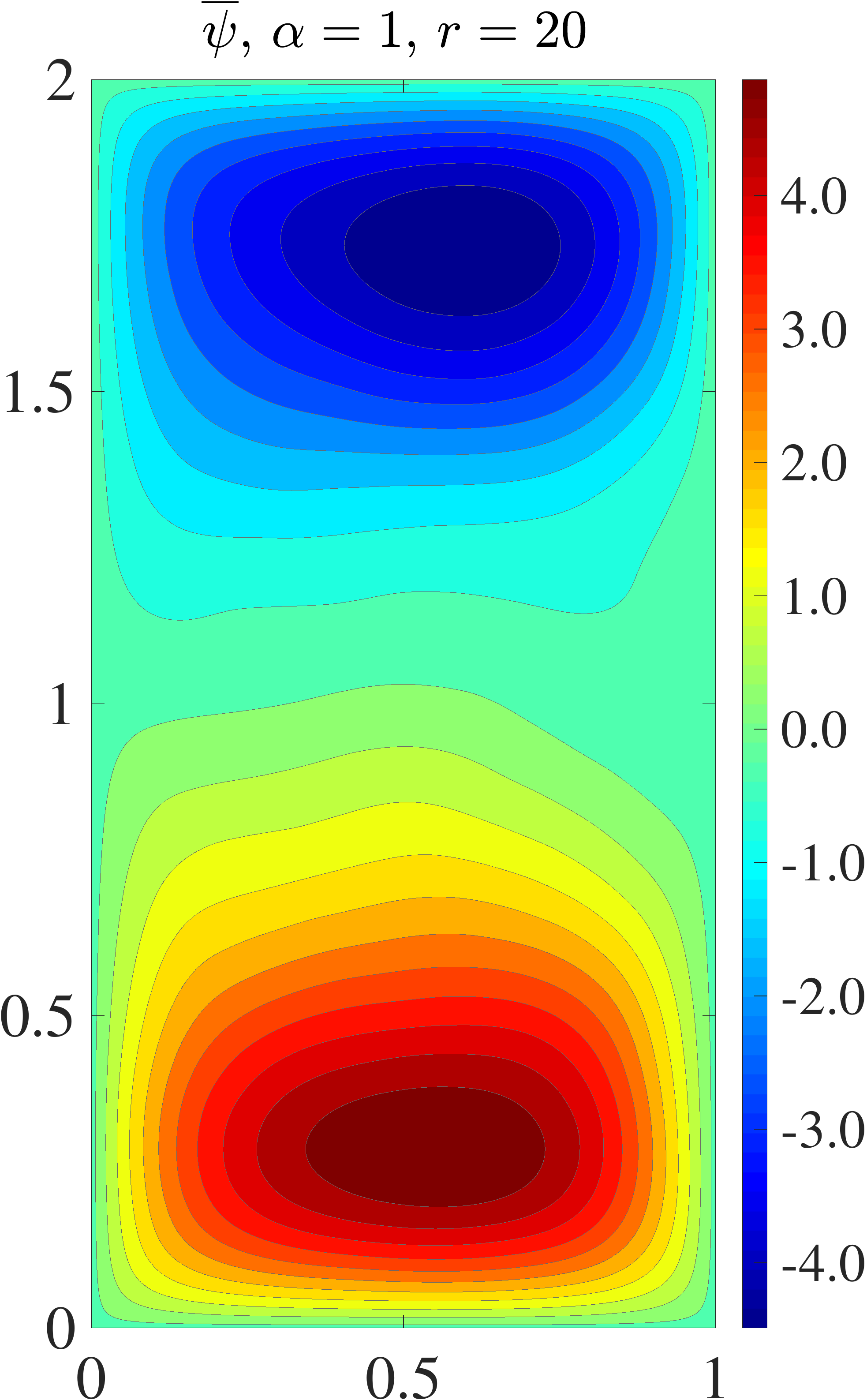}
\includegraphics[width=0.24\linewidth]{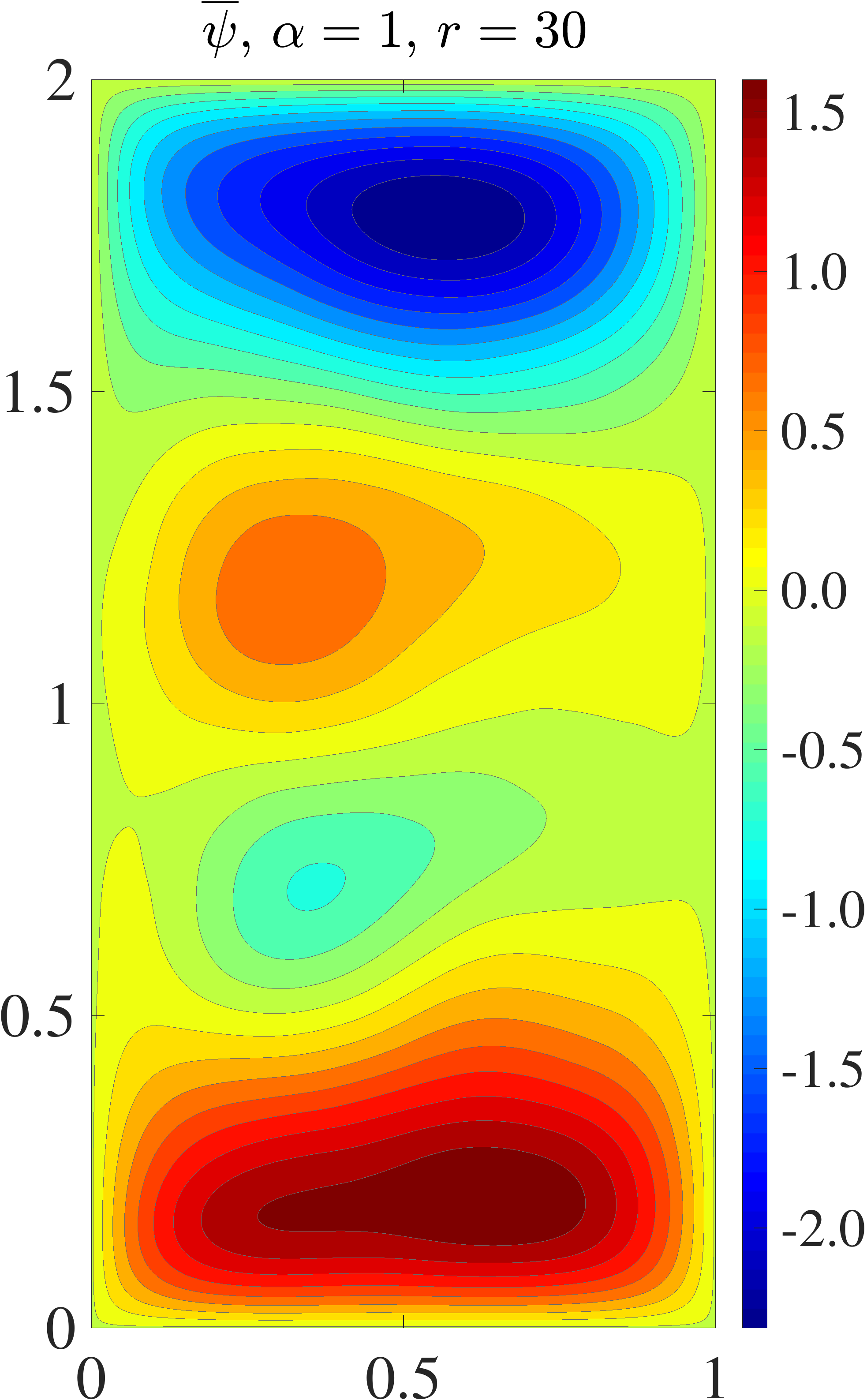}
\vspace{0.3cm}

\includegraphics[width=0.24\linewidth]{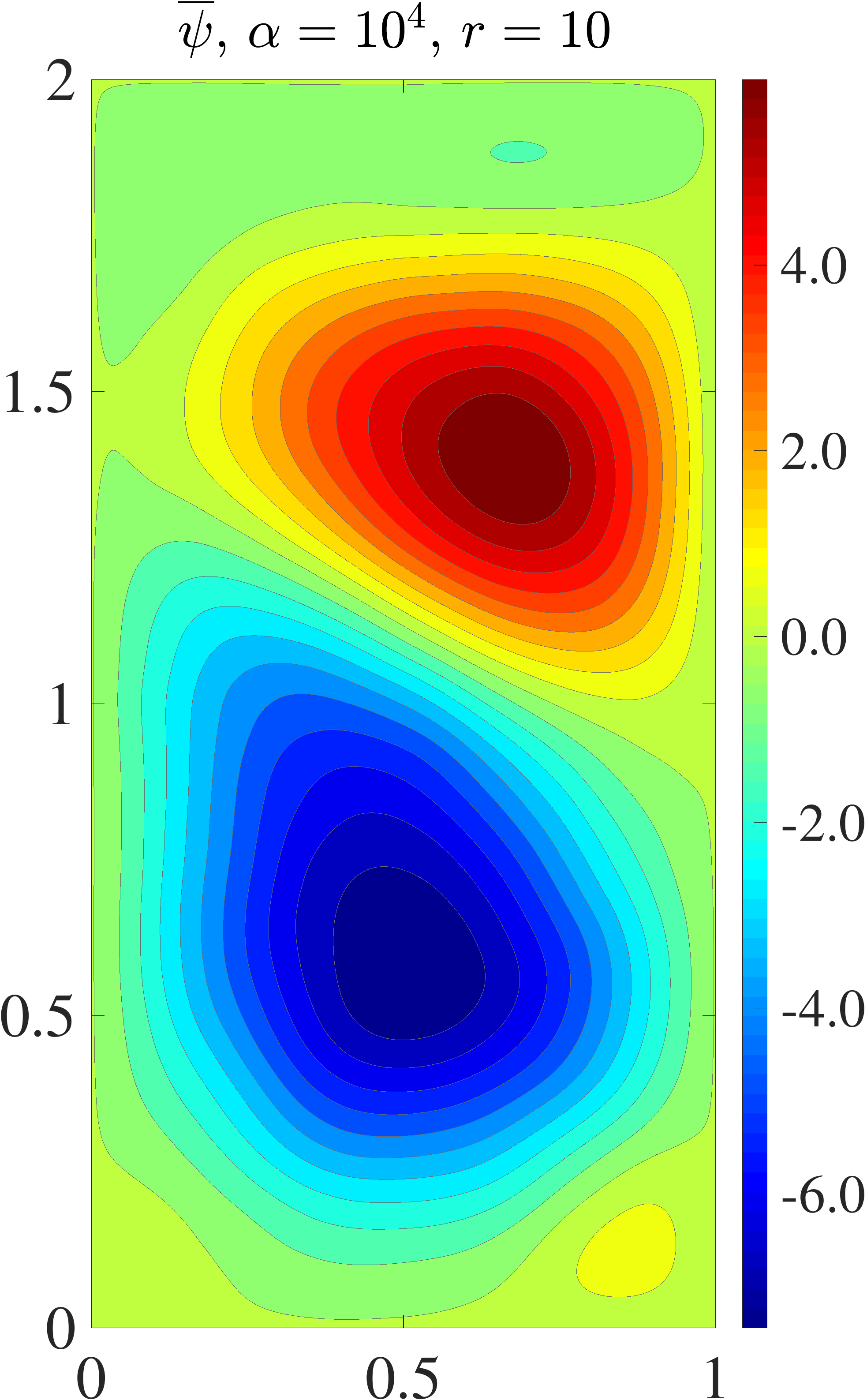}
\includegraphics[width=0.24\linewidth]{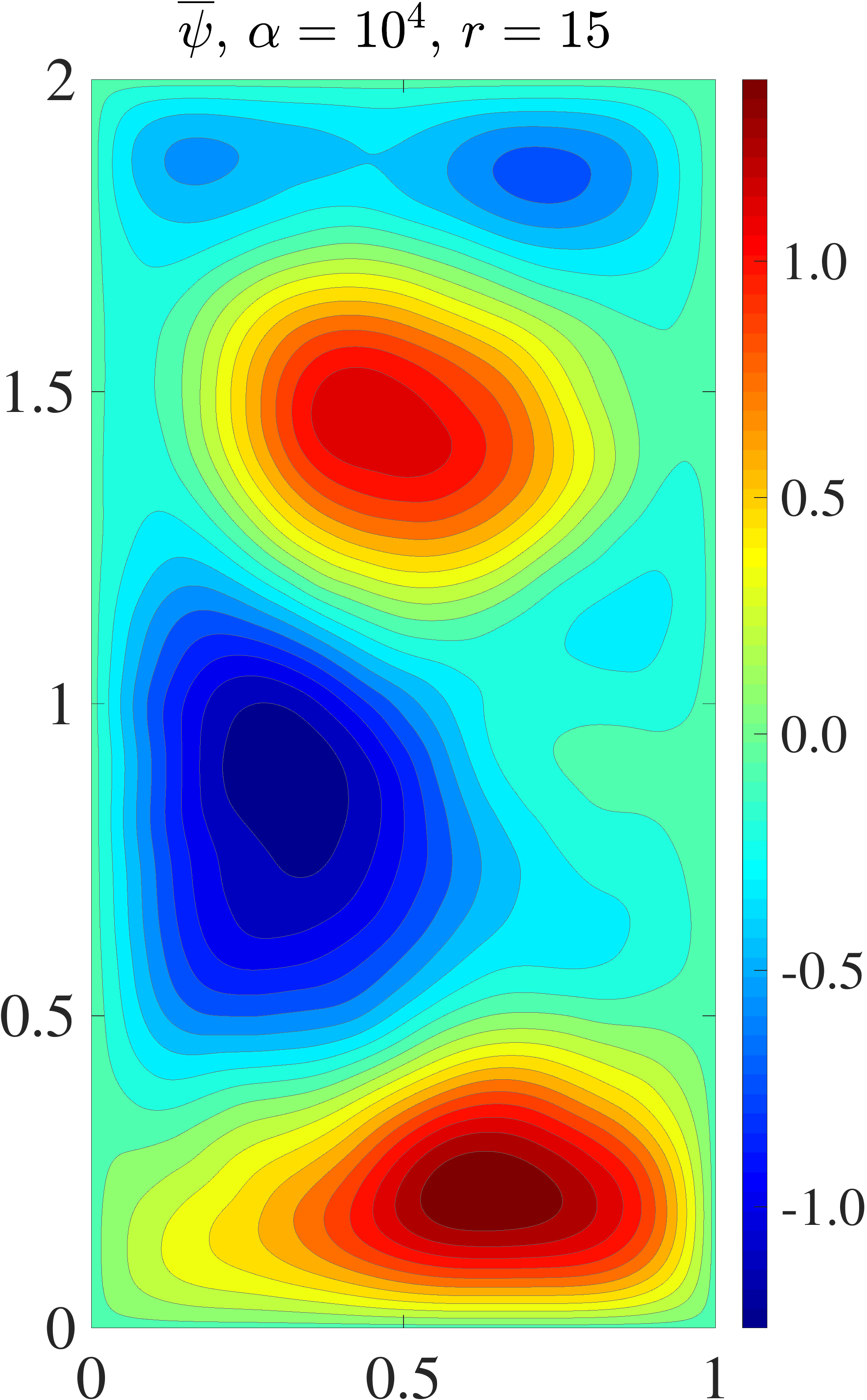}
\includegraphics[width=0.24\linewidth]{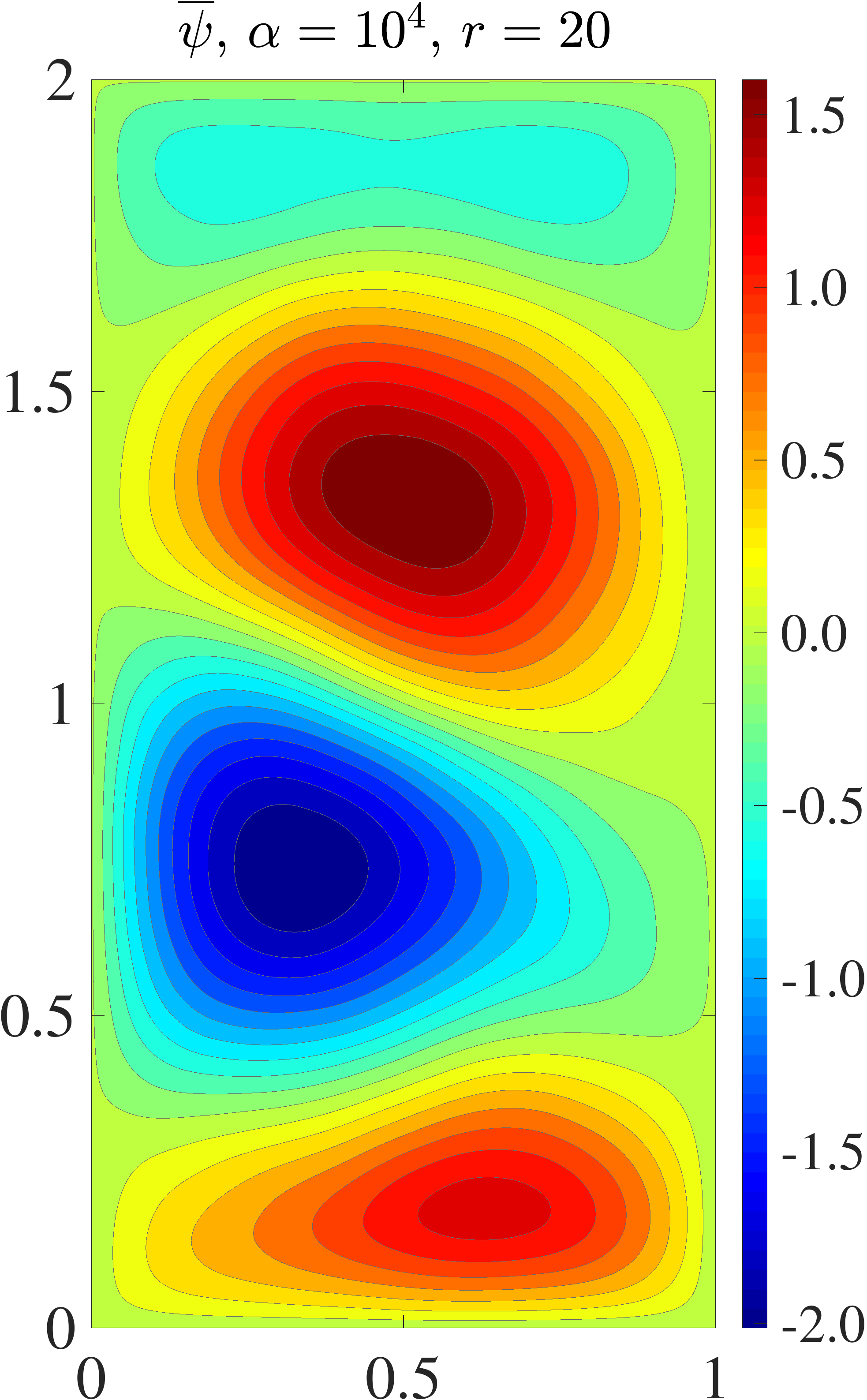}
\includegraphics[width=0.24\linewidth]{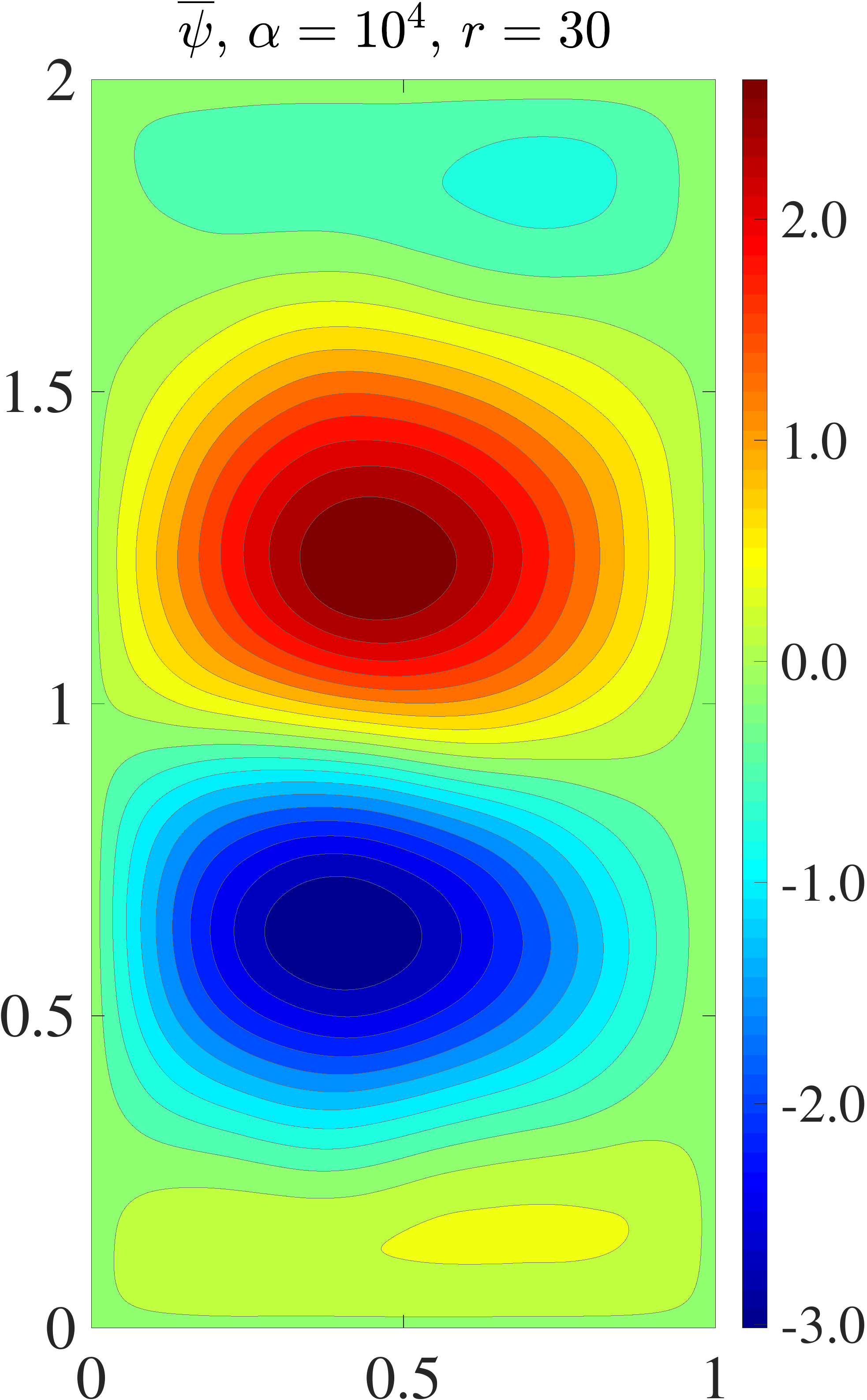}

\caption{
    Eulerian investigation, predictive regime:
	Mean streamfunction from 
	E-ROM (first row), 
	$\lambda$-ROM (second row)
	$\alpha$-ROM with $\alpha=1$ (third row), and
	$\alpha=10^{4}$ (fourth row), 
	for $r=10$ (first column), 
	$r=15$ (second column), 
	$r=20$ (third column), and 
	$r=30$ (fourth column).
	}
 \label{fig:mean-psi-meanrom-predictive}
 \end{figure}

Next, we use the qualitative Eulerian criterion to investigate the ability of the ROMs to  {\it predict} the four-gyre pattern of the time average of the streamfunction. 
In Fig.~\ref{fig:mean-psi-meanrom-predictive}, for $r=10, 15, 20$, and $30$, we plot the mean streamfunction for E-ROM, $\lambda$-ROM, and $\alpha$-ROM with $\alpha = 1$ and $\alpha = 10^{4}$. 
These results yield the following conclusions:
The E-ROM, the $\lambda$-ROM, and the $\alpha$-ROM with a low $\alpha$ value (i.e., $\alpha = 1$) yield similar results. 
These ROMs cannot recover the four-gyre pattern for $r=10,15$, and $20$, although they can capture the four-gyre pattern for $r=30$.
However, the $\alpha$-ROM with a large $\alpha$ value (i.e., $\alpha = 10^{4}$) yields dramatically better results. This ROM can clearly capture the four-gyre pattern not only for $r=30$, but also for $r=30$; for $r=15$, the pattern is somewhat captured, although not as clearly as for $r=20$ and $r=30$; finally, for $r=10$, only hints of the four-gyre pattern are present.
Overall, the plots in Fig.~\ref{fig:mean-psi-meanrom-predictive} show that the {\it Lagrangian $\alpha$-ROM with a large $\alpha$ value} (i.e., $\alpha = 10^{4}$) {\it can capture the four-gyre pattern}, whereas the {\it standard E-ROM and the Lagrangian $\lambda$-ROM cannot}.

\bigskip 

The results in Table~\ref{table:error-psi-bar-predictive} and Fig.~\ref{fig:mean-psi-meanrom-predictive} consistently show that, in the predictive regime, the {\it new Lagrangian $\alpha$-ROM with large $\alpha$ values outperforms the standard E-ROM and the Lagrangian $\lambda$-ROM} with respect to the two Eulerian metrics used in this section.
These results also support the conclusion from the reconstructive regime, i.e.,  
{\it the higher the Lagrangian data contribution} (i.e., the higher the $\alpha$ value), {\it the more accurate the results}.

\clearpage

\subsubsection{Lagrangian Investigation}
	\label{sec:lagrangian-investigation}

In this section, we perform a Lagrangian investigation of the accuracy of the two Lagrangian ROMs (i.e., the $\alpha$-ROM and $\lambda$-ROM).
We follow the same format as that used in the Eulerian investigation in Section~\ref{sec:eulerian-investigation}.

\paragraph{Reconstructive Regime:}
In Table~\ref{table:lagrangian-2}, we list the $L^2$ norm of the errors in the time-averaged FTLE~\eqref{eqn:norm-time-averaged-ftle} for E-ROM, $\lambda$-ROM, and $\alpha$-ROM for different $r$ values.
The results in Table~\ref{table:lagrangian-2} show that the $\alpha$-ROM with high $\alpha$ values (i.e., $\alpha=10^{3}$ and $\alpha=10^{4}$) consistently outperform the $\lambda$-ROM and the E-ROM for all $r$ values, but especially for the small $r$ values.
We also note that the relatively high magnitudes of the errors in Table~\ref{table:lagrangian-2} are due to the errors in the ROM velocity field approximations.
Decreasing the magnitude of the errors in the ROM velocity field approximations (e.g., by increasing the ROM dimension, $r$) would probably decrease the magnitude of the errors in the FTLE field approximation in Table~\ref{table:lagrangian-2}.


\begin{table}[H]
\centering
\caption{
    Lagrangian investigation, reconstructive regime:
	$L^2$ norm of the errors in the time-averaged FTLE~\eqref{eqn:norm-time-averaged-ftle} for E-ROM (second column) $\lambda$-ROM (third column), and $\alpha$-ROM for $\alpha=1$ (fourth column), $\alpha=10$ (fifth column), $\alpha=10^{2}$ (sixth column), $\alpha=10^{3}$ (seventh column), and $\alpha=10^{4}$ (eighth column). 	\label{table:lagrangian-2}
 	}
\begin{tabular}{|c|c|c|c|c|c|c|c|c|}
\hline
$r$  &E-ROM		&$\lambda$-ROM  &$\alpha=1$ & $\alpha=10$& $\alpha=10^2$& $\alpha=10^3$& $\alpha=10^4$ \\ \hline             
$10$	&6.0e+03&5.9e+03&5.4e+03&4.2e+03&5.2e+03&4.8e+01&2.5e+01	 \\ \hline
$15$	&4.6e+02&1.2e+03&1.2e+02&8.3e+01&7.8e+01&3.2e+01&6.1e+01\\ \hline
$20$	&9.8e+01&7.4e+01&8.5e+01&5.9e+01&2.2e+01&1.2e+01&1.0e+01 \\ \hline
$25$	&8.2e+01&6.2e+01&8.5e+01&7.6e+01&1.8e+01&1.0e+01&9.5e+00\\ \hline
$30$	&9.4e+01&8.7e+01&9.2e+01&7.0e+01&1.8e+01&1.2e+01&8.3e+00 \\ \hline
$35$	&7.8e+01&5.8e+01&7.6e+01& 5.9e+01&1.3e+01&1.1e+01&1.2e+01\\ \hline
$40$	&7.5e+01&5.3e+01&5.8e+01&5.7e+01&1.3e+01&2.0e+01&2.6e+01\\ \hline
$45$	&5.9e+01&5.0e+01&5.5e+01&4.9e+01&1.4e+01&2.4e+01&2.4e+01\\ \hline
$50$	&4.8e+01&4.4e+01&4.6e+01&3.8e+01&1.1e+01&1.4e+01&2.2e+01\\ 
\hline    
\end{tabular}
\end{table}

\paragraph{Predictive Regime:}
In Table~\ref{table:lagrangian-3}, we list the $L^2$ norm of the errors in the time-averaged FTLE~\eqref{eqn:norm-time-averaged-ftle} for E-ROM, $\lambda$-ROM, and $\alpha$-ROM for different $r$ values.
The results in Table~\ref{table:lagrangian-3} show that, as in the reconstructive regime, the $\alpha$-ROM with high $\alpha$ values (i.e., $\alpha=10^{3}$ and $\alpha=10^{4}$) consistently outperforms the $\lambda$-ROM and the E-ROM for all $r$ values.
\begin{table}[H]
\centering
\caption{    
    Lagrangian investigation, predictive regime:
$L^2$ norm of the errors in the time-averaged FTLE~\eqref{eqn:norm-time-averaged-ftle} for E-ROM (second column) $\lambda$-ROM (third column), and $\alpha$-ROM for $\alpha=1$ (fourth column), $\alpha=10$ (fifth column), $\alpha=10^{2}$ (sixth column), $\alpha=10^{3}$ (seventh column), and $\alpha=10^{4}$ (eighth column). 	
    \label{table:lagrangian-3}
 	}
\begin{tabular}{|c|c|c|c|c|c|c|c|c|}
\hline
$r$  &E-ROM		&$\lambda$-ROM  &$\alpha=1$ & $\alpha=10$& $\alpha=10^2$& $\alpha=10^3$& $\alpha=10^4$ \\ \hline             
$10$	&3.6e+03&4.4e+03&3.0e+03&4.1e+03&3.0e+03&4.4e+02& 2.4e+02\\ \hline
$15$	&5.7e+02&5.2e+02&4.1e+02& 5.0e+02&1.0e+02&1.8e+01&1.2e+01\\ \hline
$20$	&1.5e+02&8.4e+01&1.5e+02&1.7e+02&7.7e+01&1.3e+01& 8.9e+00 \\ \hline
$25$	&1.2e+02& 9.2e+01&9.5e+01&1.1e+02&5.8e+01&8.9e+00& 9.3e+00\\ \hline
$30$	&7.6e+01&7.3e+01&1.2e+02&6.5e+01&2.1e+01&6.6e+00& 6.7e+00 \\ \hline
$35$	&6.7e+01&4.6e+01&5.9e+01&4.6e+01&1.6e+01&6.6e+00&5.6e+00\\ \hline
$40$	&3.1e+01&2.8e+01&2.6e+01&3.0e+01&1.1e+01&1.3e+01& 2.0e+01\\ \hline
$45$	&2.2e+01&1.8e+01&2.0e+01&1.6e+01&9.3e+00&1.5e+01&1.9e+01\\ \hline
$50$	&1.9e+01& 1.6e+01&1.8e+01&1.5e+01&9.1e+00&1.4e+01&1.5e+01\\ 
\hline    
\end{tabular}
\end{table}

\subsection{ROM Computational Efficiency}
	\label{sec:rom-computational-efficiency}

In this section, we investigate the computational efficiency of the new Lagrangian ROMs (i.e., $\alpha$-ROM and $\lambda$-ROM).

\subsubsection{Computational Environments}

We use the following computational environments:
To generate the FOM velocity fields, we run the code on one processor (and one thread) on
a Dell workstation with a 2.00 GHZ Intel Xeon CPU running on a 64-bit Linux system.
To generate the ROM velocity fields, we use one Apple laptop with a single 2.70 GHZ CPU, running on a 64-bit Macintosh operating systems.
To generate the FTLE fields, we utilize:
(i) One computing cluster composed of 5 nodes, each node comprised of dual, quad core, hyperthreaded 2.4GHz Intel Xeon E5620 CPUs (16 processor threads), 24GB RAM, and a 40Gbps InfiniBand host card and cable; and 
(ii) Five nodes at 12 threads per node, for a total of 60 threads, and 4749mb of memory for each thread.

\subsubsection{Speed-Up Factors}

The FOM CPU time has two components: 
the CPU time of generating the velocity field; and 
the CPU time of generating the FTLE field from the velocity field. 
The ROM CPU time has three components: 
the CPU time of the offline phase (i.e., the construction of ROM operators);
the CPU time of the online phase (i.e., running the ROMs to generate the velocity field); and 
the CPU time of generating the FTLE field from the velocity field.

In this section, we investigate the CPU times of the velocity computation, since this is the main target of the proposed Lagrangian ROMs.
Thus, we first investigate the ROM speed-ups in the velocity computation and then briefly discuss the CPU times in the FTLE field computation.
Furthermore, as often done in ROM investigations, we monitor only the CPU time of the online phase of the ROMs, since the offline CPU time is offset by running the ROMs in the predictive regime, i.e., for longer time intervals (as done in this paper) or for different parameters.

To compute the computational efficiency of the new Lagrangian ROMs, we compute the ROM {\it speed-up factors ($S_{f}$)}, which are defined as follows:
\begin{eqnarray}
	S_{f}
	= \frac{\text{FOM CPU time}}{\text{ROM CPU time}} \, ,
	\label{eqn:speed-up-factor}
\end{eqnarray}
where the FOM CPU time is the CPU time of generating the velocity field and the ROM CPU time is the CPU time of the ROM online phase.

In Table~\ref{table:speed-up-factors}, for different $r$ values, we list the speed-up factors~\eqref{eqn:speed-up-factor} for E-ROM, $\lambda$-ROM (second column), and $\alpha$-ROM with $\alpha = 1, \alpha=10^{2}, \alpha=10^{3}$, and $\alpha = 10^{4}$. 
These results show that the new Lagrangian ROMs and the standard Eulerian ROM are {\it more than three orders of magnitude faster} than the FOM.

\begin{table}[htp!]
\centering
\caption{
        Speed-up factors~\eqref{eqn:speed-up-factor} for velocity field computation: 
        E-ROM (second column), $\lambda$-ROM (third column), and $\alpha$-ROM for $\alpha=1$ (fourth column), $\alpha=10$ (fifth column), $\alpha=10^{2}$ (sixth column), $\alpha=10^{3}$ (seventh column), and $\alpha=10^{4}$ (eighth column). 
	\label{table:speed-up-factors}
	}
\begin{tabular}{|c|c|c|c|c|c|c|c|c|}
\hline
$r$& E-ROM& $\lambda$-ROM& $\alpha=1$& $\alpha=10$& $\alpha=10^2$& $\alpha=10^3$& $\alpha=10^4$ \\ \hline
$10$&   1.4e+04&    1.5e+04&    1.5e+04&    1.5e+04&    1.5e+04&    1.5e+04&    1.5e+04\\ \hline
$15$&   8.4e+03&    8.2e+03&    8.2e+03&    8.3e+03&    7.8e+03&    8.0e+03&    8.0e+03\\ \hline
$20$&   5.6e+03&    5.7e+03&    5.7e+03&    5.9e+03&    6.1e+03&    5.8e+03&    5.8e+03\\ \hline
$25$&   4.8e+03&    4.6e+03&    4.7e+03&    4.8e+03&    4.5e+03&    4.6e+03&    4.6e+03\\ \hline
$30$&   3.7e+03&    3.6e+03&    3.7e+03&    3.9e+03&    3.6e+03&    3.7e+03&    3.7e+03\\ \hline
$35$&   2.9e+03&    2.9e+03&    3.0e+03&    3.0e+03&    2.9e+03&    2.9e+03&    2.9e+03\\ \hline
$40$&   2.5e+03&    2.6e+03&    2.6e+03&    2.6e+03&    2.5e+03&    2.6e+03&    2.5e+03\\ \hline
$45$&   2.0e+03&    2.0e+03&    2.1e+03&    2.1e+03&    2.0e+03&    2.0e+03&    2.0e+03\\ \hline
$50$&   1.7e+03&    1.7e+03&    1.8e+03&    1.8e+03&    1.7e+03&    1.7e+03&    1.7e+03\\  \hline
\end{tabular}
\end{table}

Although the speed-up factors for the ROM velocity computation in Table~\ref{table:speed-up-factors} are the main focus of the proposed Lagrangian ROMs, we briefly comment on the CPU time of the ROM computation of the FTLE field.
Overall, the CPU time of the ROM computation of the FTLE field is generally slightly higher than the CPU time of the FOM  computation of the FTLE field, especially when relative low $r$ values are used.
We also note that, for low $r$ values, this CPU time increase is generally lower for the new Lagrangian $\alpha$-ROM with $\alpha=10^{3}$ and $\alpha=10^{4}$ than for the standard E-ROM.
We believe that the reason for this slight CPU time increase is that, as expected, the ROM velocity accuracy is lower than the FOM velocity accuracy, which results in a slight increase in the CPU time of the FTLE field computation.
We plan to investigate this in a future study.

\bigskip

To conclude, the overall CPU time of the FTLE field computation is generally several times lower for the two new Lagrangian ROMs (as well as the standard E-ROM) than for the DNS.
Indeed, the two new Lagrangian ROMs decrease the CPU time of the FOM velocity field computation by orders of magnitude and only slightly increase the CPU time of the FOM FTLE field computation.








\section{Conclusions and Outlook}
	\label{sec:conclusions}

In this paper, we proposed  Lagrangian ROMs that use new Lagrangian inner products to build the ROM basis.
In these Lagrangian inner products, Lagrangian data ``steers" the resulting Lagrangian ROM basis  toward an accurate approximation of Lagrangian quantities, whereas Eulerian data helps the Lagrangian ROM basis yield an accurate approximation of Eulerian quantities.

For complex nonlinear systems, the low-dimensional ROMs generally need to be equipped with closure models or stabilization mechanisms~\cite{mou2020dd-vms-rom,mou2020data,xie2018data}. 
We emphasize, however, that we studied the new Lagrangian ROMs without any closure model (a challenging test) in order to separate the ROM closure problem from the ROM basis generation, which is the main focus of our paper.

We investigated the new Lagrangian ROMs in the numerical simulation of the QGE.
We considered both the reconstructive regime (in which the ROM is validated on the same time interval as the time interval used to train the ROM) and the predictive regime (in which the ROM is trained on a short time interval and validated on a longer time interval).
In both the reconstructive and predictive regimes, we showed that the new Lagrangian ROMs numerical accuracies are {\it orders of magnitude} higher than the standard Eulerian ROM accuracy in approximating both Eulerian fields (i.e., the velocity field) and Lagrangian fields (i.e., the FTLE field). 
We emphasize that, since the new Lagrangian ROMs did not employ any closure modeling, the dramatic increase in the new Lagrangian ROMs' accuracy is entirely due to the new Lagrangian inner products used to build the Lagrangian ROM basis.
Furthermore, we showed that, for the velocity field computations, the online CPU times of the new Lagrangian ROMs are orders of magnitude lower than the CPU time of the corresponding FOM.

There are numerous research directions that could provide improvements both in the efficiency and the accuracy of the new Lagrangian ROMs.
Probably the most important research avenue is the investigation of ROM closure models for the new Lagrangian ROMs.
Indeed, the new Lagrangian ROMs improved the standard Eulerian ROM's accuracy solely by using a ROM basis constructed with the new Lagrangian inner products.  
We plan to further increase the accuracy of the new Lagrangian ROMs by adding ROM closure models for the effect of the discarded ROM modes, e.g., data-driven ROM closure models~\cite{mou2020dd-vms-rom,mou2020data,xie2018data} or eddy viscosity ROM closure models~\cite{san2015stabilized}.
Another potential research direction is finding the optimal $\alpha$ value in the new Lagrangian $\alpha$-ROM.
Although the $\alpha$-ROM with higher $\alpha$ values yielded the most accurate results in our numerical investigation, finding the {\it optimal} $\alpha$ value is still an open question.
To find this optimal parameter value, one could try to extend to the Lagrangian setting the mathematical tools developed for Eulerian ROMs~\cite{KV01,xie2018numerical}.
Another research avenue is the extension of the new Lagrangian ROMs and the novel Lagrangian inner products to 
the computation of other structures that characterize transport and mixing. 
For example, instead of geometric approaches (such as the FTLE field), one could approximate probabilistic measures, such as the almost invariant sets
~\cite{DeJuLoMaPaPrRoTh2005,grover2012topological}.
Finally, although the new Lagrangian ROMs dramatically reduced the computational cost of velocity field computation, we intend to explore different approaches for speeding up the FTLE field computation from available ROM velocity data.
To this end, we plan to use the new Lagrangian ROMs in conjunction with the algorithms that have been recently proposed in~\cite{nolan2019finite,serra2020search}.

\bibliographystyle{spmpsci}      

\bibliography{traian,shane,ross_refs3,rom}

\end{document}